\documentclass[aps,prr,10pt,showpacs,tightenlines,notitlepage,nofootinbib,longbibliography,twocolumn]{revtex4-2}
\usepackage[colorlinks=true, citecolor=red, urlcolor=blue]{hyperref}
\usepackage{amsmath,amsfonts,amssymb,graphicx,hyperref, enumitem, subfigure, tabularray, bbm, bm, tikz}
\usepackage{mathtools, comment}
\usepackage{color,tikz, dsfont}
\UseTblrLibrary{diagbox}

\newcommand{\gl}[1]{\textcolor{magenta}{#1}}
\newcommand{\Mypm}{\mathbin{\tikz [x=1.4ex,y=1.4ex,line width=.1ex] \draw (0.0,0) -- (1.0,0) (0.5,0.08) -- (0.5,0.92) (0.0,0.5) -- (1.0,0.5);}}%
\usepackage[normalem]{ulem}

\begin{document}
\title{Quantum reservoir computing in Jaynes-Cummings models: Nonlinear memory and time-series prediction}
\author{Sreetama Das$^{*}$, Gian Luca Giorgi, Roberta Zambrini}
\affiliation{Institute for Cross-Disciplinary Physics and Complex Systems (IFISC) UIB-CSIC, Campus Universitat Illes Balears, 07122, Palma de Mallorca, Spain.}

\begin{abstract}
We investigate quantum reservoir computing (QRC) using a hybrid qubit-boson system described by the Jaynes–Cummings (JC) Hamiltonian and its dispersive limit (DJC). These models provide high-dimensional Hilbert spaces and intrinsic nonlinear dynamics, making them powerful substrates for temporal information processing. We systematically benchmark both reservoirs through linear and nonlinear memory tasks, demonstrating that they exhibit an unusual superior nonlinear over linear memory capacity. We further test their predictive performance on the Mackey–Glass time series, a widely used benchmark for chaotic dynamics, and show comparable forecasting ability. We also investigate how memory and prediction accuracy vary with reservoir parameters, and show the role of higher-order bosonic observables and time multiplexing in enhancing expressivity, even in minimal spin–boson configurations. Our results establish JC- and DJC-based reservoirs as versatile platforms for time-series processing and as elementary units that overcome the setting of equivalent qubit pairs and offer pathways toward tunable, high-performance quantum machine learning architectures.
\end{abstract}

\maketitle

\section{Introduction}

The Jaynes-Cummings (JC) model~\cite{jcmodel}
is a cornerstone of quantum optics, providing a minimal yet powerful framework for describing the interaction between a two-level atom and a quantized electromagnetic field. It has shaped our understanding of coherence and entanglement while driving advances across quantum optics, quantum information processing, atomic physics, and solid-state physics~\cite{Shore01071993, jcreview_2024}. The first experimental demonstration of the JC model was achieved using Rydberg atoms interacting with a single-mode optical cavity~\cite{meschede85}, giving rise to the field of cavity QED~\cite{dutra2005cavity}. Subsequent realization emerged in trapped ion systems~\cite{blockley_1992, monroe_1995}, until recently, when the progress in superconducting quantum circuits gave rise to circuit QED~\cite{Nakamura1999}, allowing the interaction of a superconducting qubit with a resonator~\cite{blais_2004, Wallraff2004, blais_2009, blais_2021}. While the JC model underpins many established quantum technologies,
its potential for emerging areas such as machine learning analog architectures and 
in neuromorphic computing ~\cite{markovic2020,mujal_2021,labay2024} has been reported only recently in pioneering experiments \cite{mcmahon2024,carles2025}. In this work, we investigate its utility for quantum reservoir computing (QRC), assessing both memory and forecasting capabilities across different operation regimes.

Reservoir computing is a supervised machine learning framework for temporal data processing~\cite{jaeger_2004, maass_2002, Verstraeten2007, TANAKA2019}, where inputs drive a complex dynamical system and are mapped into a high-dimensional space. A linear readout layer is then easily trained to target one or multiple tasks.
Crucially, the reservoir's internal parameters remain fixed, enabling a significantly simpler training process than traditional feedforward neural networks. 
In addition to its minimal training overhead, reservoir computing harnesses the intrinsic dynamical properties of physical substrates—such as optical, electronic, mechanical, and spintronic systems—for in-memory information processing and has recently been extended to operate within the quantum regime ~\cite{fuji_qrc,mujal_2021}. 
Exploiting the exponentially large Hilbert space, quantum reservoirs provide far greater degrees of freedom than classical counterparts with the same number of physical units. They are naturally suited for quantum inputs, avoiding the overhead of classical encoding \gl{\cite{Nokkala2023,Nokkala_2024}}, allowing for efficient processing of both classical and quantum data. 
Furthermore, due to the simple training process, QRC does not suffer from major training challenges encountered in variational machine learning protocols ~\cite{mcclean_barren_2018,Anschuetz2022}.

\begin{figure*}
    \centering
\includegraphics[width=0.85\textwidth]{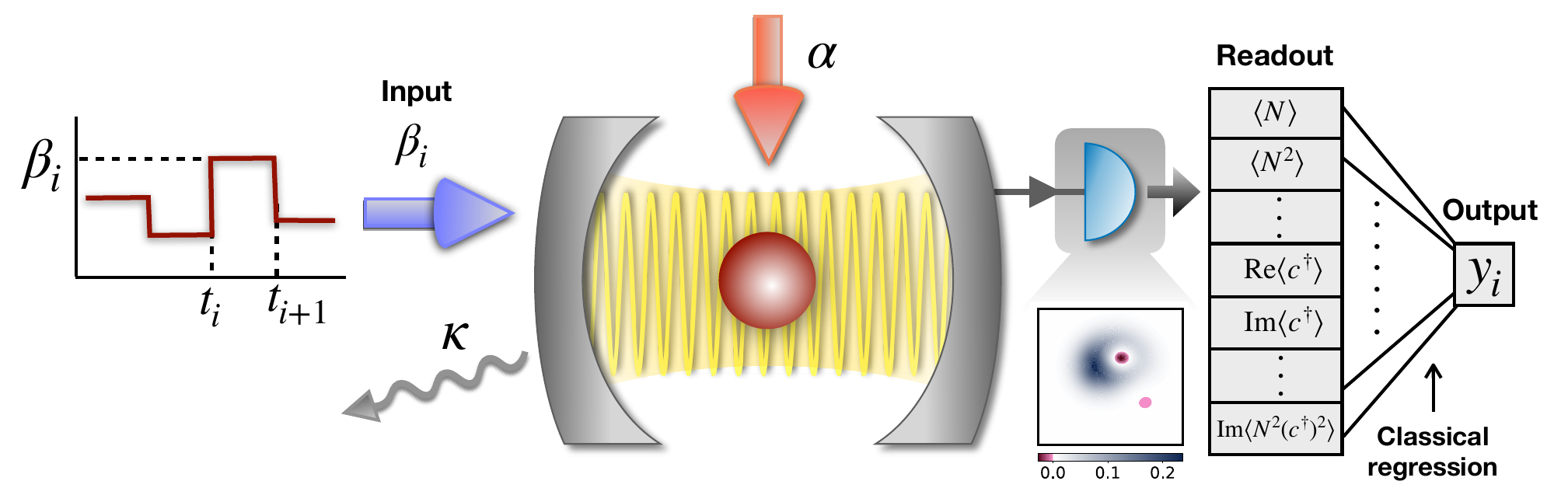}
    \caption{The physical setup for quantum reservoir computing using the JC system. The reservoir is constituted of a qubit (here shown as an atom) interacting with a single bosonic mode inside a cavity. The discrete input time-series $\{\beta_{i}\}$ is encoded in the amplitude of the cavity driving field, which is a sequence of square pulses of amplitude $\beta_{i}\equiv\beta(t_{i})$ and duration $dt$.  The qubit is driven by a classical field with fixed amplitude $\alpha$. The bosonic mode undergoes photon loss at a rate $\kappa$. The field reflected from the cavity is measured using a detector. The observables, in our case, the higher-order moments of the bosonic operators, are mapped to the target output $y_{i}$ using a linear regression. We show the Wigner distribution of the bosonic state $\rho^{b}$ in the output corresponding to a particular input in the Jaynes-Cummings reservoir. The magenta color denotes negative values of the Wigner distribution, which is a signature of nonclassicality.}
    \label{setup}
\end{figure*}

Following the seminal proposal~\cite{fuji_qrc}, a large number of works explored QRC in qubit networks ~\cite{nakajima_2019, martinez-pena_2021, Martínez-Peña2023, nakajima_2021, mujal_2023, niclas_2023, sannia_2024, li2025, hou2025}. 
The framework was later generalized to networks of quantum harmonic oscillators using Gaussian states~\cite{Nokkala2021}, followed by a variety of proposals in bosonic and fermionic systems~\cite{govia_2021, khan2021, kalfus_2022, jorge_2023, Dudas2023, llodra_2023, sannia_2025}. Pioneering implementations of QRC are based on superconducting circuits~\cite{chen_2020, kobayashi_2024,Hu_NatComm_2024}, photonic \cite{selimović2025, paparelle2025}, and atomic \cite{kornjača2024} platforms. 
The potential to enrich the input processing capabilities and to provide certain advantages over single-substrate reservoirs has started to be explored in spin-boson systems \cite{mcmahon2024,electronics2024,zhu2025,carles2025}. Considering, for example, the minimal reservoir composed of two interacting qubits, the number of features is too small to realize a complex nonlinear transformation of the inputs. Similarly, when considering two bosons in Gaussian states, the features (averages and covariance) are limited. On the other hand, replacing one of the qubits with a bosonic mode in the JC models leads to non-Gaussian effects, which are expected to have the potential to enhance the QRC expressivity and performance, as also reported for non-linear bosonic systems ~\cite{govia_2021, llodra_2025}.
A hybrid atom-field system offers rich and complex input transformation and a large set of independent observables. Furthermore, 
in most experimental platforms, the bosonic mode 
provides convenient control and readout capabilities. 
The inputs can be encoded via a classical drive, and outputs are extracted through measurements of the mode. 
Microwave signal classification has been reported in a superconducting qubit-oscillator circuit, modeled by the JC interaction in the dispersive regime~\cite{mcmahon2024}. More recently, superconducting qubit-oscillator systems have been experimentally benchmarked for traditional time-series processing, i.e., Mackey-Glass prediction and sine-square waveform classification, accounting also for an additional Kerr nonlinearity. 
In both these experimental works, the reservoir works in the dispersive regime of the JC Hamiltonian. 
While these studies indicate the suitability of these systems as reservoir computers, we note that the quantitative assessment of the linear and nonlinear memory of the reservoir in different regimes remains an open question. The performance of the reservoir is indeed  determined by the underlying dynamics of the observables, which crucially depends on the JC model operation regime and on a large number of system parameters.

In this paper, we present a systematic study of quantum reservoir computing using a qubit-boson system,  considering two distinct scenarios in which the reservoir is modeled using the JC Hamiltonian and the dispersive limit (DJC).
To construct the output layer, we use the expectation values of higher-order bosonic mode quadratures as a means to enhance the performance even in a minimum setting of one spin and one boson. Through the study of standard linear and nonlinear memory benchmarking tasks, we quantify the memory of the reservoir. We find that the reservoir has a better nonlinear memory of the input compared to linear memory benchmarking operation in different parameter regimes.
We further evaluate the reservoir's performance for the autonomous generation as well as forecasting of the Mackey-Glass task, finding no significant difference between the optimum performance of the two reservoir models. For autonomous generation, the dependence of the performance on the reservoir parameters is imperceptible. In contrast, in the delay forecasting task, we observe a clear parameter-dependent behavior, allowing us to identify the regions in parameter space that yield optimal predictive performance.

The article is organized as follows. In sec. \ref{section2}, we present the Hamiltonians for the JC and the DJC model, in the presence of an external drive. Following this, we present our scheme for QRC using these Hamiltonians. In sec. \ref{section3} we present our results, and in sec. \ref{section4} we present the conclusion and discussions.

\section{JC model for QRC}
\label{section2}
\subsection{Hamiltonians of the reservoir}
\label{subsec:hamilt}

The reservoir substrate studied in this work is a qubit with transition frequency $\omega_{a}$ interacting with a single bosonic mode of frequency $\omega_{b}$, the interaction strength being $\chi$. For generality, we adopt the term ``qubit" throughout this work, as the Jaynes-Cummings model can be realized with diverse physical platforms, including both natural and artificial two-level systems. For the purpose of illustration, in Fig. \ref{setup}, we represent the two-level system as an atom interacting with a cavity mode. The bosonic field undergoes continuous photon loss with decay rate $\kappa$. Depending on the interrelation between $\omega_{a},\, \omega_{b},\, \chi$, and $\kappa$, the dynamics can be modeled using different Hamiltonians. In this work, we address two well-known Hamiltonians, namely, the \textit{Jaynes-Cummings} model and the \textit{dispersive Jaynes-Cummings} model approximating the large detuning regime, benchmarking their respective best performance as reservoirs for QRC.

Classical driving fields encode the input in the QRC cavity, with carrier frequency $\omega_{1}$ and time-dependent amplitude $\beta(t)$  (see, for instance, the coherent microwave signal drive in  the experiments of Refs. \cite{mcmahon2024,carles2025}).
Additionally, we consider in some cases also a classical field with frequency $\omega_2$ and amplitude $\alpha $ driving the qubit. We note that the qubit drive is introduced to tune the system operation but not as an input injection. Therefore, we restrict ourselves to the simple scenario of a constant amplitude $\alpha$ instead of any complex time-dependent modulation. As we will see in the next section, whether this drive results in a meaningful advantage for QRC, will depend on the reservoir Hamiltonian.
\\ \\
\textbf{(i) Jaynes-Cummings model:} 
The JC Hamiltonian models a qubit-boson system in the \textit{strong coupling limit}, i.e., when  $\chi >\kappa$ and when the rotating wave approximation holds, i.e. $\vert \omega_{a}-\omega_{b}\vert \ll \vert\omega_{a}+\omega_{b}\vert$ and
$\chi\ll\omega_{a}, \omega_{b}$~\cite{jcreview_2024}. Considering $\hbar=1$ throughout this work, the JC Hamiltonian can be written as 
\begin{equation}
    H_{\mathrm{JC}} = \frac{\omega_{a}}{2} \sigma^{Z} + \omega_{b} c^{\dagger}c + \chi (c\sigma^{+} + c^{\dagger}\sigma^{-}).
    \label{jc_hamilt0}
\end{equation}
Here  $c$ ($c^{\dagger})$ is the bosonic annihilation (creation) operator, $\sigma^{Z}\equiv \vert e\rangle\langle e\vert-\vert g\rangle\langle g\vert$ is the Pauli-Z operator, $\sigma^{+}$ ($\sigma^{-}$) is the spin raising (lowering) operator, defined as $\sigma^{+} = \vert e\rangle\langle g\vert$ ($\sigma^{-} = \vert g\rangle\langle e\vert$); $\vert g\rangle$ and $\vert e\rangle$ are respectively the ground and the excited state of the qubit. The difference between the qubit and bosonic energies, i.e., $\delta = \omega_{a} - \omega_{b}$ is the detuning of the system. 
The cavity and qubit driving Hamiltonians can be expressed as

    \begin{eqnarray}
    &H^{b}_{d} = i\beta(t)(ce^{i\omega_{1} t} - c^{\dagger}e^{-i\omega_{1} t}),\\ \nonumber 
    &H^{a}_{d} = \alpha (\sigma^{+}e^{-i\omega_{2} t} + \sigma^{-}e^{-i\omega_{2} t}),
    \label{drive_hamilt}
\end{eqnarray}
with $\beta \in \mathbb{R}$ and $ \alpha \in \mathbb{R}$. For JC reservoir, we specifically consider the two driving fields to have the same frequency, i.e. $\omega_{1}=\omega_{2}=\omega$. In the rotating frame of the driving fields, the full Hamiltonian of the JC model is,
\begin{align}\label{jch}
    H_{\mathrm{JC}}^{\prime} &= \Delta_{a}\sigma^{Z} + \Delta_{b}c^{\dagger}c + \chi (c\sigma^{+} + c^{\dagger}\sigma^{-}) \\ \nonumber
    &+ i\beta(t)(c - c^{\dagger}) + \alpha(\sigma^{+} + \sigma^{-})\\ \nonumber
    &= (\Delta_{b} + \Delta)\sigma^{Z} + \Delta_{b}c^{\dagger}c + \chi (c\sigma^{+} + c^{\dagger}\sigma^{-})\\ \nonumber
    &+ i\beta(t)(c - c^{\dagger}) + \alpha(\sigma^{+} + \sigma^{-})
\end{align}
where $\Delta_{a} = \frac{1}{2}(\omega_{a}-\omega)$, $\Delta_{b} = \omega_{b}-\omega$ and $\Delta = \Delta_{a} - \Delta_{b}$. 
\\ \\
\textbf{(ii) Dispersive Jaynes-Cummings model:} In the limit of large qubit-boson detuning i.e. $\delta \gg \chi$, the energy exchange between qubit and boson can be neglected, and by applying Schrieffer-Wolff transformation~\cite{blais_2021, jcreview_2024} to Eq. \ref{jc_hamilt0}, the JC model can be approximated as~\cite{CARBONARO197997},
    \begin{equation}
        H_{D} \approx \frac{\omega_{a}^{\prime}}{2}\sigma^{Z} + \omega_{b} c^{\dagger}c + \chi^{\prime} c^{\dagger}c\sigma^{Z},
        \label{disp_int}
    \end{equation}
where $\omega_{a}^{\prime} = \omega_{a} + \chi^{\prime}$ and $\chi^{\prime} = \chi^{2}/\delta$. 
The Hamiltonian in Eq. \ref{disp_int} is called the dispersive JC (DJC) Hamiltonian. It is diagonal in the composite basis $\{|ng\rangle, |ne\rangle\}$, where $\{|n\rangle\}$ are the number states. It is of particular importance in cavity/circuit QED since in this regime, one can perform a non-demolition measurement of the state of the qubit by measuring the bosonic mode number operator $c^{\dagger}c$~\cite{Wallraff2004}. For our reservoir, we focus on the resonant regime, where the driving frequencies are set equal to the bare qubit and bosonic mode frequencies i.e. $\omega_{b}=\omega_{1} $ and $\omega^{\prime}_{a}=\omega_{2} $ respectively. In this case, going to the rotating frame of the driving fields, we get the DJC reservoir Hamiltonian as the following:
\begin{equation}
    H_{D}^{\prime}= \chi^{\prime} c^{\dagger}c\sigma^{Z} + i\beta(t)(c - c^{\dagger}) + \alpha(\sigma^{+} + \sigma^{-}).
    \label{djch}
\end{equation}
See Appendix \ref{derivation_djc} for a detailed derivation of Eq. \ref{djch}, and the various conditions leading to it. The driving Hamiltonians introduce transitions between different basis states and create entanglement between the qubit and the bosonic mode.

For both models, we consider a simple dissipation scenario in which the presence of (cavity) losses in the bosonic mode is accounted for by considering $\sqrt{\kappa}c$ jump operators, which leads to
the following Lindblad master equation,
    \begin{equation}
        \frac{d\rho}{dt} = -i [H, \rho] + \kappa\bigg[ c\rho c^{\dagger} - \frac{1}{2}(c^{\dagger}c\rho + \rho c^{\dagger}c )\bigg].
        \label{master_eq}
    \end{equation}
For the JC reservoir $H = H_{\mathrm{JC}}^{\prime}$, while for the DJC reservoir $H = H_{D}^{\prime}$. 
Here we have assumed the dissipation acting locally on the bosonic mode. As discussed in several previous works~\cite{Trushechkin_2016, Cattaneo_2019, cattaneo_2025}, a master equation with local dissipation channels is a valid approximation in the limit $\chi \ll \omega_{a}, \omega_{b}$, which holds true for our reservoir parameters. Furthermore, our master equation accurately captures the system's dynamics under the assumption that the qubit's intrinsic energy relaxation rate is negligible compared to the cavity decay rate $\kappa$. This approximation is well justified in numerous cavity QED experiments ~\cite{scala_2007, Scala_2007_2}, ensuring the theoretical validity of our model for such platforms. However, in a more general scenario irrespective of any particular experimental platform, the exact master equation derived from the microscopic picture should have nonlocal dissipation channels, which we briefly comment on in the Discussion section. A microscopic derivation of the master equation would also lead to the shift of the bare frequencies of the qubit and the boson. We effectively assume that these shifts have been compensated for by adjusting the natural frequencies (which are free parameters) to maintain resonance.

\subsection{Detailed setup for QRC}
A three-layer reservoir computing model consists of an input encoding layer, a high-dimensional reservoir for nonlinear dynamics, and a readout layer that extracts useful classical outputs (Fig. \ref{setup}). We consider 
a series of discrete scalar inputs  $\{s_{i}\}$ injected at different times in the evolving reservoir. 
The JC reservoir dynamics nonlinearly maps the input to a higher-dimensional output space. Besides nonlinearity, the dynamics must also satisfy the echo state and fading memory properties~\cite{jaeger:techreport2001, jaeger_2012, boyd_85}, which guarantee the independence of the initial reservoir state and of far-in-the-past input data. For each input, a fixed set of $n$ observables of the reservoir is measured at the output layer. Given $p$ inputs for training, the output $\mathbf{X}$ is a $p\times n$ dimensional array. To train the reservoir, $\mathbf{X}$ is mapped to the $p$-dimensional target output array $\mathbf{y}$ using a ridge regression (more details about the numerical simulation are provided in Appendix \ref{sec:nlevel}) with weights $\mathbf{W}$. Therefore, we can write,
\begin{equation}
    \mathbf{X}\mathbf{W}\approx \mathbf{y}.
\end{equation}
Afterward the trained weight matrix $\mathbf{W}$ can be used for predicting the test data $\mathbf{\overline{y}}$.

\begin{figure*}[t]
    \centering
    \includegraphics[width=0.99\textwidth]{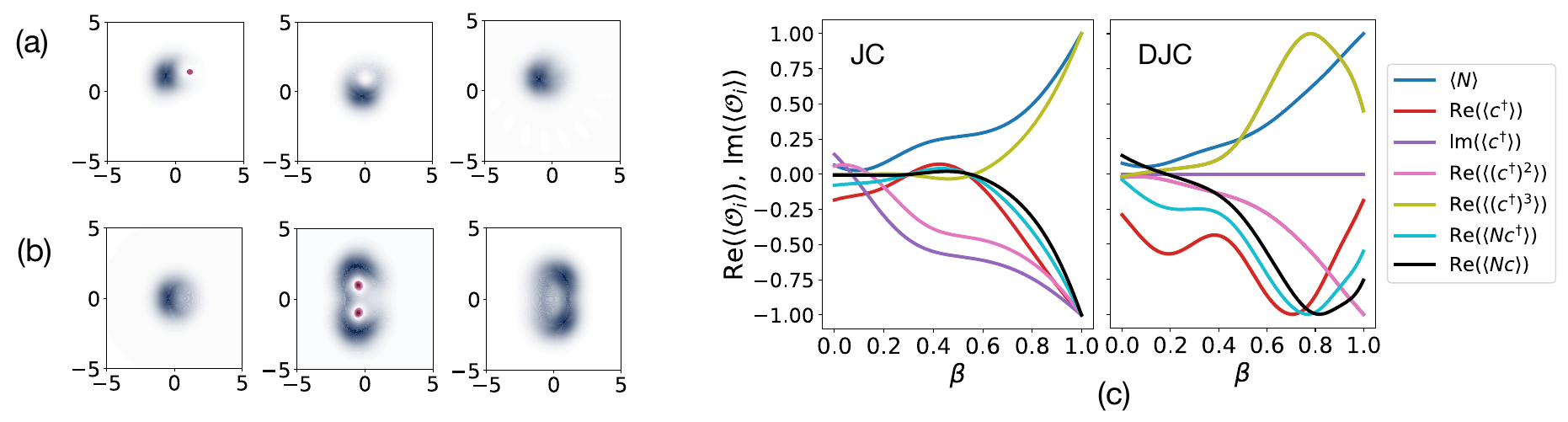}
    \caption{Wigner distribution $W(\Hat{X}, \Hat{P})$ of the bosonic reduced density matrix $\rho^{b}$ for (a) JC model and (b) DJC model, in the presence of driving and dissipation. The distribution is obtained after the washout phase for a series of uniformly sampled random inputs. The magenta color denotes negative values of $W(\Hat{X}, \Hat{P})$, which is the signature of nonclassicality of $\rho^{b}$. (c) The nonlinear response of a set of higher-order moments of the bosonic mode to input value $\beta$ for JC and DJC models. The expectation values have been scaled to lie in the range $[-1, 1]$. For all figures, $dt=10$ and $\kappa=0.1$}
    \label{wigner_nonlinearity_compare}
\end{figure*}

In our work, the input is encoded in the time-dependent amplitude $\beta(t)$ of the cavity driving field; i.e., we have $\beta(t_i) = \beta_{i} = s_{i}$. For each input, the cavity is driven during a time $dt$ with a square pulse of amplitude $\beta(t_i)$. This continuous driving strategy was shown in Ref. \cite{sannia_2024} to be more  effective than an erase-and-write map where a part of the reservoir state is first reset (erased) and then prepared (written). For both the JC and DJC reservoirs, the full dynamics is given by the master equation (\ref{master_eq}).

As for the output layer, we consider different boson mode observables. Due to the non-Gaussian nature of the bosonic state, the higher-order moments of the operators $\{c, c^{\dagger}, N\}$ can provide features that are useful for a faithful assessment of the reservoirs capability. To form the readout layer, we use both the real and imaginary parts of the expectation values of the operators,
\begin{equation}
    \mathcal{O} = \{N^{m}(c^{\dagger})^{m^{\prime}},\;N^{m}c^{m^{\prime}}\}\;\;\; \mathrm{where}\; m, m^{\prime} \geq 0,
    \label{define_obs}
\end{equation}
with $N=c^{\dagger}c$ number operator. The highest moment of $N$  we use (for $m^{\prime}=0$) is $N^{4}$, while for $c^{\dagger}$ or $c$ ($m=0$) it is $(c^{\dagger})^5$. The expectation values are linearly dependent functions of the reduced density matrix elements $\rho^{b}_{i,j}$.
They can be obtained by performing homodyne measurements of the quadrature operators $\{\hat{X}, \hat{P}\}$, and then calculating the higher-order moments.  
Our choice of readouts, triggered by the complexity of the dynamics of our models, is different from the previous works, which consider only the mode populations~\cite{Dudas2023, electronics2024}, or the low-lying moments of the qubit and bosonic observables~\cite{Nokkala2021, zhu2025}. Another recently proposed strategy is the use of the cumulative distribution
function based on the proportion of the measurement outcomes below a given threshold value~\cite{hahto2025}. A polynomial regression can be further introduced to enhance the performance in post-processing~\cite{zhu2025}. 

In all cases, we fix the total number of readouts to $n=40$. We numerically solve Eq. \ref{master_eq} to obtain the readouts, which are mapped to the target output using ridge regression. In Appendix \ref{observable2_compare}, we show the full list of operators we considered, and show their dynamical behavior for random inputs. Additionally, we present a comparison of the memory of QRC when using the elements $\rho^{b}_{i,j}$ themselves to form the readout layer. Our analysis shows that for nonlinear memory task, using the higher order bosonic operators leads to a better performance compared to using the reduced density matrix elements. For linear tasks, both sets of observables have similar performance.

In numerical simulations, the infinite Hilbert space of the bosonic mode is truncated to a finite-dimensional subspace, selected to ensure that the relevant system dynamics are faithfully preserved~\cite{kalfus_2022}. We observe that when $\kappa\leq 0.01$, the higher levels always have a small non-zero population, and it becomes numerically demanding to simulate the reservoir with a high number of bosonic levels. Therefore, we choose to work with $\kappa=0.1$, and we only work in the parameter regions in which $N=15$ levels are sufficient. Incidentally, this value of $\kappa$ is also optimum for a reasonably high truncation level of the bosonic mode, as we discuss in Appendix \ref{sec:nlevel}.

\section{Results}
\label{section3}

We will now analyze the qubit-boson dynamics and QRC performance in established tasks that allow us to benchmark different regimes. Since QRC targets complex nonlinear temporal tasks, sufficient complexity is required, and the relative performance JC models in the dispersive and resonant regimes is unknown.

The bosonic reduced density matrix $\rho^{b}$ is a highly nonlinear function of the driving $\beta$ even when we have a resonant JC model, with $\Delta_{b} = \Delta_{a} =0$  in Eq. \ref{jch}, as can be seen from the analytical study in Ref.~\cite{alsing_1992}. 
In the general driven dissipative case, where an analytical solution is difficult to find, an efficient way to visualize the complexity is through the Wigner distribution $W(\Hat{X}, \Hat{P})$ of $\rho^{b}$ in the position ($X$)-momentum ($P$) space, which we show in Fig. \ref{wigner_nonlinearity_compare}(a) and (b) for JC and DJC model respectively, for some random inputs. It is clear that the bosonic states in both cases are non-Gaussian in nature and display negativity of the W distribution, a signature of nonclassicality.

Unlike the JC model, both the free Hamiltonian and the dissipative part of the master equation for the DJC model (Eq.  \ref{disp_int}) are symmetric under the independent transformations $X \leftrightarrow -X$ and $P \leftrightarrow -P$. The driving terms, however, break these symmetries, yet the resulting Wigner distribution in Fig. 2(b) exhibits a reflection symmetry about the $X$-axis. This apparent paradox is resolved by examining the combined effect of the two drives. The dispersive interaction $\chi' c^\dagger c \sigma^z$ creates a qubit-state-dependent shift of the cavity frequency: when the qubit is in $|e\rangle$ ($|g\rangle$), the cavity experiences an effective frequency of $+\chi'$ ($-\chi'$). Considering a damped oscillator with driving  $i\beta(t)(c - c^\dagger)$ in a frame with frequencies $\pm\chi'$, one can see that the mean-field stationary solutions $\langle c_s\rangle$ for frequency (shifts) $\pm\chi'$ are actually complex conjugate (being $-\beta(\kappa\pm i\chi')/(\kappa^2-\chi'^2)$) \footnote{We also notice that these are approximately imaginary solutions for $\kappa\ll \chi'$). }
In the absence of qubit driving ($\alpha = 0$), these two scenarios would produce distinct pure states whose Wigner distributions are exact mirror images of each other with respect to the $X$-axis, but the system would be in one or the other, not both. The qubit drive $\alpha \sigma^x$ introduces  transitions between $|e\rangle$ and $|g\rangle$, entangling not trivially the qubit and cavity. Tracing out the qubit, the reduced state of the cavity becomes an incoherent mixture of the two mirror-image states described above. The overall Wigner distribution is therefore their average, which inherits and preserves the reflection symmetry about the $X$-axis. Thus, the symmetry observed in Fig. 2(b) is not a symmetry of the underlying dynamics, but rather a consequence of the statistical mixing induced by the qubit drive.

It is also interesting to look into the input dependence of the output features, as it was done in Ref.\cite{mujal_2021_b}, considering qubit networks and bosonic ones, respectively. In Fig. \ref{wigner_nonlinearity_compare}(c), we show, for both resonant and dispersive models, a rich nonlinear dependence when considering different bosonic observables for a driving amplitude $0<\beta<1$. 

We now test the linear and non-linear memory of the reservoir for a set of standard tasks, namely the short-term memory (STM) task and the parity check (PC) task. From the input series, the first $\mathcal{N}_{w}=1000$ inputs are used for the washout phase, during which the dependence on the initial reservoir state is eliminated. The next $\mathcal{N}_{\mathrm{test}}=1500$ points are used for training, and the subsequent $\mathcal{N}_{\mathrm{test}}=1000$ points are used to test the performance of the trained quantum reservoir.

\begin{figure}
    \centering
    \includegraphics[width=0.48\textwidth]{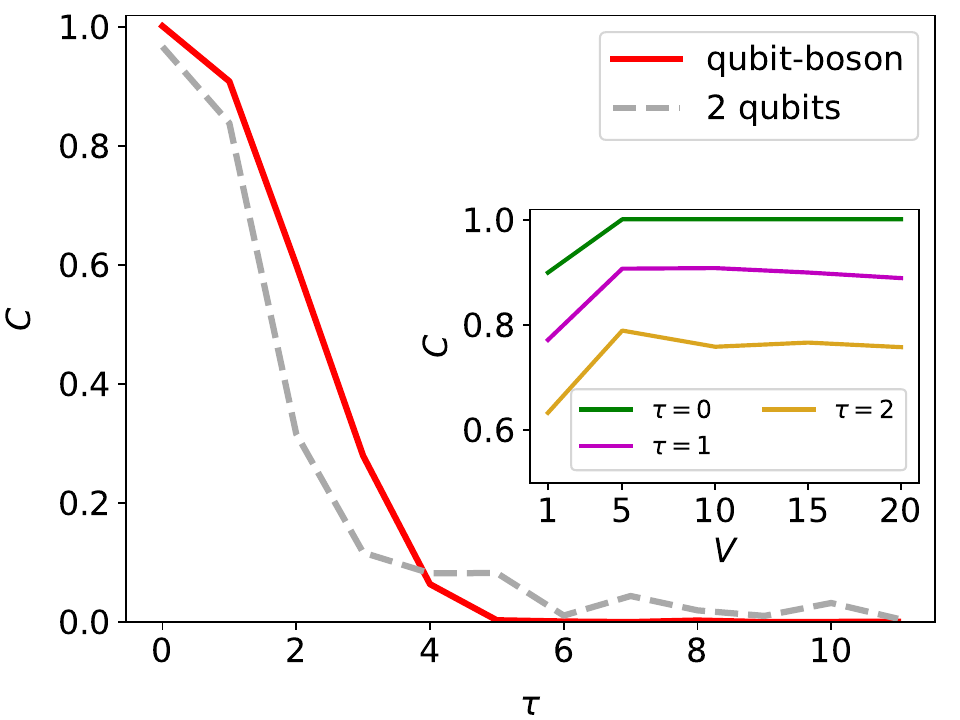}
    \caption{The memory capacity $C$ with respect to varying $\tau$ for the STM task using JC Hamiltonian (red line) when $dt=10$ and $\kappa=0.1$. The other parameters are $\Delta_{b}=1$, $\Delta=0$, $\chi=1$, and $\alpha=0$. The gray dashed line shows the capacity when using a two-qubit reservoir with the same interaction as shown in Eq. \ref{2qubit_eq2}. In the inset, we show the capacity for $\tau=0, 1, 2$ when increasing the number of virtual nodes $V$ in the readout layer. 
    }
    \label{stm_C}
\end{figure}

\subsection{Short-term memory task}
The STM task is a standard linear memory benchmarking task for reservoir computing~\cite{jaeger_2004}. The inputs $\{s_{i}\}$ are a series of uniformly generated random numbers in the range $(0, 1)$. 
The task involves reconstructing the input value at a temporal delay $\tau$, thereby providing a measure of the reservoir’s linear memory. For each random input $s_{i}$, the target output is 
\begin{equation}
    y_{i} = s_{i-\tau}.
    \label{stm_output_eqn}
\end{equation}
  
To evaluate the efficiency of memory recall, we use the memory capacity $C$ defined as the squared Pearson coefficient,
\begin{equation}
    C = \frac{\mathrm{cov}^{2}(\mathbf{y}, \overline{\mathbf{y}})}{\sigma(\mathbf{y})^{2}\sigma(\overline{\mathbf{y}})^{2}}.
    \label{capacity_def}
\end{equation}
Here $\mathbf{y}=\{y_{i}\}$ is the array of $\mathcal{N}_{\mathrm{test}}$ target outputs while $\mathbf{\overline{y}}$ denotes the array of actual predicted outputs, $\mathrm{cov}(\cdot)$ is the covariance, and $\sigma(\cdot)$ is the standard deviation. The predictive capacity can then reach the maximum value 1.

\begin{figure}
    \centering
    \includegraphics[width=0.48\textwidth]{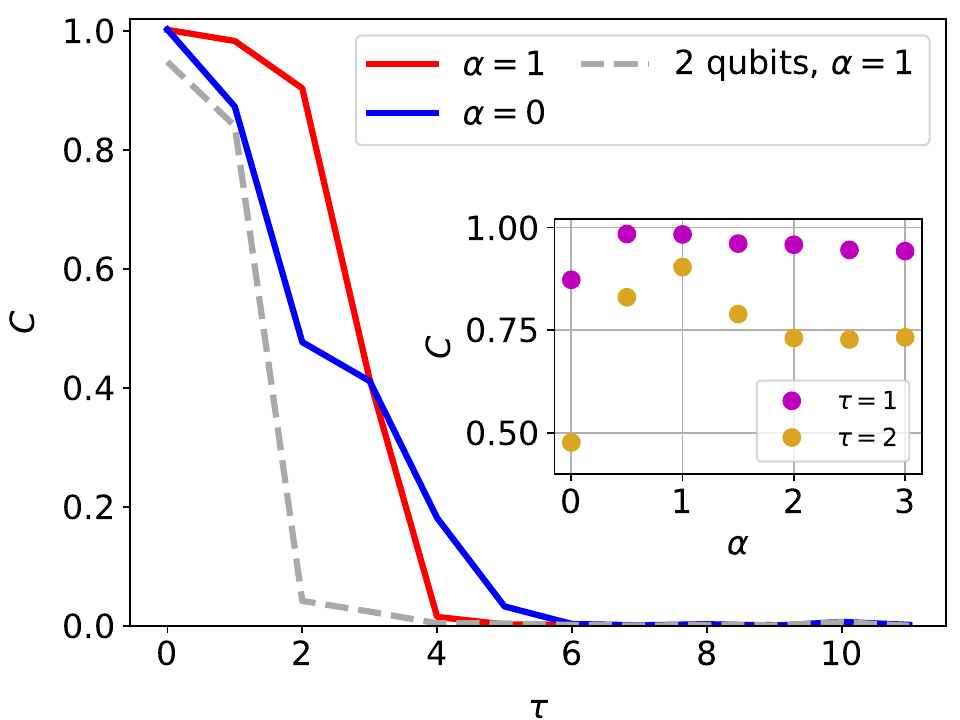}
    \caption{The $C$ vs. $\tau$ curve for the STM task using the DJC model, both with and without the qubit driving field $\alpha$. The other parameters are $dt=10$, $\kappa=0.1$, and $\chi=1$. The gray dashed line shows the capacity when using a two-qubit reservoir with the same interaction as shown in Eq. \ref{2qubit_eq3}. In the inset, we show the effect on $C$ due to varying $\alpha$, keeping the other parameters unchanged. 
    }
    \label{stm_tau_c_H3}
\end{figure}

\begin{figure*}
    \centering
    \subfigure[]{\includegraphics[width=0.48\textwidth]{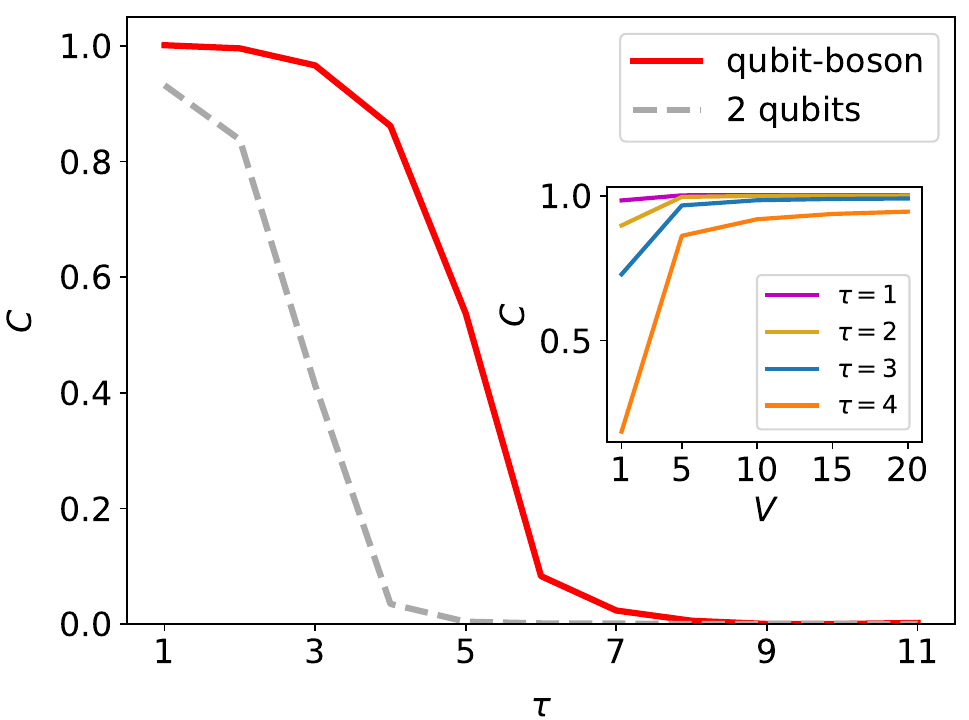}}
    \subfigure[]{\includegraphics[width=0.45\textwidth]{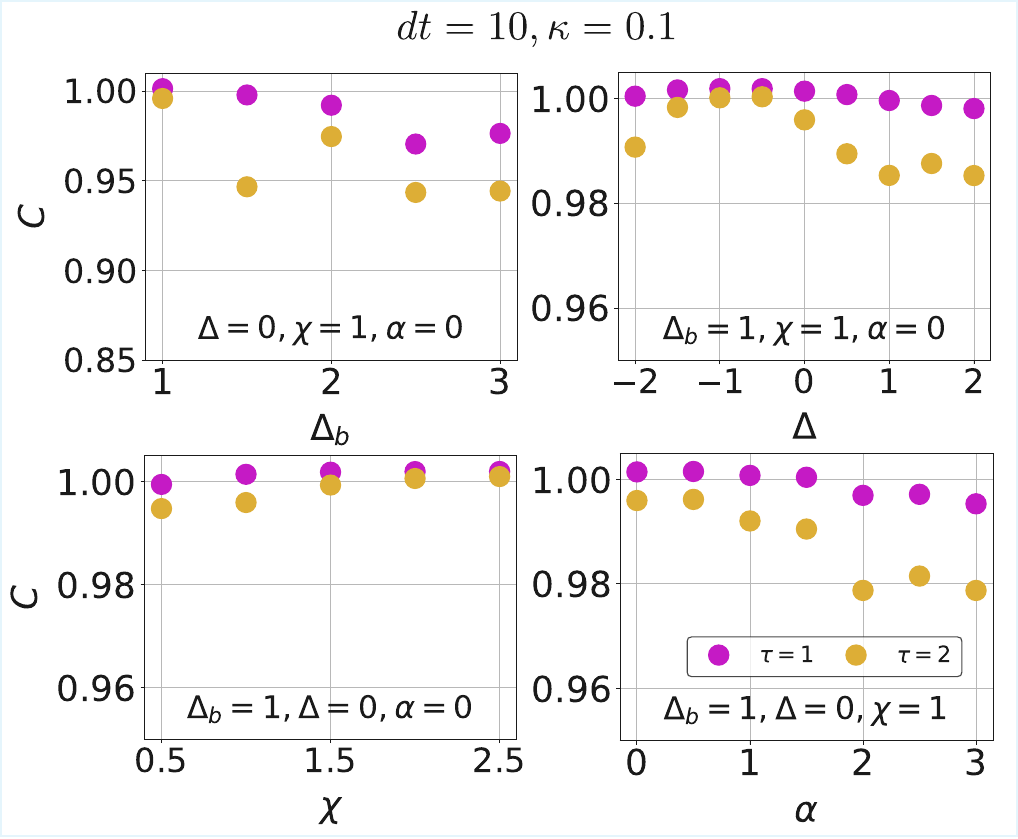}}
    \caption{(a) Capacity $C$ for PC task vs. delay $\tau$ using JC model when $dt=10$, $\kappa=0.1$ (red curve). The other parameters are the same as in Fig. \ref{stm_C}. The gray dashed line shows the capacity when using a two-qubit reservoir with similar interaction, as shown in Eq. \ref{2qubit_eq2}. In the inset, we show the capacity with respect to increasing the virtual nodes $V$ for different delays $\tau$.
    (b) The effect of varying the parameters $\Delta_{b}$, $\Delta$, $\chi$ and $\alpha$ on $C$ for $\tau=1, 2$. 
    }
    \label{H2_pc_tau_c}
\end{figure*}

For the JC model, the solution of Eq.~\ref{master_eq} is a function of the parameter set ${dt, \kappa, \Delta_{b}, \Delta, \chi, \alpha}$. While the dimensionality of this space precludes an exhaustive analysis of the QRC performance in different tasks, we find that the variation of $C$ is most strongly influenced by $dt$ and $\kappa$. Indeed, the rates of data injection and erasure (due to damping) play a prominent role in QRC \cite{sannia_2024,zhu2025}. As already mentioned before, assuming $\kappa=0.1$ guarantees both good performance and a reasonably low computational time related to the cutoff of the bosonic mode. 
By varying $dt \in \{1,10\}$ with $\kappa=0.1$, $\Delta_{b}=1$, $\Delta=0$, $\chi=1$, and $\alpha=0$, the maximum capacity is found at $dt=10$, corresponding to long driving times and moderate dissipation.

To enhance the performance, we use time-multiplexing, collecting the observable values at $V$ intermediate and equidistant points within $dt$ for one fixed input. In this case, the total number of observables at the output layer is linearly increased ($V$ times), providing more features to the linear regressor. Even a small number of virtual nodes, $V=5$, shows a  drastic improvement compared to the case when $V=1$ (no time-multiplexing). Further increase in $V$ does not significantly improve the capacity. In Fig. \ref{stm_C} (red curve), we show the trend of $C$ with respect to $\tau$ for optimum $dt$, while in the inset, we show the capacity when increasing $V$ for $\tau=0, 1, 2$.  
The capacity declines sharply with increasing $\tau$; effectively, for $\tau>2$, the reservoir's performance for the STM task is not reliable. In Appendix \ref{JC_stm_params}, we show the capacity corresponding to $\tau=1, 2$ with respect to varying the other reservoir parameters, and we can conclude that the performance for $\tau=1$ is fairly robust with respect to differing parameters, while the same varies significantly for $\tau=2$. The behavior of $C$ vs. $\tau$ shows very little dependence on the choice of the input series (we find similar results considering 10 different random input series).

\begin{figure}
    \centering
    \includegraphics[width=0.48\textwidth]{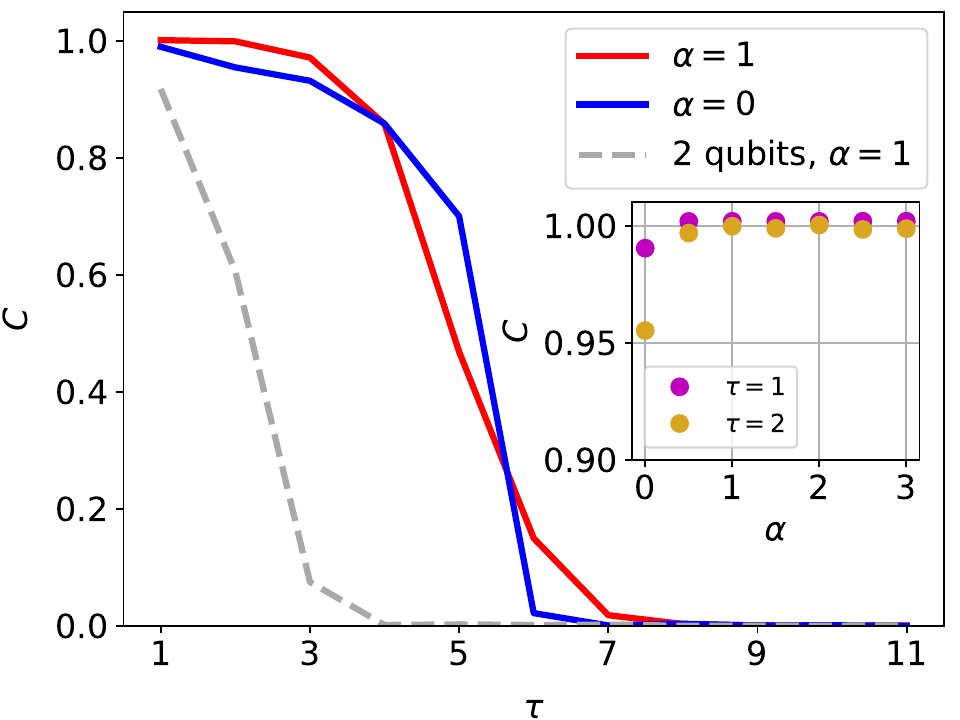}
    \caption{The $C$ vs. $\tau$ curve for PC task using DJC model, with and without the qubit driving field $\alpha$. The other parameters are the same as in Fig. \ref{stm_tau_c_H3}. The gray dashed line shows the capacity when using a two-qubit reservoir with a similar  interaction as shown in Eq. \ref{2qubit_eq3}. In the inset, we show the effect on $C$ due to varying $\alpha$, keeping the other parameters unchanged.
    }
    \label{H3_pc_tau_c}
\end{figure}

In Fig. \ref{stm_C}, we also show the capacity (gray dashed curve) when, instead of a qubit-boson reservoir, we use a two-qubit reservoir. The Hamiltonian of the two-qubit reservoir is chosen with the equivalent interaction of the JC model, as detailed in the Appendix \ref{sec:twoqubit}. Clearly, the qubit-boson reservoir has a better memory capacity; however, the difference in performance with the two-qubit reservoir is not significant in this case.

Next, we perform the STM task using the DJC Hamiltonian. In Fig. \ref{stm_tau_c_H3} we show the best $C$ versus $\tau$ curve for both $\alpha=0$ and $\alpha=1$, with $V=5$. The reservoir has fewer parameters in this case, and the optimal set of parameters is $dt=10$, $\kappa=0.1$, $\chi^{\prime}=1$, and $\alpha=1$. It is clear that the capacity improves drastically for a non-zero qubit driving amplitude $\alpha$. As explained in sec. \ref{subsec:hamilt}, a nonzero $\alpha$ creates entanglement in the otherwise separable dynamics of the DJC model. 
The gray dashed line in Fig. \ref{stm_tau_c_H3} shows the capacity from a similar two-qubit reservoir. We find that in the dispersive limit, in the absence of qubit driving, the performance of both resonant and dispersive JC is similar to what is achieved with a pair of qubits, while in the presence of qubit driving ($\alpha=1$) the DJC can sustain memory up to $\tau=3$.
In the inset of Fig. \ref{stm_tau_c_H3}, we show the capacity for $\tau=1, 2$ for varying $\alpha$, for a fixed value of $\chi^{\prime}=1$. 
The improvement due to a non-zero qubit driving is evident from this figure. We also observe a clear trend that the capacity degrades for very high values of $\alpha$. In Appendix \ref{JC_stm_params} we present the behavior of the capacity of DJC for the STM task for varying $\chi^{\prime}$.

\subsection{Parity check task}
Parity check~\cite{NIPS2004} is a standard benchmarking task for the nonlinear memory of the reservoir. The inputs are randomly generated binary digits $s_{i}\in \{0, 1\}$. The target output for the $i$th input for a fixed $\tau$ is 
\begin{equation}
    y_{i} = \sum_{j=1}^{\tau} s_{i-j}\; \mathrm{mod}\, 2.
\end{equation}
Due to the modulo division, the output is a strictly nonlinear function of the input.

For the JC model, the capacity $C$ for the parity check is shown in Fig. \ref{H2_pc_tau_c}(a) as a function of the delay $\tau$, in the presence of time-multiplexing $V=5$. The optimal capacity is obtained for $dt=10$, $\kappa=0.1$.  
For the PC task, the reservoir demonstrates a greater ability to retain past inputs than in the STM task. Its memory capacity also significantly surpasses that of a two-qubit reservoir, as indicated by the gray dashed line. Furthermore, the capacity can be increased by raising $V$, as illustrated in the inset of Fig. \ref{H2_pc_tau_c}(a).

Next, we fix the optimal parameters $dt=10$ and $\kappa=0.1$, and for the rest of the parameters $\{\Delta_{b}, \Delta, \chi, \alpha\}$, we vary each one of them while keeping the others constant. We present the results in Fig. \ref{H2_pc_tau_c}(b), which shows that in our observed parameter region, the capacity for $\tau=1, 2$ varies very little, implying a robust performance with respect to varying reservoir parameters. Note that the optimal hyperparameters depend on the task. Comparing Fig. \ref{H2_pc_tau_c}(b) with the linear memory capacity  using the JC model (Fig. \ref{H2_stm_params} in Appendix \ref{JC_stm_params}), we can appreciate the broader performance variability for the STM task.
More interestingly, comparing these two figures, we can conclude that for our observed parameter region, the capacity for the PC task always surpasses the linear memory (STM), indicating the relative robustness of the reservoir's nonlinear memory.

Similar results are also found in the dispersive limit. Fig. \ref{H3_pc_tau_c} shows the $C$ vs. $\tau$ curve for the PC task using the DJC model, 
with and without the qubit driving. In both cases, the optimum capacity is obtained for $dt=10$ and $\kappa=0.1$, and it is higher than the corresponding capacity for the STM task. For $\tau=1, 2$, the value of $C$ is higher for $\alpha=1$ than $\alpha=0$. However, when $\tau>3$, we observe a higher capacity for $\alpha=0$, similar to the observation for the STM task using the DJC model. Compared to a two-qubit reservoir, the DJC model shows a significantly improved capacity for the PC task. We note that the observed capacity for our choice of parameters is overall the same for JC and the DJC model. In the inset of Fig. \ref{H3_pc_tau_c}, we show the capacity against increasing $\alpha$ for a fixed value of $\chi^{\prime} = 1$. Similar to the JC model, we observe that the PC task capacity is more robust with respect to changes in $\alpha$ than the STM task capacity. Similar observation arises from the study of capacity with respect to varying $\chi^{\prime}$, which we show in Appendix \ref{JC_stm_params}.

\begin{table*}
\centering
\begin{tblr}{
  cells = {c},
  cell{1}{1} = {r=3}{},
  cell{1}{2} = {c=4}{},
  cell{1}{6} = {c=4}{},
  cell{2}{2} = {c=2}{},
  cell{2}{4} = {c=2}{},
  cell{2}{6} = {c=2}{},
  cell{2}{8} = {c=2}{},
  vlines,
  hline{1,4-6} = {-}{},
  hline{2-3} = {2-9}{},
}
\diagbox{Model}{Task} & {Autonomous\\generation} &                              &                   &                              & {Forecasting \\(1 step)} &                                &                   &                                 \\
                     & V=1                    &                              & V=10              &                              & V=1                      &                                & V=10              &                                 \\
                     & {Single\\segment}      & {10 segments\\avg.}          & {Single\\segment} & {10 segments\\avg.}          & {Single\\segment}        & {10 segments\\avg.}         & {Single\\segment} & {10 segments\\avg.}             \\
JC                   & 0.16                   & 0.18 $\Mypm$\; 0.06 & 0.12              & 0.15 $\Mypm$\; 0.04 & 0.024                    & 0.026 $\Mypm$\;0.0005                              & 0.010             & 0.009 $\Mypm$\; 0.0003 \\
DJC                  & 0.22                   & 0.20 $\Mypm$\; 0.01 & 0.18              & 0.16 $\Mypm$\; 0.03 & 0.033                    & 0.03 $\Mypm$\; 0.0004 & 0.019             & 0.019 $\Mypm$\;0.0004                              
\end{tblr}
\caption{The optimized RMSE for both reservoir models for the autonomous prediction as well as for the delay forecasting of the next step of the Mackey-Glass series, with and without time-multiplexing, for an arbitrary segment as well as for averaged over 10 random segments of the MG series. The values are calculated for 150 points post-training.}
\label{jc_table}
\end{table*}

\begin{figure}
    \centering
    \includegraphics[width=0.48\textwidth]{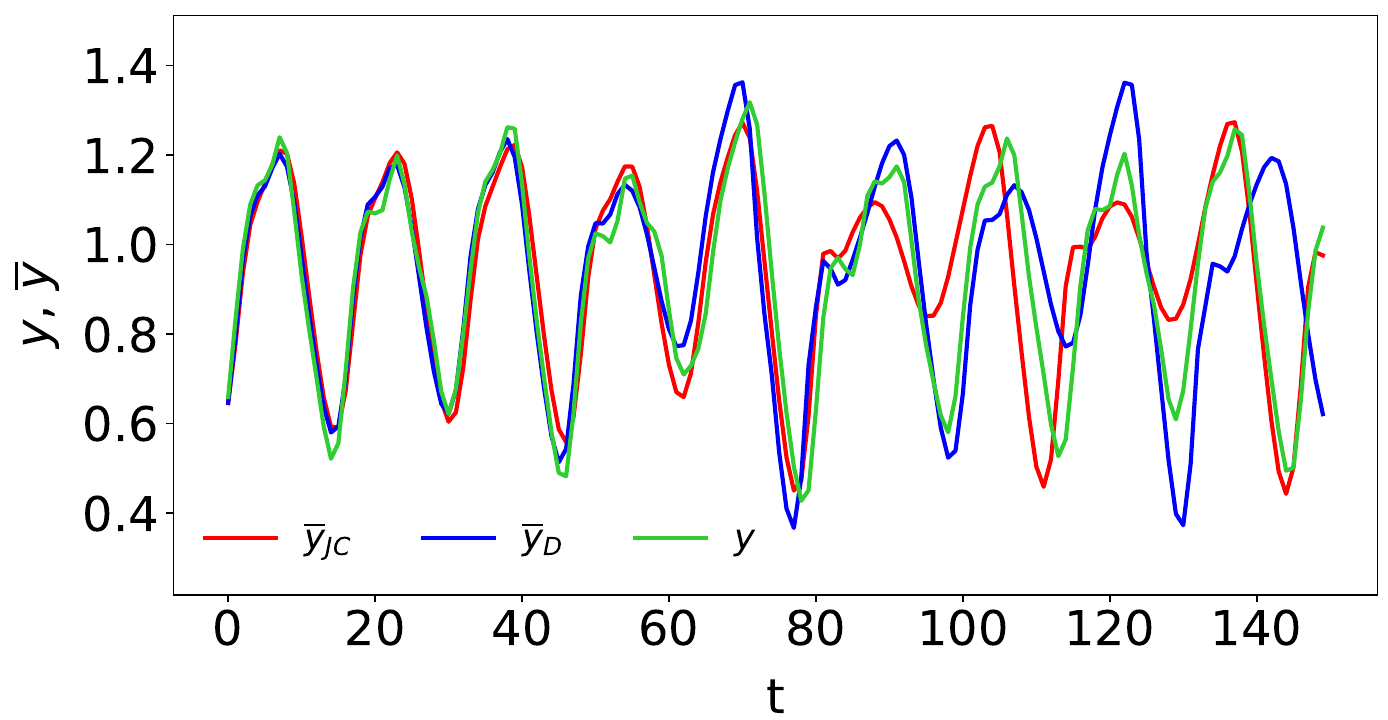}
    \caption{For the autonomous Mackey-Glass prediction task, the target $y$ and the optimum predicted output $\overline{y}_{\mathrm{JC}}$, $\overline{y}_{D}$ corresponding to the JC model and the DJC model, respectively, for 150 steps after the training. The RMSEs for 150 steps are $r=0.12$ and $r=0.16$, respectively. For JC model, the optimum parameters are $dt=10$, $\kappa=0.1$, $\Delta_{b} = 2$, $\Delta=1.5$, $\chi=0.6$, $\alpha=0$. For the DJC model, the optimum parameters are $dt=10$, $\kappa=0.1$, $\chi=1.2$, $\alpha=0$.
    }
    \label{mg_compare}
\end{figure}

\begin{figure}
    \centering
\includegraphics[width=0.47\textwidth]{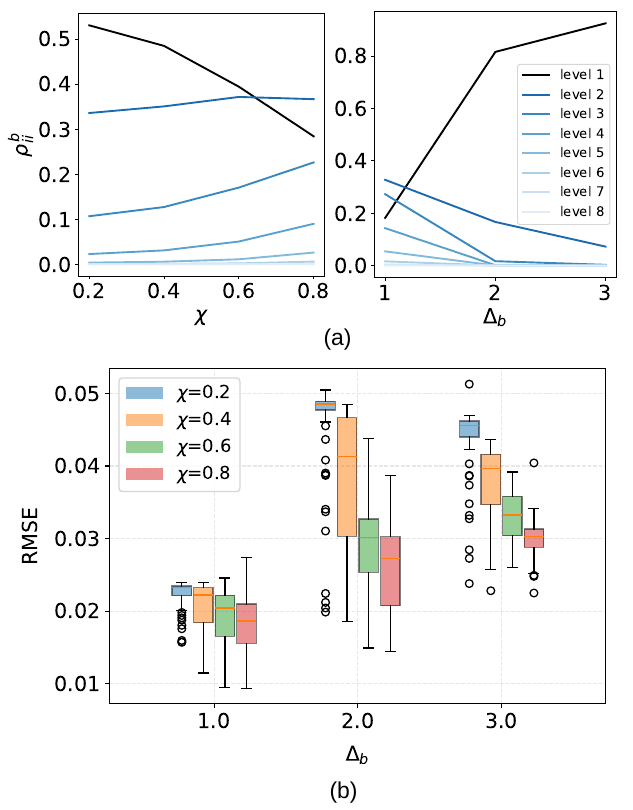}
    \caption{(a) For the JC model reservoir, the population variation of the lower $8$ levels of the bosonic mode with respect to varying $\chi$ (left) and $\Delta_{b}$ (right) for inputs drawn from the MG series. For the former, $\Delta_{b}=1$ , while for the latter, $\chi=1$. The other parameters are $\Delta=0$, $\alpha=0$. The populations correspond to a specific input after the washout phase.
    (b) Box plot representation of RMSE for varying values of $\chi$ and $\Delta_{b}$ and $V=10$, when forecasting the next steps. The boxes and the outlier points represent a set of RMSEs obtained over the range $-2 \leq \Delta \leq 2$ and $0 \leq \alpha \leq 3$. The red horizontal line inside the boxes represents the median, the upper and lower boundaries of the boxes represent $1^{\mathrm{st}}$ and $3^{\mathrm{rd}}$ quartiles, while the vertical whiskers denote the minimum and maximum range of points. The circles outside the boxes (outliers) represent the data points that are too far from the distribution.
    }
    \label{mg_params}
\end{figure}

\begin{figure}
    \centering
\includegraphics[width=0.48\textwidth]{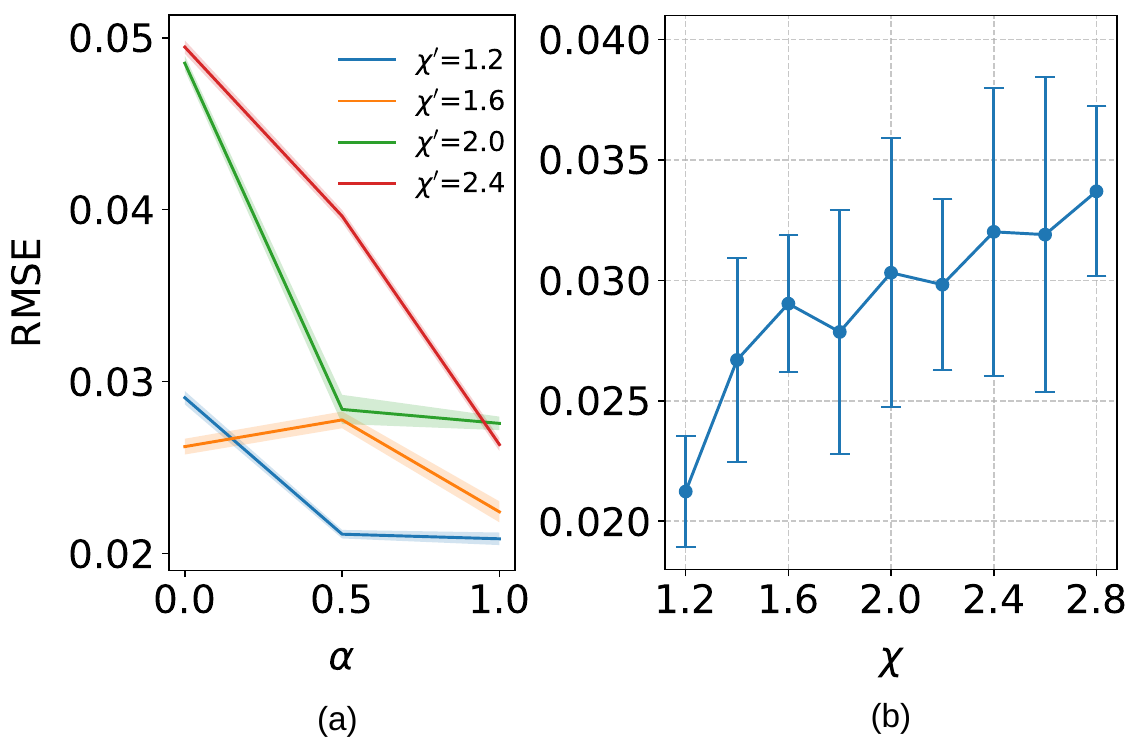}
    \caption{(a) For forecasting the next step using the DJC reservoir, the average RMSE (over 10 different segments) and standard deviation (indicated by the shaded region) for varying values of $\chi^{\prime}$ and $0 \leq \alpha \leq 1$ for $V=10$. (b) For varying values of $\chi^{\prime}$, the average and standard deviation of RMSE over different values of $\alpha$ in the range $0 \leq \alpha \leq 3$.}
    \label{H3_mg_params}
\end{figure}

\subsection{Mackey-Glass prediction}
Having gained an understanding of the linear and nonlinear memory of the reservoirs, we proceed to test their performance for predicting a complex time series. We select the Mackey-Glass (MG) series~\cite{mackey_glass_77}, which is a widely explored benchmark task in reservoir computing. The Mackey-Glass function $s(t)$ can be obtained by solving the following delay-differential equation,
\begin{equation}
    \frac{ds(t)}{dt} = -0.1s(t) + \frac{0.2s(t-\tau)}{1+s^{10}(t-\tau)}.
\end{equation}
For $\tau>16.8$, the series represents a chaotic dynamics. We use $\tau=17$, which is the most widely used value for benchmarking RC. The first 1000 inputs are used for the washout phase, and the next 1000 inputs are used for training. 

In the following, the prediction capabilities of our model will be tested in two different ways, with the reservoir working autonomously, the so-called autonomous generation, and for a 1-step forecasting. In autonomous generation, values predicted for previous time steps are employed as new inputs to forecast the values for the next time steps. This is considered to be a fairly challenging task, as the error in the output accumulates with each step. As a result, the predicted series may largely diverge from the true MG series with increasing prediction steps. The performance of an autonomous prediction of $\mathcal{N}$ such output points $\{\overline{y}_{i}\}_{i=1}^{\mathcal{N}}$ is evaluated using the scaled root mean square error (RMSE) between $\overline{y}$ and the true output $y$:
\begin{equation}
    \mathrm{RMSE}(\mathbf{y}, \overline{\mathbf{y}})=\frac{1}{y_{\mathrm{max}}-y_{\mathrm{min}}} \sqrt{\frac{\sum_{i=1}^{\mathcal{N}}(y_{i}-\overline{y}_{i})^2}{\mathcal{N}}},
\end{equation}
where $y_{\mathrm{max}}$ and $y_{\mathrm{min}}$ are the maximum and the minimum value of the target output.
In Table \ref{jc_table}, we show the optimum RMSE for both $V=1$ and $V=10$ when autonomously predicting the output of an arbitrary segment of the MG series for $150$ steps, which shows a lower RMSE for both models for $V=10$. The rising deviation between real and forecasted trajectories is illustrated in Fig. \ref{mg_compare}, where we show the target output for this temporal interval, alongside the optimum predicted output using the JC and DJC models. Although both models generate outputs that closely match the target for approximately the first $\sim 80$ steps, their trajectories diverge markedly beyond this point.

The robustness of our results is tested in $10$ randomly chosen segments of the MG series.
In the table, we present the optimum average RMSE and the standard deviation for $10$ segments of the MG series, and find comparable values for both reservoir models. The standard deviation for autonomous prediction using the JC model is quite high, which indicates that the RMSE varies largely for different segments.
The optimum performance of the JC model corresponding to Fig. \ref{mg_compare} and Table \ref{jc_table} were found by performing a grid search in the region $1 \leq \Delta_{b} \leq 3$, $-2 \leq \Delta \leq 2$, $0.2 \leq \chi \leq 0.8$, and $0 \leq \alpha \leq 3$. The same for DJC model is $1.2 \leq \chi^{\prime} \leq 3$ and $0 \leq \alpha \leq 3$. For both models, we use the parameters $dt=10$ and $\kappa=0.1$.

Beyond autonomous forecasting, Table \ref{jc_table} also presents the optimal RMSE for one-step-ahead prediction. In this task, the model forecasts the value at the next time step using the true MG series value as the current input, a method that is typically less challenging than autonomous prediction. We calculated the RMSE both for a single segment and as an average over 10 distinct segments, calculated over 150 data points following the washout and training phases.  
As expected, the RMSE is an order of magnitude lower than in the autonomous prediction scenario, and shows negligible dependence on the specific segment of the MG series used as the test dataset, indicated by the very low values of standard deviation across the segments.

The dependence of the RMSE on the reservoir parameters for the JC model is studied in Fig. \ref{mg_params}. 
Although no systematic correlation between error and parameter variation exists for autonomous generation, forecasting errors exhibit a clear inverse relationship with bosonic level population.
In Fig. \ref{mg_params}(a), for the JC reservoir, we show the population distribution across several bosonic mode levels with respect to varying $\chi$ and $\Delta_{b}$. The population in the higher levels grows with increasing $\chi$ and decreasing $\Delta_{b}$. Fig. \ref{mg_params}(b) shows the RMSE varying with $\Delta_{b}$ for several values of $\chi$. We find no significant dependence or monotonic variation of RMSE with respect to changing $\Delta$ or $\alpha$. In Fig. \ref{mg_params}(b), each box plot is used to show the aggregate behaviour of RMSEs obtained over a range of $\Delta$ and $\alpha$. Given the very low variance, averaging over different segments of the MG series is redundant here. Clearly, the median RMSE (the red horizontal line within the boxes) is lower when $\chi$ is higher and $\Delta_{b}$ is lower. 

As for the DJC model, we investigate the dependence of the RMSE on the reservoir parameters in Fig. \ref{H3_mg_params}. 
In Fig. \ref{H3_mg_params}(a), we show the RMSE with varying $\alpha$ for a set of values of $\chi^{\prime}$. The RMSE shows a non-monotonic dependence on $\alpha$ in the range $0 \leq \alpha \leq 3$, however, in most cases, RMSE drops significantly when $\alpha \neq 0$. Moreover, when $\alpha < 0.5$, a lower RMSE is more likely to occur for a lower $\chi^{\prime}$. This claim is further supported by Fig. \ref{H3_mg_params}(b), which shows the mean and standard deviation of RMSE over $\alpha$ in the range $0 \leq \alpha \leq 3$, for varying $\chi^{\prime}$.
These observations highlight the performance enhancement due to a non-zero population in the higher levels of the bosonic mode.

When compared to other physical implementations, our reservoir models perform competitively. In Ref.~\cite{sannia_2024}, using a spin-based reservoir, the authors reported an RMSE $\approx 0.1$ for autonomous generation of 150 steps with the same sampling rate. This is of the same order as the lowest RMSE obtained by us using the JC reservoir for a single segment. The autonomous prediction of the MG series using the DJC model has been theoretically studied in Ref.~\cite{electronics2024} using a different set of observables, where the authors reported RMSE as low as $\approx 0.05$.  
In Ref.~\cite{zhu2025}, the two-step delay forecasting using a three-atom Tavis-Cummings model was obtained with RMSE $=0.05$. In our case, the JC model reservoir using one atom achieves the two-step forecasting with RMSE $\approx 0.06$. More recently, an experimental study~\cite{carles2025} employing a circuit QED reservoir based on the DJC Hamiltonian reported RMSE $\approx 0.1$ for one-step delay forecasting, which is significantly higher than the optimum theoretical value of RMSE $\approx 0.03$ that we obtained using a DJC reservoir with $V=1$. It is important to note, however, that different studies employ different sets of observables and sampling rates for the Mackey–Glass series, both of which strongly influence prediction performance, making direct comparison of different reservoirs inherently difficult. While our results are in line with state of the art performance, the possibility to explore experimentally regions with higher bosonic excitations (that are numerically costly) could lead to improved performance. Indeed, based on our observations, we anticipate that for higher bosonic level occupation, the RMSE for autonomous generation could be further reduced, benefiting from the richer, higher-dimensional input mapping provided by the bosonic mode.

\section{Discussion}
\label{section4}
In this work, we implemented quantum reservoir computing in a hybrid qubit-boson system described using Jaynes-Cummings and dispersive Jaynes-Cummings Hamiltonian in the presence of driving and dissipation. We presented numerical evidence that both reservoirs are capable of complex nonlinear input processing. The outputs are extracted using higher-order moments of the bosonic mode, which captures its complex non-Gaussian dynamics. With the proper operational conditions (by tuning hyperparameters) both can achieve similar performance in different tasks. Similar results are found when considering the bosonic state component in the Fock basis. Using the short-term memory and the parity-check tasks, we demonstrated that the reservoir has a very limited memory of the injected input sequences  for linear tasks but a fairly good memory for nonlinear tasks. Note that this is in contrast to the observations in previously studied quantum reservoirs, for which the linear memory capacity is typically better than the nonlinear capacity~\cite{fuji_qrc, sannia_2024, llodra2024}. Whereas the former does not improve with time multiplexing, 
we observe that the latter can be improved by increasing the number of virtual nodes $V$. The expressivity in the JC and DJC model is also benchmarked with a two-qubit QRC (addressing interactions equivalent to both resonant and dispersive regimes). We showed that both the reservoirs have a superior nonlinear memory capacity compared to an equivalent two-qubit reservoir, also allowing for chaotic series forecasting.
Both the reservoir models display robust performance against variations in their internal parameters, while the presence of the qubit driving ($\alpha$) is shown to be essential in the dispersive regime.

In the evaluation of the reservoirs on autonomous generation of the Mackey–Glass series, the JC and DJC models yielded comparable performance with optimum RMSE values in the range  $10\% < \mathrm{RMSE} < 22\%$. In contrast, for 1-step forecasting, we achieved a minimum RMSE of approximately $1\%$,
indicating strong predictive accuracy. In this setting, the RMSE showed a clear dependence on reservoir parameters: increasing the population of higher bosonic levels consistently reduced the error, providing a promising guideline for experimental implementations.
It can also be further investigated if the expressiveness and the performance of the DJC reservoir can be improved through an input encoding that breaks the symmetry observed in the corresponding Wigner distributions. 

In general, a microscopic derivation of the master equation will have nonlocal decay channels. In the presence of interaction with both bosonic and qubit environment and considering the interaction between the qubit and the boson, a global approach with a consistently derived master equation in the so-called partial secular approximation should be considered \cite{Cattaneo_2019}.  We have chosen to work with a local master equation with only bosonic spontaneous decay, under the assumption of weak coupling between the qubit and the mode~\cite{Trushechkin_2016, Cattaneo_2019, cattaneo_2025}, and a slower decay rate of the qubit. 
It is to be noted that in the dispersive limit of the JC model, the cavity's dissipation channel can, in principle, induce an effective qubit decay. In this work we have not included this effect that can be effectively mitigated in state-of-the-art circuit-QED setups through the use of Purcell filters, which suppress the density of electromagnetic states at the qubit frequency, thereby preserving the qubit's coherence \cite{reed_purcell1_2010, bronn_2015}.

A natural extension of this work is to assess theoretically whether the memory capacity and the predictive performance of both our reservoirs improve when taking into account the nonlinear Kerr effects in the qubit-boson system or different monitoring strategies. As explored in the experimental work \cite{carles2025}, such an improvement may depend on the complex interplay between Kerr strength and the other reservoir parameters. On the other hand, a theoretical analysis to assess the effect of the parity measurement strategy implemented in Ref. \cite{mcmahon2024} on the memory  and the role of the state non-Gaussianity are challenging open questions. Furthermore, this QRC system can be scaled up, considering not only coupled JC units, but also multiphoton JC interaction with one atom~\cite{laha_2024}, or even multiple atoms interacting with one single mode~\cite{zhu2025}. Finally, exploring tailored measurement strategies presents a promising avenue not only to assess potential limitations in online processing  but also for enhancing the complexity of the reservoir dynamics.

\begin{acknowledgments}


We acknowledge the Spanish State Research Agency, through the Maríia de Maeztu project CEX2021-001164-M,  funded by MICIU/AEI/10.13039/501100011033;  through the COQUSY project PID2022-140506NB-C21 and -C22 funded by MICIU/AEI/10.13039/50110001103 and by ERDF, EU; and through the QuantCom project CNS2024-154720, funded by MICIU/AEI/10.13039/501100011033 and cofunded by the European Union; the project is funded under the Quantera II program that has received funding from the EU's H2020 research and innovation program under Grant Agreement No. 101017733, and from the Spanish State Research Agency  (project Coquadis PCI2024-153446) funded by MICIU/ AEI/10.13039/50110001103; MINECO through the QUANTUM SPAIN project, and EU through the RTRP - NextGenerationEU within the framework of the Digital Spain 2025 Agenda; and CSIC’s Quantum Technologies Platform (QTEP).

\end{acknowledgments}

\appendix

\section{Dispersive limit of driven Jaynes-Cummings model}
\label{derivation_djc}
From Eq. (\ref{jc_hamilt0}) and Eq. (\ref{drive_hamilt}), the full Hamiltonian of the driven JC model in the laboratory frame is,
\begin{equation}
    H_{\mathrm{JC}}^{(1)} = H_{\mathrm{JC}} + H_{d}^{b} + H_{d}^{a}.
    \label{driven_jc_lab}
\end{equation}
In the dispersive regime, we have $\delta \gg \chi$. To expand $H_{\mathrm{JC}}$ as a perturbative series of the small parameter $\chi/\delta$, we apply the Schrieffer-Wolff transformation on the Hamiltonian $H_{\mathrm{JC}}^{(1)}$ to obtain
\begin{equation}
    H_{\mathrm{JC}}^{(2)} = e^{S}H_{\mathrm{JC}}^{(1)}e^{-S},
\end{equation}
where $S=\frac{\chi}{\delta}(c^{\dagger}\sigma^{-}-c\sigma^{+})$. From the first term of Eq. \ref{driven_jc_lab} we get,
\begin{align}
    e^{S}H_{\mathrm{JC}}e^{-S}&=H_{\mathrm{JC}} + [S, H_{\mathrm{JC}}] +\frac{1}{2!}[S, [S, H_{\mathrm{JC}}]] + \;...\\ \nonumber
    &\approx \omega_{b}c^{\dagger}c + \big(\omega_{a}+\chi^{\prime}\big)\frac{\sigma^{Z}}{2} + \chi^{\prime}c^{\dagger}c\sigma^{Z},
\end{align}
where $\chi^{\prime}=\chi^{2}/\delta$ is the dispersive shift of the qubit frequency. In this expansion, we ignore terms which are $\textit{O}(\chi^{3}/\delta^{2})$ or higher.

From the transformation of the bosonic driving field, we similarly obtain,
\begin{eqnarray}
    e^{S}H_{d}^{b}e^{-S} \approx H_{d}^{b} -i\frac{\beta\chi}{\delta}(\sigma^{-}e^{i\omega_{1}t}-\sigma^{+}e^{-i\omega_{1}t}),
\end{eqnarray}
where we have ignored terms proportional to $\big(\frac{\chi}{\delta}\big)^{2}$ or higher, since they are very small in the dispersive regime.
Therefore, the bosonic drive induces an effective drive on the qubit. Similarly, transforming the qubit drive, we obtain,
\begin{equation}
    e^{S}H_{d}^{a}e^{-S} \approx H_{d}^{a}-\frac{\alpha\chi}{\delta}(c^{\dagger}e^{-i\omega_{2}t}+ce^{i\omega_{2}t})\sigma^{Z}.
\end{equation}
Summing up, we get,
\begin{align}
\label{still_not_final}
    H_{\mathrm{JC}}^{(2)}=&\omega_{b}c^{\dagger}c + (\omega_{a}+\chi^{\prime})\frac{\sigma^{Z}}{2}+\chi^{\prime}c^{\dagger}c\sigma^{Z}+H_{d}^{b} +H_{d}^{a}\\ \nonumber
    & -i\frac{\beta\chi}{\delta}(\sigma^{-}e^{i\omega_{1}t}-\sigma^{+}e^{-i\omega_{1}t})\\ \nonumber
    &-\frac{\alpha\chi}{\delta}(c^{\dagger}e^{-i\omega_{2}t}+ce^{i\omega_{2}t})\sigma^{Z}.
\end{align}

Finally, we go the rotating frame of the two driving fields by applying the transformation $U=\mathrm{exp}(-i(\omega_{1}c^{\dagger}c+\frac{\omega_{2}}{2}\sigma^{Z})t)$, thereby obtaining,
\begin{align}
\label{almost_final}
H_{\mathrm{JC}}^{(3)} &= (\omega_{b}-\omega_{1})c^{\dagger}c + (\omega_{a}+\chi^{\prime}-\omega_{2})\frac{\sigma^{Z}}{2}+\chi^{\prime}c^{\dagger}c\sigma^{Z} \\ \nonumber
&-i\beta(c-c^{\dagger}) +\alpha\sigma^{X} -i\frac{\beta\chi}{\delta}(\sigma^{-}e^{i\Delta_{12} t}-\sigma^{+}e^{-i\Delta_{12} t}) \\ \nonumber
&-\frac{\alpha\chi}{\delta}(c^{\dagger}e^{-i\Delta_{12} t}+ce^{i\Delta_{12} t}),
\end{align}
where $\Delta_{12} = \omega_{1}-\omega_{2}$.

Now, in a circuit QED setup, the physical values of the parameters are $\chi\sim \mathrm{MHz}$, $\delta\sim \mathrm{GHz}$, which implies that $\chi/\delta \ll \chi^{2}/\delta=\chi^{\prime}$. For numerical purpose, we have used the scaled values $\chi{\prime}=1$, $\mathrm{max}\,\beta(t)=1$ for STM and PC task, while $0\leq \alpha \leq 3$. Therefore, the scalar factors in front of the last two terms in Eq. \ref{almost_final} are much smaller than $\chi^{\prime}$, and we can safely drop them.
Moreover, these terms are associated with phase factors which oscillate with a frequency $\Delta_{12}$. Since we consider resonance driving in the dispersive regime, we have $\omega_{1}=\omega_{b}$ and $\omega_{2}\approx \omega_{a}$, which implies $\Delta_{12} \approx \delta$. Therefore, the highly oscillating last two terms can also be ignored due to the rotating wave approximation. Finally, after considering the resonance condition, we obtain the Hamiltonian of the DJC reservoir as the following,
\begin{equation}
    H_{\mathrm{DJC}} \approx \chi^{\prime}c^{\dagger}c\sigma^{Z}-i\beta(c-c^{\dagger}) +\alpha\sigma^{X}.
\end{equation}

\section{Details of numerical simulation}
\label{sec:nlevel}
\begin{figure}[t]
    \centering
    \includegraphics[width=0.4\textwidth]{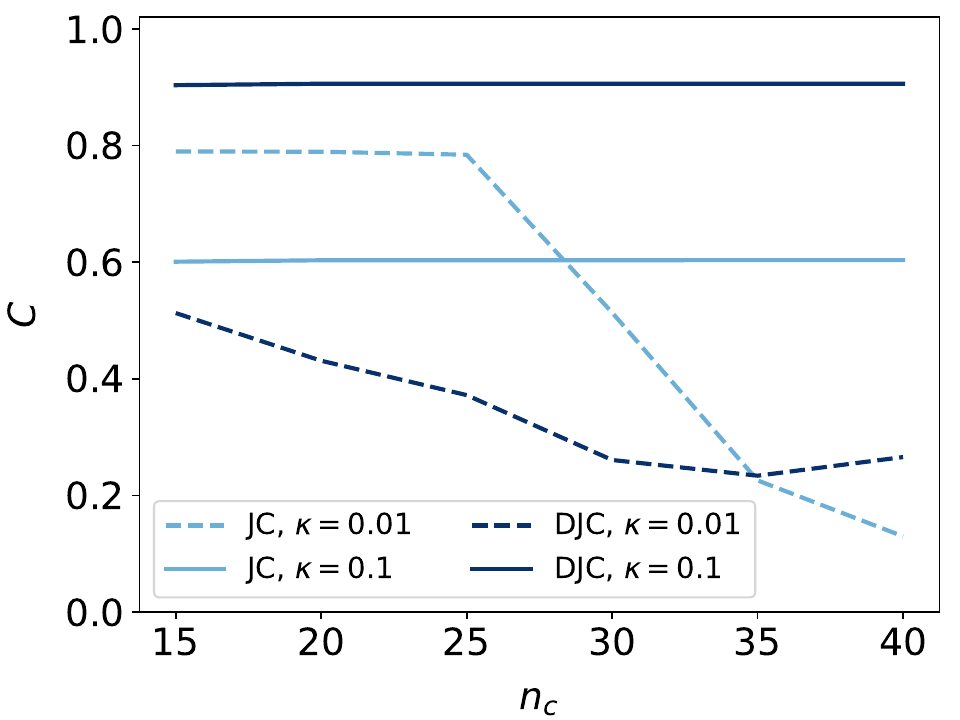}
    \caption{Capacity of the STM task with increasing number of levels in the truncated bosonic mode for $dt=10$, $\kappa=0.01$ and $\kappa=0.1$.}
    \label{nlevel_compare}
\end{figure}

We numerically solve Eq. \ref{master_eq} using the QuTiP library~\cite{qutip1, qutip2, qutip3} to get the expectation values. The readout layer is mapped to the target output using Scikit ridge regression with regularization strength $0.05$. In Fig. \ref{nlevel_compare}, we show how the capacity for the STM task varies with the number of levels $n_c$ of the bosonic mode for $\kappa=0.01$ and $\kappa=0.1$. We could not simulate the dynamics for $n_{c}>40$ due to significantly large computation time. For the JC model and for $\kappa=0.01$, the capacity seems to saturate when increasing $n_{c}$ up to 25, but degrades when $n_{c}>25$. For the DJC model, the capacity for $\kappa=0.01$ monotonically decreases with increasing $n_{c}$. This implies that the dynamics of the reservoir populates the higher levels of the bosonic mode, which induces error when truncating the mode to a lower value of $n_{c}$. However, when $\kappa=0.1$, for both models the higher levels remain inaccessible to the reservoir dynamics, and the capacity remains unchanged when using a high value of $n_{c}$. The capacity in this case is also higher than that corresponding to $\kappa=0.01$ for all values of $n_{c}$ examined by us. Therefore, we use $\kappa=0.1$ for all our studies, and we use $15$ levels of the bosonic mode, where truncation does not hinder the accuracy in the dynamics simulation. We verify that, when studying the capacity of the STM and PC tasks with respect to varying reservoir parameters, the levels higher than $15$ are never populated. For inputs corresponding to the  MG series, the higher levels are populated for a significant region of the parameter space. In this case, we constrain the JC model parameters in the region $1 \leq \Delta_{b} \leq 3$, $-2 \leq \Delta \leq 2$, $0.2 \leq \chi \leq 0.8$, $0 \leq \alpha \leq 3$, and the DJC model parameters in the region $1.2 \leq \chi^{\prime} \leq 3$, $0 \leq \alpha \leq 3$. In this region, higher than the $15^{\mathrm{th}}$ level are unpopulated, and the RMSE values do not change when the truncation level is increased. As a note, if the truncation is done below the $15^{\mathrm{th}}$ level for this parameter region, then the incorrect RMSE can either be more or less than the correct one. In other words, in general the incorrect RMSE does not vary in a strictly monotonic with the truncation error.

\begin{figure}[t]
    \centering
    \includegraphics[width=0.49\textwidth]{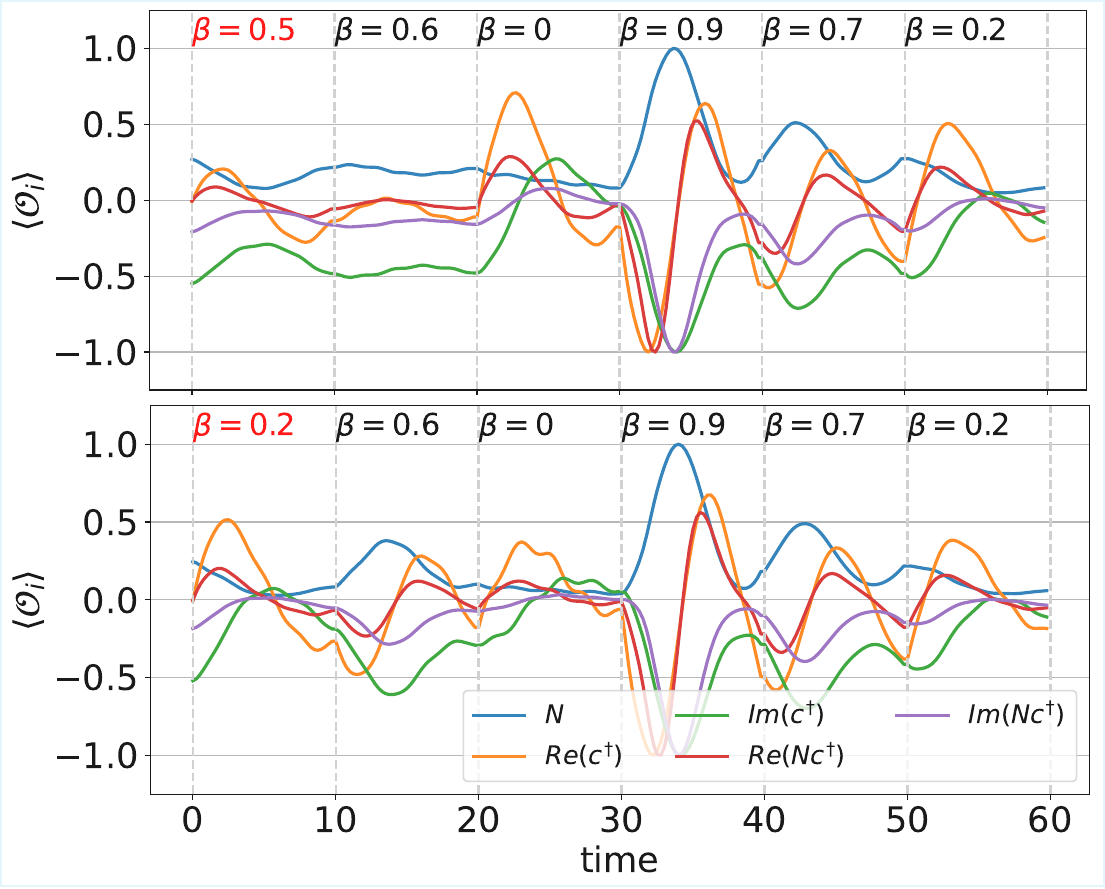}
    \caption{Dynamics of the expectation values of our observables for the JC model, $dt=10$ and $\kappa=0.1$. For presentation purposes, we have scaled the expectation values in the range $[-1, 1]$. The vertical dashed lines denote instances when a new input is introduced. The top and bottom plots correspond to different inputs in the first step, respectively $\beta=0.5$ and $\beta=0.2$. The other parameters are $\Delta_{b} = 1$, $\Delta = 0$, $\chi = 1$, $\alpha=0$.}
    \label{show_dynamics}
\end{figure}

\section{Comparison with a different set of observables}
\label{observable2_compare}
The operators that we use to form the readout layer in our work are the following,
\begin{align}
\label{list_ops}
    \mathcal{O}=&\{N, N^2, N^3, N^4, c^{\dagger}, (c^{\dagger})^2, (c^{\dagger})^3, (c^{\dagger})^4, \\ \nonumber
    &(c^{\dagger})^5, Nc^{\dagger}, Nc, N(c^{\dagger})^2, Nc^2, N(c^{\dagger})^3, Nc^3, \\ \nonumber
    &N(c^{\dagger})^4, Nc^4, N(c^{\dagger})^5, Nc^5,  N^{2}c^{\dagger}, N^{2}c,  N^{2}(c^{\dagger})^2\}.
\end{align}
Counting separately the real and imaginary parts of the expectation values of these operators gives 40 readouts (few can be redundant). The first four moments of the number operator are functions of the diagonal elements $\rho^{b}_{i,i}$ ($i=1, 2, .., 15$) of the reduced density matrix, whereas the rest of them are functions of the off-diagonal elements, for instance, $\langle c\rangle = f(\rho^{b}_{0,1}, \rho^{b}_{1,2}, ..)$ whereas $\langle c^2\rangle = f(\rho_{0, 2}, \rho_{1, 3}, ..)$ etc.

The dynamical behavior of a subset of observables subject to random inputs is shown in Fig. \ref{show_dynamics} to provide a visual confirmation of the reservoir's fading memory.  
Upon completion of the washout phase, we obtain the dynamics for the next 6 inputs in the series. Then we repeat the simulation from the start with the same input sequence; however, after the washout phase, we change only the first input to observe how far the effect of the changed input propagates to the future. We see that a change of input results in a drastic change of the oscillation amplitudes in that time-step, as well as in the next one or two time-steps. However, the observables in the far future remain almost unchanged. This observation holds true for both reservoir models and gives us an idea about the extent of the fading memory of our reservoir.

\begin{figure}[t]
    \centering
    \subfigure[]{\includegraphics[width=0.232\textwidth]{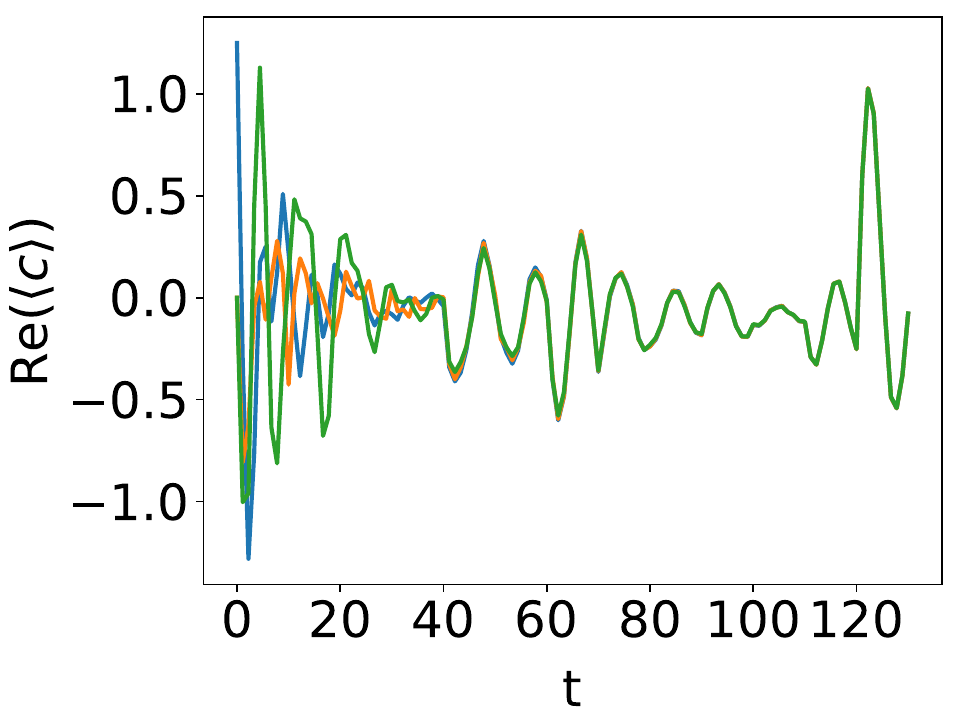}}
    \subfigure[]{\includegraphics[width=0.232\textwidth]{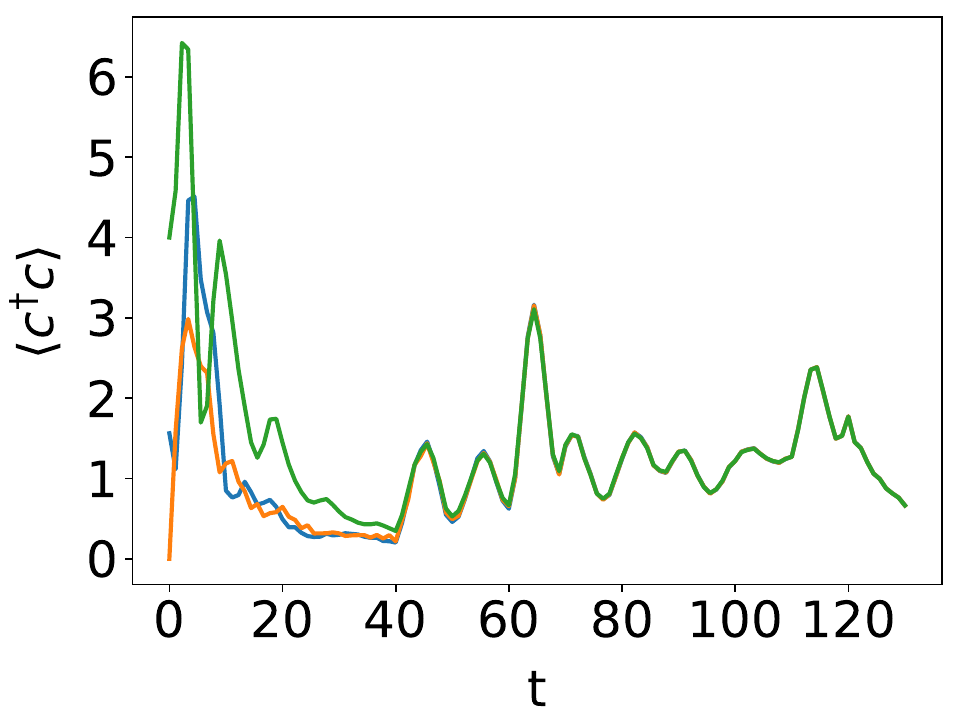}}\vspace{-1em}
    \subfigure[]{\includegraphics[width=0.232\textwidth]{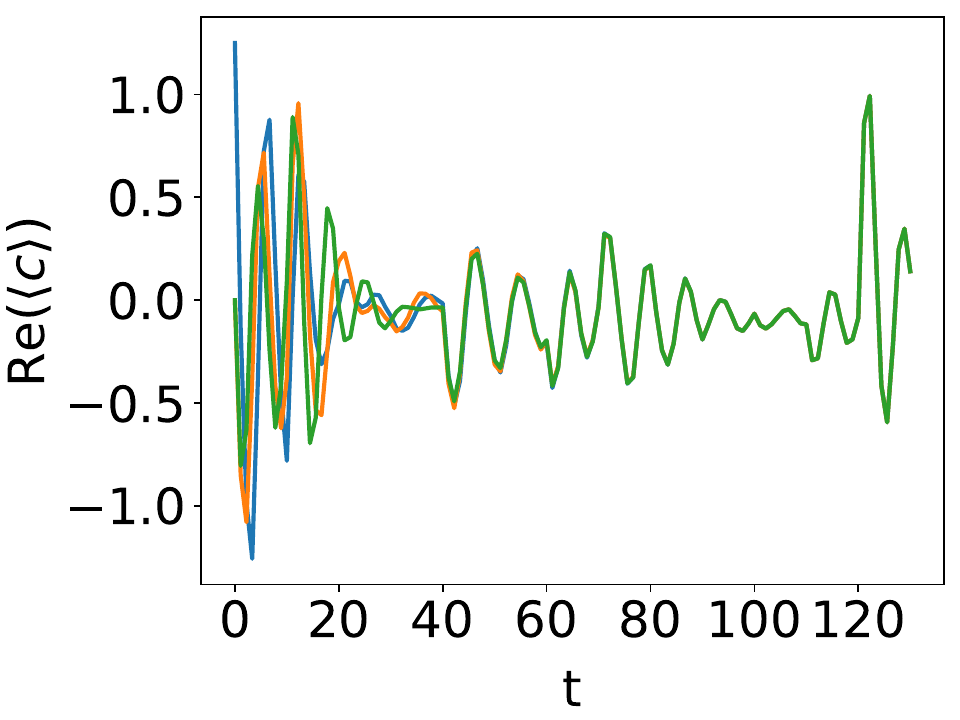}}
    \subfigure[]{\includegraphics[width=0.232\textwidth]{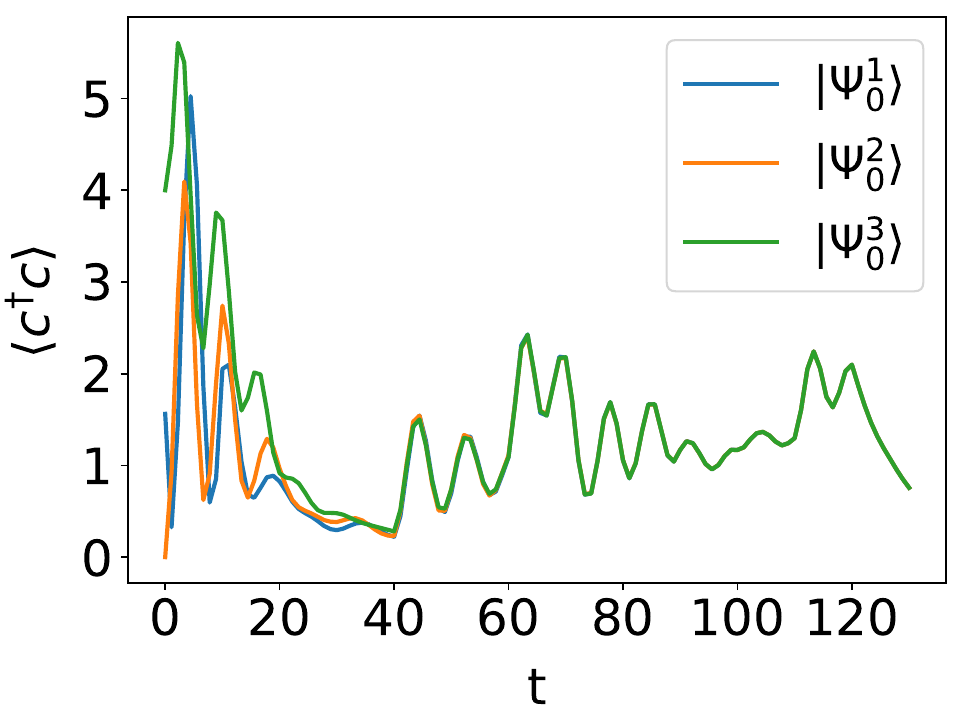}}
    \caption{Dynamics of $\mathrm{Re}(\langle c\rangle)$, and $\langle c^{\dagger}c\rangle$ for a given random input sequence, for three different initial states $\vert\Psi_{0}^{1}\rangle$, $\vert\Psi_{0}^{2}\rangle$ and $\vert\Psi_{0}^{3}\rangle$, as described in Appendix \ref{observable2_compare}. Panels (a) and (b) correspond to JC reservoir, whereas panels (c) and (d) correspond to DJC reservoir. The dynamics of $\mathrm{Im}(\langle c\rangle)$ converges as well, which we omit here for a clear presentation purpose.}
    \label{echo_state}
\end{figure}

We also verify that both the reservoirs have the echo state property, i.e. after a sufficient number of input injections, the state of the reservoir at any time does not depend on the initial state. In Fig. \ref{echo_state}, we show the behavior of $\mathrm{Re}(\langle c\rangle)$, $\mathrm{Im}(\langle c\rangle)$ and $\langle c^{\dagger}c\rangle$ with respect to number of input injections, starting from three different initial states $\vert\Psi_{0}^{1}\rangle$, $\vert\Psi_{0}^{2}\rangle$ and $\vert\Psi_{0}^{3}\rangle$, which are the following,
\begin{align}
    &\vert\Psi_{0}^{1}\rangle = D(1.25)\vert 0\rangle \otimes \vert e\rangle,\\ \nonumber
    &\vert\Psi_{0}^{2}\rangle = \vert 0\rangle \otimes \vert e\rangle, \\ \nonumber
    &\vert\Psi_{0}^{3}\rangle = \vert 4\rangle \otimes \frac{(\vert g\rangle +\vert e\rangle)}{\sqrt{2}}.
\end{align}
Here $D(\cdot)$ is the displacement operator. For both reservoirs, the observables converge for different initial states after a few tens of input injections, therefore confirming the echo state property. It is important to note here that when $\alpha=0$ for the DJC reservoir, as discussed in Sec. \ref{section3}A, the dynamics of the qubit and the bosonic mode are separable. The corresponding Hamiltonian as well as the master equation conserves the expectation value $\langle\sigma^{Z}\rangle$ of the qubit state. The boson evolves in time independent of the qubit, and acts as the reservoir in this case. Therefore, even though initial qubit states with different $\langle \sigma^{Z}\rangle$ never converge in time, the echo state property is satisfied for different initial states of the bosonic mode.

\begin{figure}[t]
    \centering
    \subfigure[]{\includegraphics[width=0.235\textwidth]{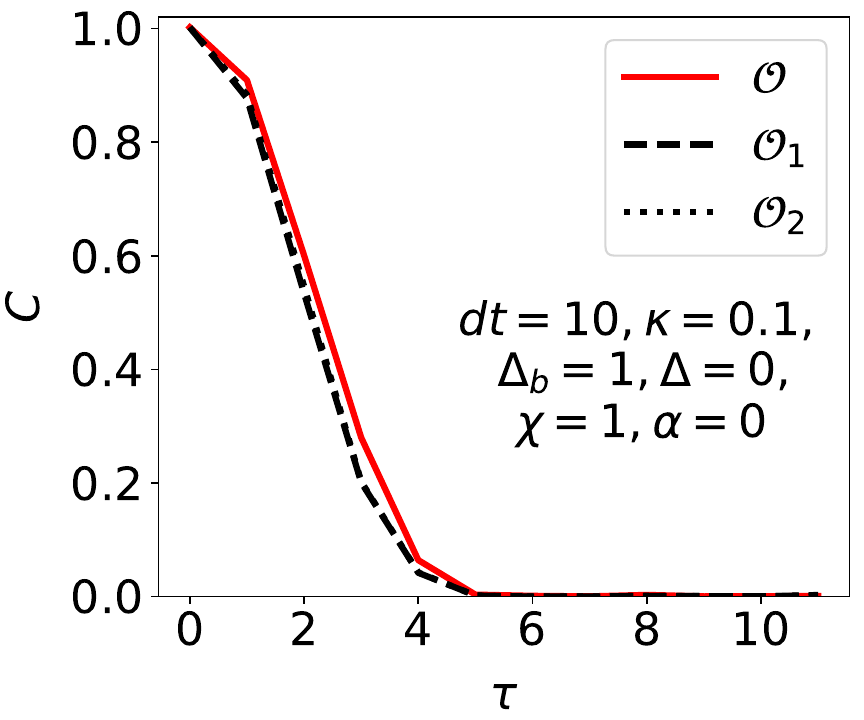}}
    \subfigure[]{\includegraphics[width=0.23\textwidth]{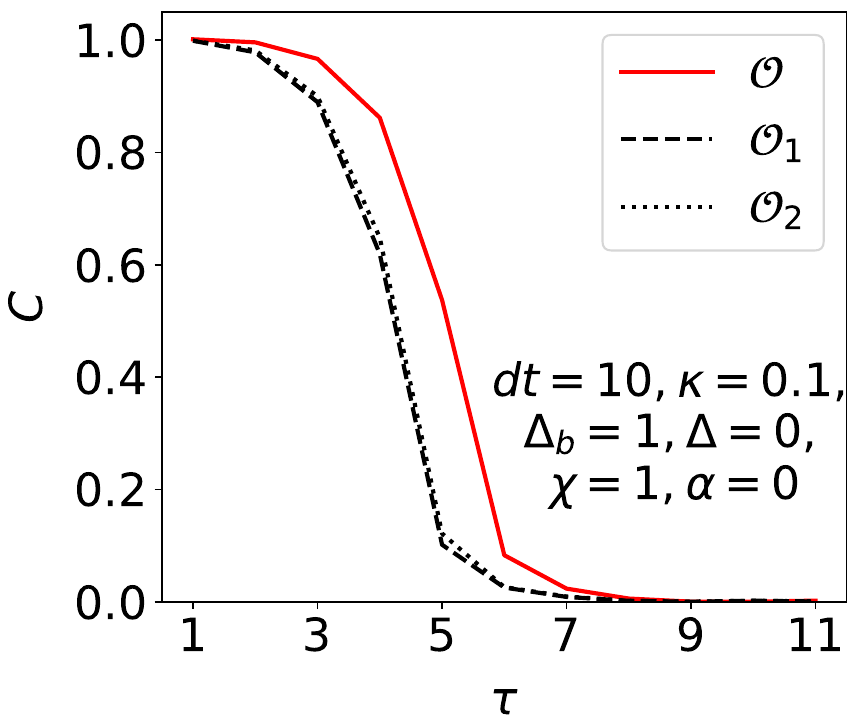}}\vspace{-1em}
    \subfigure[]{\includegraphics[width=0.232\textwidth]{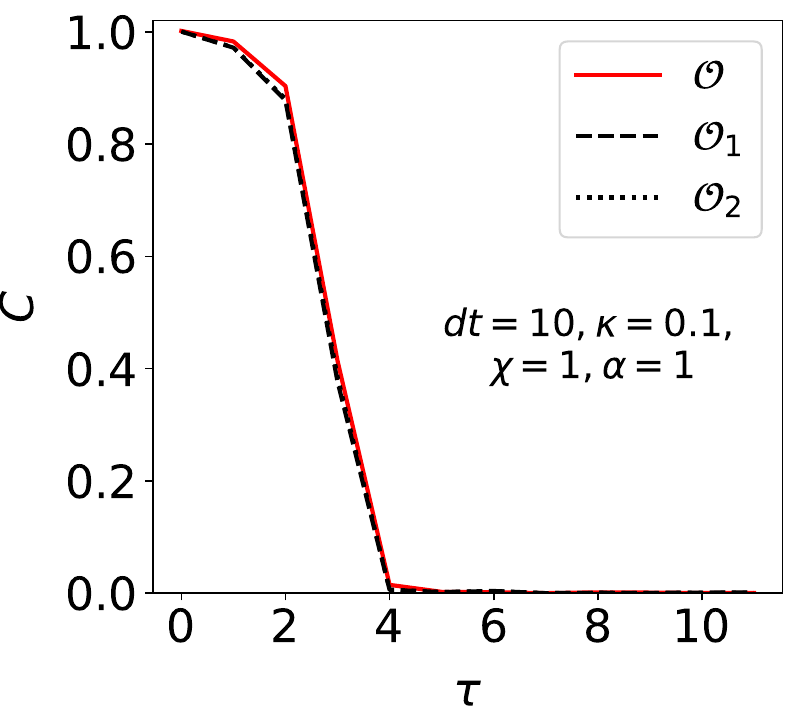}}
    \subfigure[]{\includegraphics[width=0.237\textwidth]{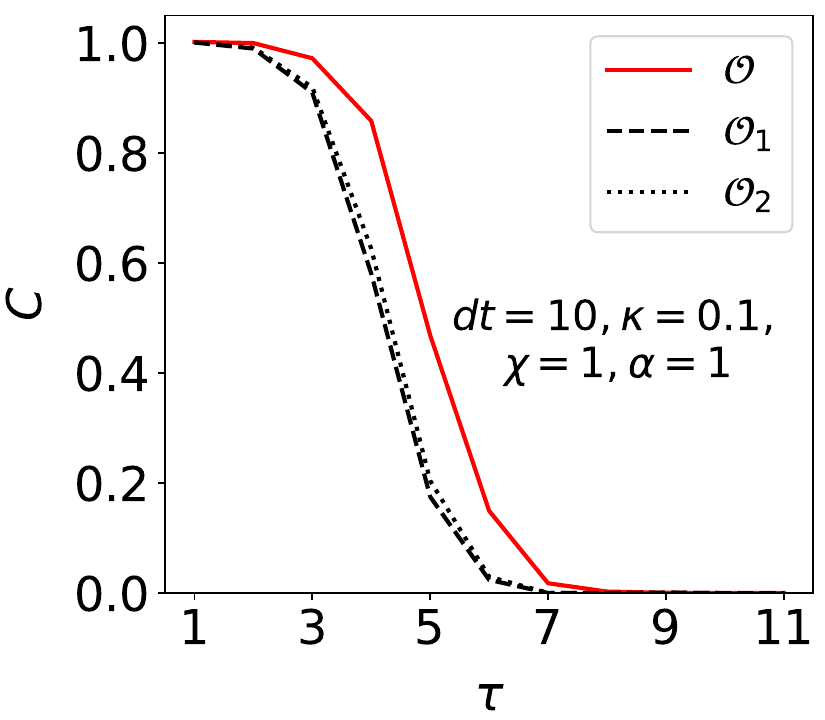}}
    \caption{Comparison between the set of observables $\mathcal{O}$, $\mathcal{O}_{1}$ and $\mathcal{O}_{2}$. Panels (a) and (b) are respectively for STM and PC task using JC model, whereas panels (c) and (d) are respectively for the STM and PC task using the DJC model.}
    \label{rdm_compare}
\end{figure}

Next, we discuss the performance of QRC when directly using the reduced density matrix elements $\rho^{b}_{i,j}$ to form the readout layer, and compare it with the performance obtained using the operators in Eq. \ref{list_ops}. We consider the two following sets of observables.
\begin{equation}
    \mathcal{O}_1=\{ \mathrm{Re}(\rho^{b}_{i,j}) (1\leq i,j \leq 10),  \mathrm{Im}(\rho^{b}_{i,j}) (1\leq i,j \leq 6) \},
\end{equation}
which gives us 40 readouts in total, and 
\begin{equation}
    \mathcal{O}_2=\{ \mathrm{Re}(\rho^{b}_{i,j}), \mathrm{Im}(\rho^{b}_{i,j}) \}, \;  1\leq i,j \leq 15,
\end{equation} 
which gives 225 readouts in total.

In Fig. \ref{rdm_compare}, we show the results for STM and PC tasks using the JC and DJC models. We find that despite a significantly larger size of the readout set for $\mathcal{O}_{2}$, it does not result in an improvement of the capacity over $\mathcal{O}_{1}$. For the STM task, all three sets of readouts $\mathcal{O}, \mathcal{O}_{1}, \mathcal{O}_{2}$ with the same reservoir parameters result in almost the same capacity. For the PC task, the readouts $\mathcal{O}$ lead to a better capacity.

\section{Parameter dependency of the JC model for the STM task}
\label{JC_stm_params}
\begin{figure}[]
    \centering
    \subfigure[]{\includegraphics[width=0.45\textwidth]{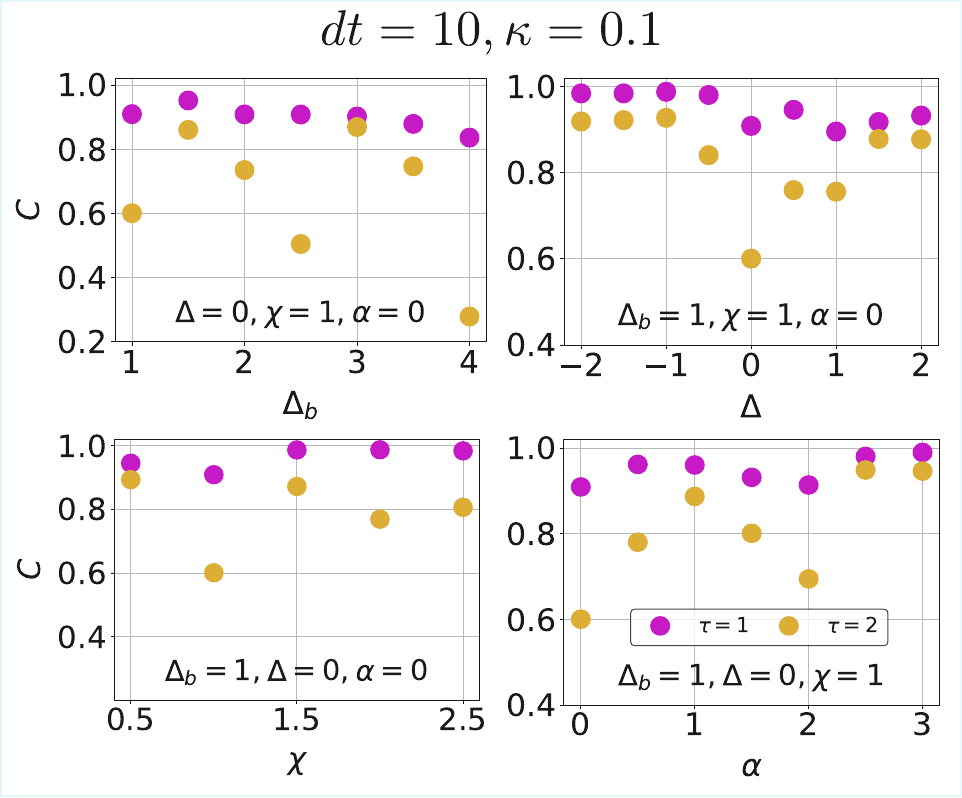}}
    \subfigure[]{\includegraphics[width=0.23\textwidth]{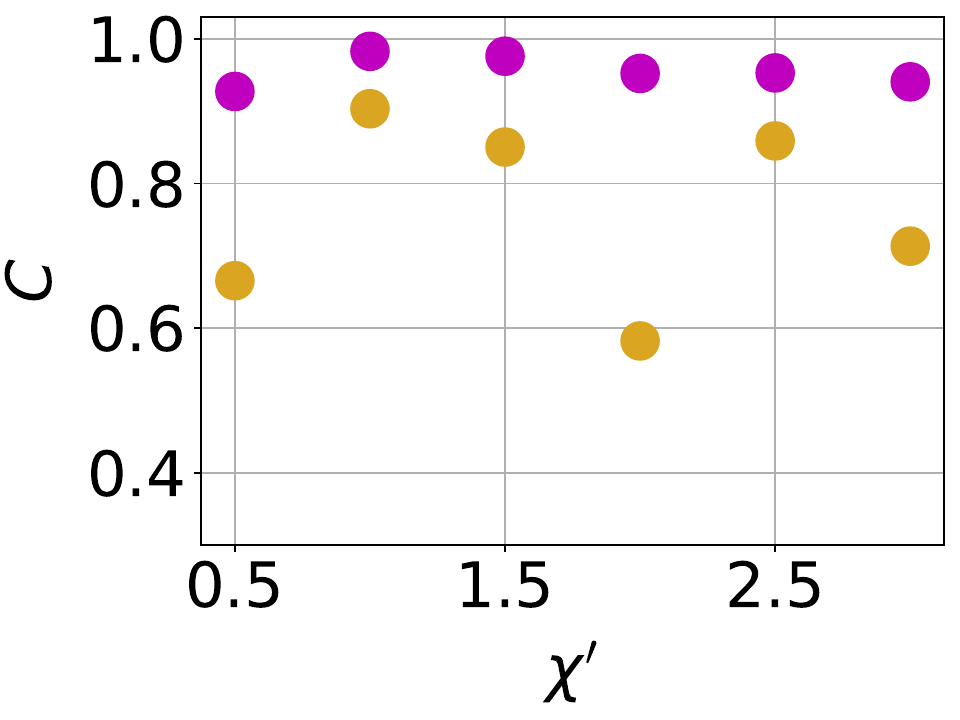}}
    \subfigure[]{\includegraphics[width=0.23\textwidth]{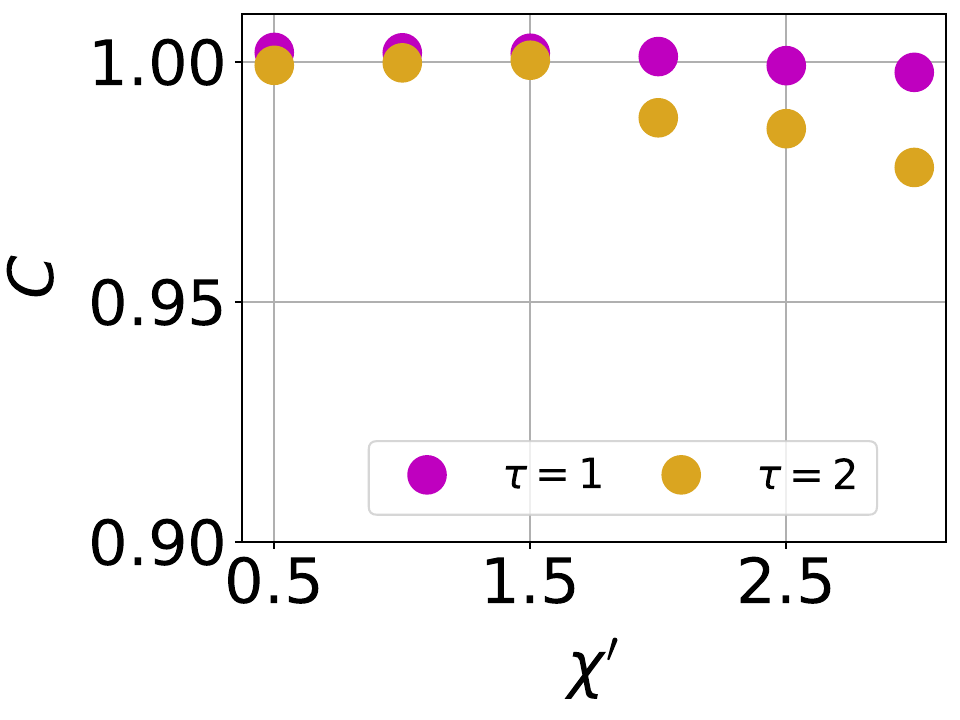}}
    \caption{(a) The capacity of the JC model for the STM task with respect to varying reservoir parameters for $dt=10$ and $\kappa=0.1$. (b) Capacity of DJC model for STM task with varying $\chi^{\prime}$ and (c) capacity of the DJC model for the PC task with varying $\chi^{\prime}$. For both cases, $\alpha=1$.}
    \label{H2_stm_params}
\end{figure}

In Fig. \ref{H2_stm_params}(a), we show the capacity corresponding to $\tau=1, 2$ of the STM task with respect to varying reservoir parameters of the JC model. For $\tau=1$ the capacity shows a higher degree of robustness than for $\tau=2$. In Fig. \ref{H2_stm_params}(b) and (c), we show the capacities of DJC model corresponding to STM and PC task with respect to varying $\chi^{\prime}$. For the former, we again observe an oscillating behavior, and greater robustness when $\tau=1$. For the latter, similar to JC model for PC task, we see a very robust capacity with respect to varying $\chi^{\prime}$.

\section{Hamiltonians of the two-qubit reservoir}
\label{sec:twoqubit}
To demonstrate the improved performance of QRC due to replacing a qubit with a bosonic mode, we compare our qubit-boson model with an analogous two-qubit reservoir. To obtain the corresponding Hamiltonians, we do the replacements $c^{\dagger}\rightarrow \sigma^{+}$, $c\rightarrow \sigma^{-}$, and $c^{\dagger}c\rightarrow (\sigma^{Z}+\mathds{1})/2$. Here $\mathds{1}$ is the identity operator for a qubit. Therefore, for the JC-type of interaction, the Hamiltonian of the two-qubit reservoir is,
\begin{equation}
\label{2qubit_eq1}
    H_{qJC} = \frac{\omega_{a}}{2}\sigma^{Z}_{1} + \frac{\sigma^{Z}_{2} + \mathds{1}}{2}\omega_{b} + \chi(\sigma^{+}_{1}\sigma^{-}_{2} + \sigma^{-}_{1}\sigma^{+}_{2}),
\end{equation}
where $\sigma^{i}_{1}$ and $\sigma^{i}_{2}$ are the operators corresponding to the first and the second qubit. Instead of the bosonic driving, we drive the second qubit. Thus, the full Hamiltonian of the driven two-qubit reservoir in the interaction picture is,
\begin{align}
\label{2qubit_eq2}
    H^{\prime}_{qJC} =\; &\Delta_{a}\sigma^{Z}_{1} + \frac{\Delta_{b}}{2}\sigma^{Z}_{2} + \chi(\sigma^{+}_{1}\sigma^{-}_{2} + \sigma^{-}_{1}\sigma^{+}_{2})\\ \nonumber
    &+ \alpha\sigma^{X}_{1} + \beta(t)\sigma^{X}_{2}. 
\end{align}
Similarly, for the resonantly driven DJC model, we get the interaction picture Hamiltonian,
\begin{align}
\label{2qubit_eq3}
    H^{\prime}_{qDJC} = &\chi^{\prime}\sigma^{Z}_{1}\bigg(\frac{\sigma^{Z}_{2}+\mathds{1}}{2}\bigg) +\alpha\sigma^{X}_{1} + \beta\sigma^{X}_{2}.
\end{align}
The dissipation in the second qubit is modelled using the Lindblad operator $a = \sigma^{-}_{2}$. Using the above Hamiltonians and the Lindblad operator, we solve Eq. \ref{master_eq}. To form the readout layer, we use the real and imaginary parts of $\langle\sigma^{Z}_{2}\rangle$, $\langle\sigma^{+}_{2}\rangle$, and $\langle\sigma^{-}_{2}\rangle$. We keep the other parameters of the reservoir and the hyperparameters of the numerical simulation the same for all reservoirs.

\bibliography{main}

\begin{thebibliography}{73}%
\makeatletter
\providecommand \@ifxundefined [1]{%
 \@ifx{#1\undefined}
}%
\providecommand \@ifnum [1]{%
 \ifnum #1\expandafter \@firstoftwo
 \else \expandafter \@secondoftwo
 \fi
}%
\providecommand \@ifx [1]{%
 \ifx #1\expandafter \@firstoftwo
 \else \expandafter \@secondoftwo
 \fi
}%
\providecommand \natexlab [1]{#1}%
\providecommand \enquote  [1]{``#1''}%
\providecommand \bibnamefont  [1]{#1}%
\providecommand \bibfnamefont [1]{#1}%
\providecommand \citenamefont [1]{#1}%
\providecommand \href@noop [0]{\@secondoftwo}%
\providecommand \href [0]{\begingroup \@sanitize@url \@href}%
\providecommand \@href[1]{\@@startlink{#1}\@@href}%
\providecommand \@@href[1]{\endgroup#1\@@endlink}%
\providecommand \@sanitize@url [0]{\catcode `\\12\catcode `\$12\catcode
  `\&12\catcode `\#12\catcode `\^12\catcode `\_12\catcode `\%12\relax}%
\providecommand \@@startlink[1]{}%
\providecommand \@@endlink[0]{}%
\providecommand \url  [0]{\begingroup\@sanitize@url \@url }%
\providecommand \@url [1]{\endgroup\@href {#1}{\urlprefix }}%
\providecommand \urlprefix  [0]{URL }%
\providecommand \Eprint [0]{\href }%
\providecommand \doibase [0]{https://doi.org/}%
\providecommand \selectlanguage [0]{\@gobble}%
\providecommand \bibinfo  [0]{\@secondoftwo}%
\providecommand \bibfield  [0]{\@secondoftwo}%
\providecommand \translation [1]{[#1]}%
\providecommand \BibitemOpen [0]{}%
\providecommand \bibitemStop [0]{}%
\providecommand \bibitemNoStop [0]{.\EOS\space}%
\providecommand \EOS [0]{\spacefactor3000\relax}%
\providecommand \BibitemShut  [1]{\csname bibitem#1\endcsname}%
\let\auto@bib@innerbib\@empty
\bibitem [{\citenamefont {Jaynes}\ and\ \citenamefont
  {Cummings}(1963)}]{jcmodel}%
  \BibitemOpen
  \bibfield  {author} {\bibinfo {author} {\bibfnamefont {E.}~\bibnamefont
  {Jaynes}}\ and\ \bibinfo {author} {\bibfnamefont {F.}~\bibnamefont
  {Cummings}},\ }\bibfield  {title} {\bibinfo {title} {Comparison of quantum
  and semiclassical radiation theories with application to the beam maser},\
  }\href {https://doi.org/10.1109/PROC.1963.1664} {\bibfield  {journal}
  {\bibinfo  {journal} {Proceedings of the IEEE}\ }\textbf {\bibinfo {volume}
  {51}},\ \bibinfo {pages} {89} (\bibinfo {year} {1963})}\BibitemShut {NoStop}%
\bibitem [{\citenamefont {Shore}\ and\ \citenamefont
  {Knight}(1993)}]{Shore01071993}%
  \BibitemOpen
  \bibfield  {author} {\bibinfo {author} {\bibfnamefont {B.~W.}\ \bibnamefont
  {Shore}}\ and\ \bibinfo {author} {\bibfnamefont {P.~L.}\ \bibnamefont
  {Knight}},\ }\bibfield  {title} {\bibinfo {title} {The jaynes-cummings
  model},\ }\href {https://doi.org/10.1080/09500349314551321} {\bibfield
  {journal} {\bibinfo  {journal} {Journal of Modern Optics}\ }\textbf {\bibinfo
  {volume} {40}},\ \bibinfo {pages} {1195} (\bibinfo {year}
  {1993})}\BibitemShut {NoStop}%
\bibitem [{\citenamefont {Larson}\ and\ \citenamefont
  {Mavrogordatos}(2024)}]{jcreview_2024}%
  \BibitemOpen
  \bibfield  {author} {\bibinfo {author} {\bibfnamefont {J.}~\bibnamefont
  {Larson}}\ and\ \bibinfo {author} {\bibfnamefont {T.}~\bibnamefont
  {Mavrogordatos}},\ }\href {https://doi.org/10.1088/978-0-7503-6452-2} {\emph
  {\bibinfo {title} {The Jaynes–Cummings Model and its Descendants (Second
  Edition)}}},\ 2053-2563\ (\bibinfo  {publisher} {IOP Publishing},\ \bibinfo
  {year} {2024})\BibitemShut {NoStop}%
\bibitem [{\citenamefont {Meschede}\ \emph {et~al.}(1985)\citenamefont
  {Meschede}, \citenamefont {Walther},\ and\ \citenamefont
  {M\"uller}}]{meschede85}%
  \BibitemOpen
  \bibfield  {author} {\bibinfo {author} {\bibfnamefont {D.}~\bibnamefont
  {Meschede}}, \bibinfo {author} {\bibfnamefont {H.}~\bibnamefont {Walther}},\
  and\ \bibinfo {author} {\bibfnamefont {G.}~\bibnamefont {M\"uller}},\
  }\bibfield  {title} {\bibinfo {title} {One-atom maser},\ }\href
  {https://doi.org/10.1103/PhysRevLett.54.551} {\bibfield  {journal} {\bibinfo
  {journal} {Phys. Rev. Lett.}\ }\textbf {\bibinfo {volume} {54}},\ \bibinfo
  {pages} {551} (\bibinfo {year} {1985})}\BibitemShut {NoStop}%
\bibitem [{\citenamefont {Dutra}(2005)}]{dutra2005cavity}%
  \BibitemOpen
  \bibfield  {author} {\bibinfo {author} {\bibfnamefont {S.}~\bibnamefont
  {Dutra}},\ }\href {https://books.google.es/books?id=ZPCa9yM4QzIC} {\emph
  {\bibinfo {title} {Cavity Quantum Electrodynamics: The Strange Theory of
  Light in a Box}}},\ Wiley Series in Lasers and Applications\ (\bibinfo
  {publisher} {Wiley},\ \bibinfo {address} {Hoboken, NJ},\ \bibinfo {year}
  {2005})\BibitemShut {NoStop}%
\bibitem [{\citenamefont {Blockley}\ \emph {et~al.}(1992)\citenamefont
  {Blockley}, \citenamefont {Walls},\ and\ \citenamefont
  {Risken}}]{blockley_1992}%
  \BibitemOpen
  \bibfield  {author} {\bibinfo {author} {\bibfnamefont {C.~A.}\ \bibnamefont
  {Blockley}}, \bibinfo {author} {\bibfnamefont {D.~F.}\ \bibnamefont
  {Walls}},\ and\ \bibinfo {author} {\bibfnamefont {H.}~\bibnamefont
  {Risken}},\ }\bibfield  {title} {\bibinfo {title} {Quantum collapses and
  revivals in a quantized trap},\ }\href
  {https://doi.org/10.1209/0295-5075/17/6/006} {\bibfield  {journal} {\bibinfo
  {journal} {Europhysics Letters}\ }\textbf {\bibinfo {volume} {17}},\ \bibinfo
  {pages} {509} (\bibinfo {year} {1992})}\BibitemShut {NoStop}%
\bibitem [{\citenamefont {Monroe}\ \emph {et~al.}(1995)\citenamefont {Monroe},
  \citenamefont {Meekhof}, \citenamefont {King}, \citenamefont {Itano},\ and\
  \citenamefont {Wineland}}]{monroe_1995}%
  \BibitemOpen
  \bibfield  {author} {\bibinfo {author} {\bibfnamefont {C.}~\bibnamefont
  {Monroe}}, \bibinfo {author} {\bibfnamefont {D.~M.}\ \bibnamefont {Meekhof}},
  \bibinfo {author} {\bibfnamefont {B.~E.}\ \bibnamefont {King}}, \bibinfo
  {author} {\bibfnamefont {W.~M.}\ \bibnamefont {Itano}},\ and\ \bibinfo
  {author} {\bibfnamefont {D.~J.}\ \bibnamefont {Wineland}},\ }\bibfield
  {title} {\bibinfo {title} {Demonstration of a fundamental quantum logic
  gate},\ }\href {https://doi.org/10.1103/PhysRevLett.75.4714} {\bibfield
  {journal} {\bibinfo  {journal} {Phys. Rev. Lett.}\ }\textbf {\bibinfo
  {volume} {75}},\ \bibinfo {pages} {4714} (\bibinfo {year}
  {1995})}\BibitemShut {NoStop}%
\bibitem [{\citenamefont {Nakamura}\ \emph {et~al.}(1999)\citenamefont
  {Nakamura}, \citenamefont {Pashkin},\ and\ \citenamefont
  {Tsai}}]{Nakamura1999}%
  \BibitemOpen
  \bibfield  {author} {\bibinfo {author} {\bibfnamefont {Y.}~\bibnamefont
  {Nakamura}}, \bibinfo {author} {\bibfnamefont {Y.~A.}\ \bibnamefont
  {Pashkin}},\ and\ \bibinfo {author} {\bibfnamefont {J.~S.}\ \bibnamefont
  {Tsai}},\ }\bibfield  {title} {\bibinfo {title} {Coherent control of
  macroscopic quantum states in a single-cooper-pair box},\ }\href
  {https://doi.org/10.1038/19718} {\bibfield  {journal} {\bibinfo  {journal}
  {Nature}\ }\textbf {\bibinfo {volume} {398}},\ \bibinfo {pages} {786}
  (\bibinfo {year} {1999})}\BibitemShut {NoStop}%
\bibitem [{\citenamefont {Blais}\ \emph {et~al.}(2004)\citenamefont {Blais},
  \citenamefont {Huang}, \citenamefont {Wallraff}, \citenamefont {Girvin},\
  and\ \citenamefont {Schoelkopf}}]{blais_2004}%
  \BibitemOpen
  \bibfield  {author} {\bibinfo {author} {\bibfnamefont {A.}~\bibnamefont
  {Blais}}, \bibinfo {author} {\bibfnamefont {R.-S.}\ \bibnamefont {Huang}},
  \bibinfo {author} {\bibfnamefont {A.}~\bibnamefont {Wallraff}}, \bibinfo
  {author} {\bibfnamefont {S.~M.}\ \bibnamefont {Girvin}},\ and\ \bibinfo
  {author} {\bibfnamefont {R.~J.}\ \bibnamefont {Schoelkopf}},\ }\bibfield
  {title} {\bibinfo {title} {Cavity quantum electrodynamics for superconducting
  electrical circuits: An architecture for quantum computation},\ }\href
  {https://doi.org/10.1103/PhysRevA.69.062320} {\bibfield  {journal} {\bibinfo
  {journal} {Phys. Rev. A}\ }\textbf {\bibinfo {volume} {69}},\ \bibinfo
  {pages} {062320} (\bibinfo {year} {2004})}\BibitemShut {NoStop}%
\bibitem [{\citenamefont {Wallraff}\ \emph {et~al.}(2004)\citenamefont
  {Wallraff}, \citenamefont {Schuster}, \citenamefont {Blais}, \citenamefont
  {Frunzio}, \citenamefont {Huang}, \citenamefont {Majer}, \citenamefont
  {Kumar}, \citenamefont {Girvin},\ and\ \citenamefont
  {Schoelkopf}}]{Wallraff2004}%
  \BibitemOpen
  \bibfield  {author} {\bibinfo {author} {\bibfnamefont {A.}~\bibnamefont
  {Wallraff}}, \bibinfo {author} {\bibfnamefont {D.~I.}\ \bibnamefont
  {Schuster}}, \bibinfo {author} {\bibfnamefont {A.}~\bibnamefont {Blais}},
  \bibinfo {author} {\bibfnamefont {L.}~\bibnamefont {Frunzio}}, \bibinfo
  {author} {\bibfnamefont {R.-.~S.}\ \bibnamefont {Huang}}, \bibinfo {author}
  {\bibfnamefont {J.}~\bibnamefont {Majer}}, \bibinfo {author} {\bibfnamefont
  {S.}~\bibnamefont {Kumar}}, \bibinfo {author} {\bibfnamefont {S.~M.}\
  \bibnamefont {Girvin}},\ and\ \bibinfo {author} {\bibfnamefont {R.~J.}\
  \bibnamefont {Schoelkopf}},\ }\bibfield  {title} {\bibinfo {title} {Strong
  coupling of a single photon to a superconducting qubit using circuit quantum
  electrodynamics},\ }\href {https://doi.org/10.1038/nature02851} {\bibfield
  {journal} {\bibinfo  {journal} {Nature}\ }\textbf {\bibinfo {volume} {431}},\
  \bibinfo {pages} {162} (\bibinfo {year} {2004})}\BibitemShut {NoStop}%
\bibitem [{\citenamefont {Boissonneault}\ \emph {et~al.}(2009)\citenamefont
  {Boissonneault}, \citenamefont {Gambetta},\ and\ \citenamefont
  {Blais}}]{blais_2009}%
  \BibitemOpen
  \bibfield  {author} {\bibinfo {author} {\bibfnamefont {M.}~\bibnamefont
  {Boissonneault}}, \bibinfo {author} {\bibfnamefont {J.~M.}\ \bibnamefont
  {Gambetta}},\ and\ \bibinfo {author} {\bibfnamefont {A.}~\bibnamefont
  {Blais}},\ }\bibfield  {title} {\bibinfo {title} {Dispersive regime of
  circuit qed: Photon-dependent qubit dephasing and relaxation rates},\ }\href
  {https://doi.org/10.1103/PhysRevA.79.013819} {\bibfield  {journal} {\bibinfo
  {journal} {Phys. Rev. A}\ }\textbf {\bibinfo {volume} {79}},\ \bibinfo
  {pages} {013819} (\bibinfo {year} {2009})}\BibitemShut {NoStop}%
\bibitem [{\citenamefont {Blais}\ \emph {et~al.}(2021)\citenamefont {Blais},
  \citenamefont {Grimsmo}, \citenamefont {Girvin},\ and\ \citenamefont
  {Wallraff}}]{blais_2021}%
  \BibitemOpen
  \bibfield  {author} {\bibinfo {author} {\bibfnamefont {A.}~\bibnamefont
  {Blais}}, \bibinfo {author} {\bibfnamefont {A.~L.}\ \bibnamefont {Grimsmo}},
  \bibinfo {author} {\bibfnamefont {S.~M.}\ \bibnamefont {Girvin}},\ and\
  \bibinfo {author} {\bibfnamefont {A.}~\bibnamefont {Wallraff}},\ }\bibfield
  {title} {\bibinfo {title} {Circuit quantum electrodynamics},\ }\href
  {https://doi.org/10.1103/RevModPhys.93.025005} {\bibfield  {journal}
  {\bibinfo  {journal} {Rev. Mod. Phys.}\ }\textbf {\bibinfo {volume} {93}},\
  \bibinfo {pages} {025005} (\bibinfo {year} {2021})}\BibitemShut {NoStop}%
\bibitem [{\citenamefont {Marković}\ and\ \citenamefont
  {Grollier}(2020)}]{markovic2020}%
  \BibitemOpen
  \bibfield  {author} {\bibinfo {author} {\bibfnamefont {D.}~\bibnamefont
  {Marković}}\ and\ \bibinfo {author} {\bibfnamefont {J.}~\bibnamefont
  {Grollier}},\ }\bibfield  {title} {\bibinfo {title} {Quantum neuromorphic
  computing},\ }\href {https://doi.org/10.1063/5.0020014} {\bibfield  {journal}
  {\bibinfo  {journal} {Applied Physics Letters}\ }\textbf {\bibinfo {volume}
  {117}},\ \bibinfo {pages} {150501} (\bibinfo {year} {2020})}\BibitemShut
  {NoStop}%
\bibitem [{\citenamefont {Mujal}\ \emph
  {et~al.}(2021{\natexlab{a}})\citenamefont {Mujal}, \citenamefont
  {Martínez-Peña}, \citenamefont {Nokkala}, \citenamefont {García-Beni},
  \citenamefont {Giorgi}, \citenamefont {Soriano},\ and\ \citenamefont
  {Zambrini}}]{mujal_2021}%
  \BibitemOpen
  \bibfield  {author} {\bibinfo {author} {\bibfnamefont {P.}~\bibnamefont
  {Mujal}}, \bibinfo {author} {\bibfnamefont {R.}~\bibnamefont
  {Martínez-Peña}}, \bibinfo {author} {\bibfnamefont {J.}~\bibnamefont
  {Nokkala}}, \bibinfo {author} {\bibfnamefont {J.}~\bibnamefont
  {García-Beni}}, \bibinfo {author} {\bibfnamefont {G.~L.}\ \bibnamefont
  {Giorgi}}, \bibinfo {author} {\bibfnamefont {M.~C.}\ \bibnamefont
  {Soriano}},\ and\ \bibinfo {author} {\bibfnamefont {R.}~\bibnamefont
  {Zambrini}},\ }\bibfield  {title} {\bibinfo {title} {Opportunities in quantum
  reservoir computing and extreme learning machines},\ }\href
  {https://doi.org/https://doi.org/10.1002/qute.202100027} {\bibfield
  {journal} {\bibinfo  {journal} {Advanced Quantum Technologies}\ }\textbf
  {\bibinfo {volume} {4}},\ \bibinfo {pages} {2100027} (\bibinfo {year}
  {2021}{\natexlab{a}})}\BibitemShut {NoStop}%
\bibitem [{\citenamefont {Labay-Mora}\ \emph {et~al.}(2024)\citenamefont
  {Labay-Mora}, \citenamefont {García-Beni}, \citenamefont {Giorgi},
  \citenamefont {Soriano},\ and\ \citenamefont {Zambrini}}]{labay2024}%
  \BibitemOpen
  \bibfield  {author} {\bibinfo {author} {\bibfnamefont {A.}~\bibnamefont
  {Labay-Mora}}, \bibinfo {author} {\bibfnamefont {J.}~\bibnamefont
  {García-Beni}}, \bibinfo {author} {\bibfnamefont {G.~L.}\ \bibnamefont
  {Giorgi}}, \bibinfo {author} {\bibfnamefont {M.~C.}\ \bibnamefont
  {Soriano}},\ and\ \bibinfo {author} {\bibfnamefont {R.}~\bibnamefont
  {Zambrini}},\ }\bibfield  {title} {\bibinfo {title} {Neural networks with
  quantum states of light},\ }\href {https://doi.org/10.1098/rsta.2023.0346}
  {\bibfield  {journal} {\bibinfo  {journal} {Philosophical Transactions of the
  Royal Society A: Mathematical, Physical and Engineering Sciences}\ }\textbf
  {\bibinfo {volume} {382}},\ \bibinfo {pages} {20230346} (\bibinfo {year}
  {2024})}\BibitemShut {NoStop}%
\bibitem [{\citenamefont {Senanian}\ \emph {et~al.}(2024)\citenamefont
  {Senanian}, \citenamefont {Prabhu}, \citenamefont {Kremenetski},
  \citenamefont {Roy}, \citenamefont {Cao}, \citenamefont {Kline},
  \citenamefont {Onodera}, \citenamefont {Wright}, \citenamefont {Wu},
  \citenamefont {Fatemi},\ and\ \citenamefont {McMahon}}]{mcmahon2024}%
  \BibitemOpen
  \bibfield  {author} {\bibinfo {author} {\bibfnamefont {A.}~\bibnamefont
  {Senanian}}, \bibinfo {author} {\bibfnamefont {S.}~\bibnamefont {Prabhu}},
  \bibinfo {author} {\bibfnamefont {V.}~\bibnamefont {Kremenetski}}, \bibinfo
  {author} {\bibfnamefont {S.}~\bibnamefont {Roy}}, \bibinfo {author}
  {\bibfnamefont {Y.}~\bibnamefont {Cao}}, \bibinfo {author} {\bibfnamefont
  {J.}~\bibnamefont {Kline}}, \bibinfo {author} {\bibfnamefont
  {T.}~\bibnamefont {Onodera}}, \bibinfo {author} {\bibfnamefont {L.~G.}\
  \bibnamefont {Wright}}, \bibinfo {author} {\bibfnamefont {X.}~\bibnamefont
  {Wu}}, \bibinfo {author} {\bibfnamefont {V.}~\bibnamefont {Fatemi}},\ and\
  \bibinfo {author} {\bibfnamefont {P.~L.}\ \bibnamefont {McMahon}},\
  }\bibfield  {title} {\bibinfo {title} {Microwave signal processing using an
  analog quantum reservoir computer},\ }\href
  {https://doi.org/10.1038/s41467-024-51161-8} {\bibfield  {journal} {\bibinfo
  {journal} {Nature Communications}\ }\textbf {\bibinfo {volume} {15}},\
  \bibinfo {pages} {7490} (\bibinfo {year} {2024})}\BibitemShut {NoStop}%
\bibitem [{\citenamefont {Carles}\ \emph {et~al.}(2025)\citenamefont {Carles},
  \citenamefont {Dudas}, \citenamefont {Balembois}, \citenamefont {Grollier},\
  and\ \citenamefont {Marković}}]{carles2025}%
  \BibitemOpen
  \bibfield  {author} {\bibinfo {author} {\bibfnamefont {B.}~\bibnamefont
  {Carles}}, \bibinfo {author} {\bibfnamefont {J.}~\bibnamefont {Dudas}},
  \bibinfo {author} {\bibfnamefont {L.}~\bibnamefont {Balembois}}, \bibinfo
  {author} {\bibfnamefont {J.}~\bibnamefont {Grollier}},\ and\ \bibinfo
  {author} {\bibfnamefont {D.}~\bibnamefont {Marković}},\ }\href
  {https://arxiv.org/abs/2506.22016} {\bibinfo {title} {Experimental quantum
  reservoir computing with a circuit quantum electrodynamics system}} (\bibinfo
  {year} {2025}),\ \Eprint {https://arxiv.org/abs/2506.22016} {arXiv:2506.22016
  [quant-ph]} \BibitemShut {NoStop}%
\bibitem [{\citenamefont {Jaeger}\ and\ \citenamefont
  {Haas}(2004)}]{jaeger_2004}%
  \BibitemOpen
  \bibfield  {author} {\bibinfo {author} {\bibfnamefont {H.}~\bibnamefont
  {Jaeger}}\ and\ \bibinfo {author} {\bibfnamefont {H.}~\bibnamefont {Haas}},\
  }\bibfield  {title} {\bibinfo {title} {Harnessing nonlinearity: Predicting
  chaotic systems and saving energy in wireless communication},\ }\href
  {https://doi.org/10.1126/science.1091277} {\bibfield  {journal} {\bibinfo
  {journal} {Science}\ }\textbf {\bibinfo {volume} {304}},\ \bibinfo {pages}
  {78} (\bibinfo {year} {2004})}\BibitemShut {NoStop}%
\bibitem [{\citenamefont {Maass}\ \emph {et~al.}(2002)\citenamefont {Maass},
  \citenamefont {Natschl{\"a}ger},\ and\ \citenamefont {Markram}}]{maass_2002}%
  \BibitemOpen
  \bibfield  {author} {\bibinfo {author} {\bibfnamefont {W.}~\bibnamefont
  {Maass}}, \bibinfo {author} {\bibfnamefont {T.}~\bibnamefont
  {Natschl{\"a}ger}},\ and\ \bibinfo {author} {\bibfnamefont {H.}~\bibnamefont
  {Markram}},\ }\bibfield  {title} {\bibinfo {title} {Real-time computing
  without stable states: a new framework for neural computation based on
  perturbations},\ }\href@noop {} {\bibfield  {journal} {\bibinfo  {journal}
  {Neural Comput}\ }\textbf {\bibinfo {volume} {14}},\ \bibinfo {pages} {2531}
  (\bibinfo {year} {2002})}\BibitemShut {NoStop}%
\bibitem [{\citenamefont {Verstraeten}\ \emph {et~al.}(2007)\citenamefont
  {Verstraeten}, \citenamefont {Schrauwen}, \citenamefont {D'Haene},\ and\
  \citenamefont {Stroobandt}}]{Verstraeten2007}%
  \BibitemOpen
  \bibfield  {author} {\bibinfo {author} {\bibfnamefont {D.}~\bibnamefont
  {Verstraeten}}, \bibinfo {author} {\bibfnamefont {B.}~\bibnamefont
  {Schrauwen}}, \bibinfo {author} {\bibfnamefont {M.}~\bibnamefont {D'Haene}},\
  and\ \bibinfo {author} {\bibfnamefont {D.}~\bibnamefont {Stroobandt}},\
  }\bibfield  {title} {\bibinfo {title} {An experimental unification of
  reservoir computing methods},\ }\href@noop {} {\bibfield  {journal} {\bibinfo
   {journal} {Neural Networks}\ }\textbf {\bibinfo {volume} {20}},\ \bibinfo
  {pages} {391} (\bibinfo {year} {2007})}\BibitemShut {NoStop}%
\bibitem [{\citenamefont {Tanaka}\ \emph {et~al.}(2019)\citenamefont {Tanaka},
  \citenamefont {Yamane}, \citenamefont {Héroux}, \citenamefont {Nakane},
  \citenamefont {Kanazawa}, \citenamefont {Takeda}, \citenamefont {Numata},
  \citenamefont {Nakano},\ and\ \citenamefont {Hirose}}]{TANAKA2019}%
  \BibitemOpen
  \bibfield  {author} {\bibinfo {author} {\bibfnamefont {G.}~\bibnamefont
  {Tanaka}}, \bibinfo {author} {\bibfnamefont {T.}~\bibnamefont {Yamane}},
  \bibinfo {author} {\bibfnamefont {J.~B.}\ \bibnamefont {Héroux}}, \bibinfo
  {author} {\bibfnamefont {R.}~\bibnamefont {Nakane}}, \bibinfo {author}
  {\bibfnamefont {N.}~\bibnamefont {Kanazawa}}, \bibinfo {author}
  {\bibfnamefont {S.}~\bibnamefont {Takeda}}, \bibinfo {author} {\bibfnamefont
  {H.}~\bibnamefont {Numata}}, \bibinfo {author} {\bibfnamefont
  {D.}~\bibnamefont {Nakano}},\ and\ \bibinfo {author} {\bibfnamefont
  {A.}~\bibnamefont {Hirose}},\ }\bibfield  {title} {\bibinfo {title} {Recent
  advances in physical reservoir computing: A review},\ }\href
  {https://doi.org/https://doi.org/10.1016/j.neunet.2019.03.005} {\bibfield
  {journal} {\bibinfo  {journal} {Neural Networks}\ }\textbf {\bibinfo {volume}
  {115}},\ \bibinfo {pages} {100} (\bibinfo {year} {2019})}\BibitemShut
  {NoStop}%
\bibitem [{\citenamefont {Fujii}\ and\ \citenamefont
  {Nakajima}(2017)}]{fuji_qrc}%
  \BibitemOpen
  \bibfield  {author} {\bibinfo {author} {\bibfnamefont {K.}~\bibnamefont
  {Fujii}}\ and\ \bibinfo {author} {\bibfnamefont {K.}~\bibnamefont
  {Nakajima}},\ }\bibfield  {title} {\bibinfo {title} {Harnessing
  disordered-ensemble quantum dynamics for machine learning},\ }\href
  {https://doi.org/10.1103/PhysRevApplied.8.024030} {\bibfield  {journal}
  {\bibinfo  {journal} {Phys. Rev. Appl.}\ }\textbf {\bibinfo {volume} {8}},\
  \bibinfo {pages} {024030} (\bibinfo {year} {2017})}\BibitemShut {NoStop}%
\bibitem [{\citenamefont {Nokkala}(2023)}]{Nokkala2023}%
  \BibitemOpen
  \bibfield  {author} {\bibinfo {author} {\bibfnamefont {J.}~\bibnamefont
  {Nokkala}},\ }\bibfield  {title} {\bibinfo {title} {Online quantum time
  series processing with random oscillator networks},\ }\href
  {https://doi.org/10.1038/s41598-023-34811-7} {\bibfield  {journal} {\bibinfo
  {journal} {Scientific Reports}\ }\textbf {\bibinfo {volume} {13}},\ \bibinfo
  {pages} {7694} (\bibinfo {year} {2023})}\BibitemShut {NoStop}%
\bibitem [{\citenamefont {Nokkala}\ \emph {et~al.}(2024)\citenamefont
  {Nokkala}, \citenamefont {Giorgi},\ and\ \citenamefont
  {Zambrini}}]{Nokkala_2024}%
  \BibitemOpen
  \bibfield  {author} {\bibinfo {author} {\bibfnamefont {J.}~\bibnamefont
  {Nokkala}}, \bibinfo {author} {\bibfnamefont {G.~L.}\ \bibnamefont
  {Giorgi}},\ and\ \bibinfo {author} {\bibfnamefont {R.}~\bibnamefont
  {Zambrini}},\ }\bibfield  {title} {\bibinfo {title} {Retrieving past quantum
  features with deep hybrid classical-quantum reservoir computing},\ }\href
  {https://doi.org/10.1088/2632-2153/ad5f12} {\bibfield  {journal} {\bibinfo
  {journal} {Machine Learning: Science and Technology}\ }\textbf {\bibinfo
  {volume} {5}},\ \bibinfo {pages} {035022} (\bibinfo {year}
  {2024})}\BibitemShut {NoStop}%
\bibitem [{\citenamefont {McClean}\ \emph {et~al.}(2018)\citenamefont
  {McClean}, \citenamefont {Boixo}, \citenamefont {Smelyanskiy}, \citenamefont
  {Babbush},\ and\ \citenamefont {Neven}}]{mcclean_barren_2018}%
  \BibitemOpen
  \bibfield  {author} {\bibinfo {author} {\bibfnamefont {J.~R.}\ \bibnamefont
  {McClean}}, \bibinfo {author} {\bibfnamefont {S.}~\bibnamefont {Boixo}},
  \bibinfo {author} {\bibfnamefont {V.~N.}\ \bibnamefont {Smelyanskiy}},
  \bibinfo {author} {\bibfnamefont {R.}~\bibnamefont {Babbush}},\ and\ \bibinfo
  {author} {\bibfnamefont {H.}~\bibnamefont {Neven}},\ }\bibfield  {title}
  {\bibinfo {title} {Barren plateaus in quantum neural network training
  landscapes},\ }\href {https://doi.org/10.1038/s41467-018-07090-4} {\bibfield
  {journal} {\bibinfo  {journal} {Nature Communications}\ }\textbf {\bibinfo
  {volume} {9}},\ \bibinfo {pages} {4812} (\bibinfo {year} {2018})}\BibitemShut
  {NoStop}%
\bibitem [{\citenamefont {Anschuetz}\ and\ \citenamefont
  {Kiani}(2022)}]{Anschuetz2022}%
  \BibitemOpen
  \bibfield  {author} {\bibinfo {author} {\bibfnamefont {E.~R.}\ \bibnamefont
  {Anschuetz}}\ and\ \bibinfo {author} {\bibfnamefont {B.~T.}\ \bibnamefont
  {Kiani}},\ }\bibfield  {title} {\bibinfo {title} {Quantum variational
  algorithms are swamped with traps},\ }\href
  {https://doi.org/10.1038/s41467-022-35364-5} {\bibfield  {journal} {\bibinfo
  {journal} {Nature Communications}\ }\textbf {\bibinfo {volume} {13}},\
  \bibinfo {pages} {7760} (\bibinfo {year} {2022})}\BibitemShut {NoStop}%
\bibitem [{\citenamefont {Nakajima}\ \emph {et~al.}(2019)\citenamefont
  {Nakajima}, \citenamefont {Fujii}, \citenamefont {Negoro}, \citenamefont
  {Mitarai},\ and\ \citenamefont {Kitagawa}}]{nakajima_2019}%
  \BibitemOpen
  \bibfield  {author} {\bibinfo {author} {\bibfnamefont {K.}~\bibnamefont
  {Nakajima}}, \bibinfo {author} {\bibfnamefont {K.}~\bibnamefont {Fujii}},
  \bibinfo {author} {\bibfnamefont {M.}~\bibnamefont {Negoro}}, \bibinfo
  {author} {\bibfnamefont {K.}~\bibnamefont {Mitarai}},\ and\ \bibinfo {author}
  {\bibfnamefont {M.}~\bibnamefont {Kitagawa}},\ }\bibfield  {title} {\bibinfo
  {title} {Boosting computational power through spatial multiplexing in quantum
  reservoir computing},\ }\href
  {https://doi.org/10.1103/PhysRevApplied.11.034021} {\bibfield  {journal}
  {\bibinfo  {journal} {Phys. Rev. Appl.}\ }\textbf {\bibinfo {volume} {11}},\
  \bibinfo {pages} {034021} (\bibinfo {year} {2019})}\BibitemShut {NoStop}%
\bibitem [{\citenamefont {Mart\'{\i}nez-Pe\~na}\ \emph
  {et~al.}(2021)\citenamefont {Mart\'{\i}nez-Pe\~na}, \citenamefont {Giorgi},
  \citenamefont {Nokkala}, \citenamefont {Soriano},\ and\ \citenamefont
  {Zambrini}}]{martinez-pena_2021}%
  \BibitemOpen
  \bibfield  {author} {\bibinfo {author} {\bibfnamefont {R.}~\bibnamefont
  {Mart\'{\i}nez-Pe\~na}}, \bibinfo {author} {\bibfnamefont {G.~L.}\
  \bibnamefont {Giorgi}}, \bibinfo {author} {\bibfnamefont {J.}~\bibnamefont
  {Nokkala}}, \bibinfo {author} {\bibfnamefont {M.~C.}\ \bibnamefont
  {Soriano}},\ and\ \bibinfo {author} {\bibfnamefont {R.}~\bibnamefont
  {Zambrini}},\ }\bibfield  {title} {\bibinfo {title} {Dynamical phase
  transitions in quantum reservoir computing},\ }\href
  {https://doi.org/10.1103/PhysRevLett.127.100502} {\bibfield  {journal}
  {\bibinfo  {journal} {Phys. Rev. Lett.}\ }\textbf {\bibinfo {volume} {127}},\
  \bibinfo {pages} {100502} (\bibinfo {year} {2021})}\BibitemShut {NoStop}%
\bibitem [{\citenamefont {Mart{\'i}nez-Pe{\~{n}}a}\ \emph
  {et~al.}(2023)\citenamefont {Mart{\'i}nez-Pe{\~{n}}a}, \citenamefont
  {Nokkala}, \citenamefont {Giorgi}, \citenamefont {Zambrini},\ and\
  \citenamefont {Soriano}}]{Martínez-Peña2023}%
  \BibitemOpen
  \bibfield  {author} {\bibinfo {author} {\bibfnamefont {R.}~\bibnamefont
  {Mart{\'i}nez-Pe{\~{n}}a}}, \bibinfo {author} {\bibfnamefont
  {J.}~\bibnamefont {Nokkala}}, \bibinfo {author} {\bibfnamefont {G.~L.}\
  \bibnamefont {Giorgi}}, \bibinfo {author} {\bibfnamefont {R.}~\bibnamefont
  {Zambrini}},\ and\ \bibinfo {author} {\bibfnamefont {M.~C.}\ \bibnamefont
  {Soriano}},\ }\bibfield  {title} {\bibinfo {title} {Information processing
  capacity of spin-based quantum reservoir computing systems},\ }\href
  {https://doi.org/10.1007/s12559-020-09772-y} {\bibfield  {journal} {\bibinfo
  {journal} {Cognitive Computation}\ }\textbf {\bibinfo {volume} {15}},\
  \bibinfo {pages} {1440} (\bibinfo {year} {2023})}\BibitemShut {NoStop}%
\bibitem [{\citenamefont {Tran}\ and\ \citenamefont
  {Nakajima}(2021)}]{nakajima_2021}%
  \BibitemOpen
  \bibfield  {author} {\bibinfo {author} {\bibfnamefont {Q.~H.}\ \bibnamefont
  {Tran}}\ and\ \bibinfo {author} {\bibfnamefont {K.}~\bibnamefont
  {Nakajima}},\ }\bibfield  {title} {\bibinfo {title} {Learning temporal
  quantum tomography},\ }\href {https://doi.org/10.1103/PhysRevLett.127.260401}
  {\bibfield  {journal} {\bibinfo  {journal} {Phys. Rev. Lett.}\ }\textbf
  {\bibinfo {volume} {127}},\ \bibinfo {pages} {260401} (\bibinfo {year}
  {2021})}\BibitemShut {NoStop}%
\bibitem [{\citenamefont {Mujal}\ \emph {et~al.}(2023)\citenamefont {Mujal},
  \citenamefont {Mart{\'i}nez-Pe{\~{n}}a}, \citenamefont {Giorgi},
  \citenamefont {Soriano},\ and\ \citenamefont {Zambrini}}]{mujal_2023}%
  \BibitemOpen
  \bibfield  {author} {\bibinfo {author} {\bibfnamefont {P.}~\bibnamefont
  {Mujal}}, \bibinfo {author} {\bibfnamefont {R.}~\bibnamefont
  {Mart{\'i}nez-Pe{\~{n}}a}}, \bibinfo {author} {\bibfnamefont {G.~L.}\
  \bibnamefont {Giorgi}}, \bibinfo {author} {\bibfnamefont {M.~C.}\
  \bibnamefont {Soriano}},\ and\ \bibinfo {author} {\bibfnamefont
  {R.}~\bibnamefont {Zambrini}},\ }\bibfield  {title} {\bibinfo {title}
  {Time-series quantum reservoir computing with weak and projective
  measurements},\ }\href {https://doi.org/10.1038/s41534-023-00682-z}
  {\bibfield  {journal} {\bibinfo  {journal} {npj Quantum Information}\
  }\textbf {\bibinfo {volume} {9}},\ \bibinfo {pages} {16} (\bibinfo {year}
  {2023})}\BibitemShut {NoStop}%
\bibitem [{\citenamefont {G\"otting}\ \emph {et~al.}(2023)\citenamefont
  {G\"otting}, \citenamefont {Lohof},\ and\ \citenamefont
  {Gies}}]{niclas_2023}%
  \BibitemOpen
  \bibfield  {author} {\bibinfo {author} {\bibfnamefont {N.}~\bibnamefont
  {G\"otting}}, \bibinfo {author} {\bibfnamefont {F.}~\bibnamefont {Lohof}},\
  and\ \bibinfo {author} {\bibfnamefont {C.}~\bibnamefont {Gies}},\ }\bibfield
  {title} {\bibinfo {title} {Exploring quantumness in quantum reservoir
  computing},\ }\href {https://doi.org/10.1103/PhysRevA.108.052427} {\bibfield
  {journal} {\bibinfo  {journal} {Phys. Rev. A}\ }\textbf {\bibinfo {volume}
  {108}},\ \bibinfo {pages} {052427} (\bibinfo {year} {2023})}\BibitemShut
  {NoStop}%
\bibitem [{\citenamefont {Sannia}\ \emph {et~al.}(2024)\citenamefont {Sannia},
  \citenamefont {Mart{\'{i}}nez-Pe{\~{n}}a}, \citenamefont {Soriano},
  \citenamefont {Giorgi},\ and\ \citenamefont {Zambrini}}]{sannia_2024}%
  \BibitemOpen
  \bibfield  {author} {\bibinfo {author} {\bibfnamefont {A.}~\bibnamefont
  {Sannia}}, \bibinfo {author} {\bibfnamefont {R.}~\bibnamefont
  {Mart{\'{i}}nez-Pe{\~{n}}a}}, \bibinfo {author} {\bibfnamefont {M.~C.}\
  \bibnamefont {Soriano}}, \bibinfo {author} {\bibfnamefont {G.~L.}\
  \bibnamefont {Giorgi}},\ and\ \bibinfo {author} {\bibfnamefont
  {R.}~\bibnamefont {Zambrini}},\ }\bibfield  {title} {\bibinfo {title}
  {Dissipation as a resource for {Q}uantum {R}eservoir {C}omputing},\ }\href
  {https://doi.org/10.22331/q-2024-03-20-1291} {\bibfield  {journal} {\bibinfo
  {journal} {{Quantum}}\ }\textbf {\bibinfo {volume} {8}},\ \bibinfo {pages}
  {1291} (\bibinfo {year} {2024})}\BibitemShut {NoStop}%
\bibitem [{\citenamefont {Li}\ \emph {et~al.}(2026)\citenamefont {Li},
  \citenamefont {Mukhopadhyay}, \citenamefont {Bayat},\ and\ \citenamefont
  {Habibnia}}]{li2025}%
  \BibitemOpen
  \bibfield  {author} {\bibinfo {author} {\bibfnamefont {Q.}~\bibnamefont
  {Li}}, \bibinfo {author} {\bibfnamefont {C.}~\bibnamefont {Mukhopadhyay}},
  \bibinfo {author} {\bibfnamefont {A.}~\bibnamefont {Bayat}},\ and\ \bibinfo
  {author} {\bibfnamefont {A.}~\bibnamefont {Habibnia}},\ }\bibfield  {title}
  {\bibinfo {title} {Quantum reservoir computing for realized volatility
  forecasting},\ }\href {https://doi.org/10.1103/rbj7-4wnq} {\bibfield
  {journal} {\bibinfo  {journal} {Phys. Rev. Res.}\ }\textbf {\bibinfo {volume}
  {8}},\ \bibinfo {pages} {023028} (\bibinfo {year} {2026})}\BibitemShut
  {NoStop}%
\bibitem [{\citenamefont {Hou}\ \emph {et~al.}(2026)\citenamefont {Hou},
  \citenamefont {Hua}, \citenamefont {Wu}, \citenamefont {Xia}, \citenamefont
  {Chen}, \citenamefont {Li}, \citenamefont {Li}, \citenamefont {Peng},\ and\
  \citenamefont {Du}}]{hou2025}%
  \BibitemOpen
  \bibfield  {author} {\bibinfo {author} {\bibfnamefont {Y.}~\bibnamefont
  {Hou}}, \bibinfo {author} {\bibfnamefont {J.}~\bibnamefont {Hua}}, \bibinfo
  {author} {\bibfnamefont {Z.}~\bibnamefont {Wu}}, \bibinfo {author}
  {\bibfnamefont {W.}~\bibnamefont {Xia}}, \bibinfo {author} {\bibfnamefont
  {Y.}~\bibnamefont {Chen}}, \bibinfo {author} {\bibfnamefont {X.}~\bibnamefont
  {Li}}, \bibinfo {author} {\bibfnamefont {Z.}~\bibnamefont {Li}}, \bibinfo
  {author} {\bibfnamefont {X.}~\bibnamefont {Peng}},\ and\ \bibinfo {author}
  {\bibfnamefont {J.}~\bibnamefont {Du}},\ }\bibfield  {title} {\bibinfo
  {title} {High-accuracy temporal prediction via experimental quantum reservoir
  computing in correlated spins},\ }\href {https://doi.org/10.1103/r8ww-qw7j}
  {\bibfield  {journal} {\bibinfo  {journal} {Phys. Rev. Lett.}\ }\textbf
  {\bibinfo {volume} {136}},\ \bibinfo {pages} {120602} (\bibinfo {year}
  {2026})}\BibitemShut {NoStop}%
\bibitem [{\citenamefont {Nokkala}\ \emph {et~al.}(2021)\citenamefont
  {Nokkala}, \citenamefont {Mart{\'i}nez-Pe{\~{n}}a}, \citenamefont {Giorgi},
  \citenamefont {Parigi}, \citenamefont {Soriano},\ and\ \citenamefont
  {Zambrini}}]{Nokkala2021}%
  \BibitemOpen
  \bibfield  {author} {\bibinfo {author} {\bibfnamefont {J.}~\bibnamefont
  {Nokkala}}, \bibinfo {author} {\bibfnamefont {R.}~\bibnamefont
  {Mart{\'i}nez-Pe{\~{n}}a}}, \bibinfo {author} {\bibfnamefont {G.~L.}\
  \bibnamefont {Giorgi}}, \bibinfo {author} {\bibfnamefont {V.}~\bibnamefont
  {Parigi}}, \bibinfo {author} {\bibfnamefont {M.~C.}\ \bibnamefont
  {Soriano}},\ and\ \bibinfo {author} {\bibfnamefont {R.}~\bibnamefont
  {Zambrini}},\ }\bibfield  {title} {\bibinfo {title} {Gaussian states of
  continuous-variable quantum systems provide universal and versatile reservoir
  computing},\ }\href {https://doi.org/10.1038/s42005-021-00556-w} {\bibfield
  {journal} {\bibinfo  {journal} {Communications Physics}\ }\textbf {\bibinfo
  {volume} {4}},\ \bibinfo {pages} {53} (\bibinfo {year} {2021})}\BibitemShut
  {NoStop}%
\bibitem [{\citenamefont {Govia}\ \emph {et~al.}(2021)\citenamefont {Govia},
  \citenamefont {Ribeill}, \citenamefont {Rowlands}, \citenamefont {Krovi},\
  and\ \citenamefont {Ohki}}]{govia_2021}%
  \BibitemOpen
  \bibfield  {author} {\bibinfo {author} {\bibfnamefont {L.~C.~G.}\
  \bibnamefont {Govia}}, \bibinfo {author} {\bibfnamefont {G.~J.}\ \bibnamefont
  {Ribeill}}, \bibinfo {author} {\bibfnamefont {G.~E.}\ \bibnamefont
  {Rowlands}}, \bibinfo {author} {\bibfnamefont {H.~K.}\ \bibnamefont
  {Krovi}},\ and\ \bibinfo {author} {\bibfnamefont {T.~A.}\ \bibnamefont
  {Ohki}},\ }\bibfield  {title} {\bibinfo {title} {Quantum reservoir computing
  with a single nonlinear oscillator},\ }\href
  {https://doi.org/10.1103/PhysRevResearch.3.013077} {\bibfield  {journal}
  {\bibinfo  {journal} {Phys. Rev. Res.}\ }\textbf {\bibinfo {volume} {3}},\
  \bibinfo {pages} {013077} (\bibinfo {year} {2021})}\BibitemShut {NoStop}%
\bibitem [{\citenamefont {Khan}\ \emph {et~al.}(2021)\citenamefont {Khan},
  \citenamefont {Hu}, \citenamefont {Angelatos},\ and\ \citenamefont
  {Türeci}}]{khan2021}%
  \BibitemOpen
  \bibfield  {author} {\bibinfo {author} {\bibfnamefont {S.~A.}\ \bibnamefont
  {Khan}}, \bibinfo {author} {\bibfnamefont {F.}~\bibnamefont {Hu}}, \bibinfo
  {author} {\bibfnamefont {G.}~\bibnamefont {Angelatos}},\ and\ \bibinfo
  {author} {\bibfnamefont {H.~E.}\ \bibnamefont {Türeci}},\ }\href
  {https://arxiv.org/abs/2110.13849} {\bibinfo {title} {Physical reservoir
  computing using finitely-sampled quantum systems}} (\bibinfo {year} {2021}),\
  \Eprint {https://arxiv.org/abs/2110.13849} {arXiv:2110.13849 [quant-ph]}
  \BibitemShut {NoStop}%
\bibitem [{\citenamefont {Kalfus}\ \emph {et~al.}(2022)\citenamefont {Kalfus},
  \citenamefont {Ribeill}, \citenamefont {Rowlands}, \citenamefont {Krovi},
  \citenamefont {Ohki},\ and\ \citenamefont {Govia}}]{kalfus_2022}%
  \BibitemOpen
  \bibfield  {author} {\bibinfo {author} {\bibfnamefont {W.~D.}\ \bibnamefont
  {Kalfus}}, \bibinfo {author} {\bibfnamefont {G.~J.}\ \bibnamefont {Ribeill}},
  \bibinfo {author} {\bibfnamefont {G.~E.}\ \bibnamefont {Rowlands}}, \bibinfo
  {author} {\bibfnamefont {H.~K.}\ \bibnamefont {Krovi}}, \bibinfo {author}
  {\bibfnamefont {T.~A.}\ \bibnamefont {Ohki}},\ and\ \bibinfo {author}
  {\bibfnamefont {L.~C.~G.}\ \bibnamefont {Govia}},\ }\bibfield  {title}
  {\bibinfo {title} {Hilbert space as a computational resource in reservoir
  computing},\ }\href {https://doi.org/10.1103/PhysRevResearch.4.033007}
  {\bibfield  {journal} {\bibinfo  {journal} {Phys. Rev. Res.}\ }\textbf
  {\bibinfo {volume} {4}},\ \bibinfo {pages} {033007} (\bibinfo {year}
  {2022})}\BibitemShut {NoStop}%
\bibitem [{\citenamefont {Garc\'{\i}a-Beni}\ \emph {et~al.}(2023)\citenamefont
  {Garc\'{\i}a-Beni}, \citenamefont {Giorgi}, \citenamefont {Soriano},\ and\
  \citenamefont {Zambrini}}]{jorge_2023}%
  \BibitemOpen
  \bibfield  {author} {\bibinfo {author} {\bibfnamefont {J.}~\bibnamefont
  {Garc\'{\i}a-Beni}}, \bibinfo {author} {\bibfnamefont {G.~L.}\ \bibnamefont
  {Giorgi}}, \bibinfo {author} {\bibfnamefont {M.~C.}\ \bibnamefont
  {Soriano}},\ and\ \bibinfo {author} {\bibfnamefont {R.}~\bibnamefont
  {Zambrini}},\ }\bibfield  {title} {\bibinfo {title} {Scalable photonic
  platform for real-time quantum reservoir computing},\ }\href
  {https://doi.org/10.1103/PhysRevApplied.20.014051} {\bibfield  {journal}
  {\bibinfo  {journal} {Phys. Rev. Appl.}\ }\textbf {\bibinfo {volume} {20}},\
  \bibinfo {pages} {014051} (\bibinfo {year} {2023})}\BibitemShut {NoStop}%
\bibitem [{\citenamefont {Dudas}\ \emph {et~al.}(2023)\citenamefont {Dudas},
  \citenamefont {Carles}, \citenamefont {Plouet}, \citenamefont {Mizrahi},
  \citenamefont {Grollier},\ and\ \citenamefont {Markovi{\'{c}}}}]{Dudas2023}%
  \BibitemOpen
  \bibfield  {author} {\bibinfo {author} {\bibfnamefont {J.}~\bibnamefont
  {Dudas}}, \bibinfo {author} {\bibfnamefont {B.}~\bibnamefont {Carles}},
  \bibinfo {author} {\bibfnamefont {E.}~\bibnamefont {Plouet}}, \bibinfo
  {author} {\bibfnamefont {F.~A.}\ \bibnamefont {Mizrahi}}, \bibinfo {author}
  {\bibfnamefont {J.}~\bibnamefont {Grollier}},\ and\ \bibinfo {author}
  {\bibfnamefont {D.}~\bibnamefont {Markovi{\'{c}}}},\ }\bibfield  {title}
  {\bibinfo {title} {Quantum reservoir computing implementation on coherently
  coupled quantum oscillators},\ }\href
  {https://doi.org/10.1038/s41534-023-00734-4} {\bibfield  {journal} {\bibinfo
  {journal} {npj Quantum Information}\ }\textbf {\bibinfo {volume} {9}},\
  \bibinfo {pages} {64} (\bibinfo {year} {2023})}\BibitemShut {NoStop}%
\bibitem [{\citenamefont {Llodrà}\ \emph {et~al.}(2023)\citenamefont
  {Llodrà}, \citenamefont {Charalambous}, \citenamefont {Giorgi},\ and\
  \citenamefont {Zambrini}}]{llodra_2023}%
  \BibitemOpen
  \bibfield  {author} {\bibinfo {author} {\bibfnamefont {G.}~\bibnamefont
  {Llodrà}}, \bibinfo {author} {\bibfnamefont {C.}~\bibnamefont
  {Charalambous}}, \bibinfo {author} {\bibfnamefont {G.~L.}\ \bibnamefont
  {Giorgi}},\ and\ \bibinfo {author} {\bibfnamefont {R.}~\bibnamefont
  {Zambrini}},\ }\bibfield  {title} {\bibinfo {title} {Benchmarking the role of
  particle statistics in quantum reservoir computing},\ }\href
  {https://doi.org/https://doi.org/10.1002/qute.202200100} {\bibfield
  {journal} {\bibinfo  {journal} {Advanced Quantum Technologies}\ }\textbf
  {\bibinfo {volume} {6}},\ \bibinfo {pages} {2200100} (\bibinfo {year}
  {2023})}\BibitemShut {NoStop}%
\bibitem [{\citenamefont {Sannia}\ \emph {et~al.}(2025)\citenamefont {Sannia},
  \citenamefont {Giorgi}, \citenamefont {Longhi},\ and\ \citenamefont
  {Zambrini}}]{sannia_2025}%
  \BibitemOpen
  \bibfield  {author} {\bibinfo {author} {\bibfnamefont {A.}~\bibnamefont
  {Sannia}}, \bibinfo {author} {\bibfnamefont {G.~L.}\ \bibnamefont {Giorgi}},
  \bibinfo {author} {\bibfnamefont {S.}~\bibnamefont {Longhi}},\ and\ \bibinfo
  {author} {\bibfnamefont {R.}~\bibnamefont {Zambrini}},\ }\bibfield  {title}
  {\bibinfo {title} {Liouvillian skin effect in quantum neural networks},\
  }\href {https://doi.org/10.1364/OPTICAQ.541744} {\bibfield  {journal}
  {\bibinfo  {journal} {Optica Quantum}\ }\textbf {\bibinfo {volume} {3}},\
  \bibinfo {pages} {189} (\bibinfo {year} {2025})}\BibitemShut {NoStop}%
\bibitem [{\citenamefont {Chen}\ \emph {et~al.}(2020)\citenamefont {Chen},
  \citenamefont {Nurdin},\ and\ \citenamefont {Yamamoto}}]{chen_2020}%
  \BibitemOpen
  \bibfield  {author} {\bibinfo {author} {\bibfnamefont {J.}~\bibnamefont
  {Chen}}, \bibinfo {author} {\bibfnamefont {H.~I.}\ \bibnamefont {Nurdin}},\
  and\ \bibinfo {author} {\bibfnamefont {N.}~\bibnamefont {Yamamoto}},\
  }\bibfield  {title} {\bibinfo {title} {Temporal information processing on
  noisy quantum computers},\ }\href
  {https://doi.org/10.1103/PhysRevApplied.14.024065} {\bibfield  {journal}
  {\bibinfo  {journal} {Phys. Rev. Appl.}\ }\textbf {\bibinfo {volume} {14}},\
  \bibinfo {pages} {024065} (\bibinfo {year} {2020})}\BibitemShut {NoStop}%
\bibitem [{\citenamefont {Kobayashi}\ \emph {et~al.}(2024)\citenamefont
  {Kobayashi}, \citenamefont {Fujii},\ and\ \citenamefont
  {Yamamoto}}]{kobayashi_2024}%
  \BibitemOpen
  \bibfield  {author} {\bibinfo {author} {\bibfnamefont {K.}~\bibnamefont
  {Kobayashi}}, \bibinfo {author} {\bibfnamefont {K.}~\bibnamefont {Fujii}},\
  and\ \bibinfo {author} {\bibfnamefont {N.}~\bibnamefont {Yamamoto}},\
  }\bibfield  {title} {\bibinfo {title} {Feedback-driven quantum reservoir
  computing for time-series analysis},\ }\href
  {https://doi.org/10.1103/PRXQuantum.5.040325} {\bibfield  {journal} {\bibinfo
   {journal} {PRX Quantum}\ }\textbf {\bibinfo {volume} {5}},\ \bibinfo {pages}
  {040325} (\bibinfo {year} {2024})}\BibitemShut {NoStop}%
\bibitem [{\citenamefont {Hu}\ \emph {et~al.}(2024)\citenamefont {Hu},
  \citenamefont {Khan}, \citenamefont {Bronn}, \citenamefont {Angelatos},
  \citenamefont {Rowlands}, \citenamefont {Ribeill},\ and\ \citenamefont
  {T{\"u}reci}}]{Hu_NatComm_2024}%
  \BibitemOpen
  \bibfield  {author} {\bibinfo {author} {\bibfnamefont {F.}~\bibnamefont
  {Hu}}, \bibinfo {author} {\bibfnamefont {S.~A.}\ \bibnamefont {Khan}},
  \bibinfo {author} {\bibfnamefont {N.~T.}\ \bibnamefont {Bronn}}, \bibinfo
  {author} {\bibfnamefont {G.}~\bibnamefont {Angelatos}}, \bibinfo {author}
  {\bibfnamefont {G.~E.}\ \bibnamefont {Rowlands}}, \bibinfo {author}
  {\bibfnamefont {G.~J.}\ \bibnamefont {Ribeill}},\ and\ \bibinfo {author}
  {\bibfnamefont {H.~E.}\ \bibnamefont {T{\"u}reci}},\ }\bibfield  {title}
  {\bibinfo {title} {Overcoming the coherence time barrier in quantum machine
  learning on temporal data},\ }\href
  {https://doi.org/10.1038/s41467-024-51162-7} {\bibfield  {journal} {\bibinfo
  {journal} {Nature Communications}\ }\textbf {\bibinfo {volume} {15}},\
  \bibinfo {pages} {7491} (\bibinfo {year} {2024})}\BibitemShut {NoStop}%
\bibitem [{\citenamefont {Selimović}\ \emph {et~al.}(2025)\citenamefont
  {Selimović}, \citenamefont {Agresti}, \citenamefont {Siemaszko},
  \citenamefont {Morris}, \citenamefont {Dakić}, \citenamefont {Albiero},
  \citenamefont {Crespi}, \citenamefont {Ceccarelli}, \citenamefont {Osellame},
  \citenamefont {Stobińska},\ and\ \citenamefont {Walther}}]{selimović2025}%
  \BibitemOpen
  \bibfield  {author} {\bibinfo {author} {\bibfnamefont {M.}~\bibnamefont
  {Selimović}}, \bibinfo {author} {\bibfnamefont {I.}~\bibnamefont {Agresti}},
  \bibinfo {author} {\bibfnamefont {M.}~\bibnamefont {Siemaszko}}, \bibinfo
  {author} {\bibfnamefont {J.}~\bibnamefont {Morris}}, \bibinfo {author}
  {\bibfnamefont {B.}~\bibnamefont {Dakić}}, \bibinfo {author} {\bibfnamefont
  {R.}~\bibnamefont {Albiero}}, \bibinfo {author} {\bibfnamefont
  {A.}~\bibnamefont {Crespi}}, \bibinfo {author} {\bibfnamefont
  {F.}~\bibnamefont {Ceccarelli}}, \bibinfo {author} {\bibfnamefont
  {R.}~\bibnamefont {Osellame}}, \bibinfo {author} {\bibfnamefont
  {M.}~\bibnamefont {Stobińska}},\ and\ \bibinfo {author} {\bibfnamefont
  {P.}~\bibnamefont {Walther}},\ }\href {https://arxiv.org/abs/2504.18694}
  {\bibinfo {title} {Experimental neuromorphic computing based on quantum
  memristor}} (\bibinfo {year} {2025}),\ \Eprint
  {https://arxiv.org/abs/2504.18694} {arXiv:2504.18694 [quant-ph]} \BibitemShut
  {NoStop}%
\bibitem [{\citenamefont {Paparelle}\ \emph {et~al.}(2026)\citenamefont
  {Paparelle}, \citenamefont {Henaff}, \citenamefont {Garc{\'i}a-Beni},
  \citenamefont {Gillet}, \citenamefont {Montesinos}, \citenamefont {Giorgi},
  \citenamefont {Soriano}, \citenamefont {Zambrini},\ and\ \citenamefont
  {Parigi}}]{paparelle2025}%
  \BibitemOpen
  \bibfield  {author} {\bibinfo {author} {\bibfnamefont {I.}~\bibnamefont
  {Paparelle}}, \bibinfo {author} {\bibfnamefont {J.}~\bibnamefont {Henaff}},
  \bibinfo {author} {\bibfnamefont {J.}~\bibnamefont {Garc{\'i}a-Beni}},
  \bibinfo {author} {\bibfnamefont {{\'E}.}~\bibnamefont {Gillet}}, \bibinfo
  {author} {\bibfnamefont {D.}~\bibnamefont {Montesinos}}, \bibinfo {author}
  {\bibfnamefont {G.~L.}\ \bibnamefont {Giorgi}}, \bibinfo {author}
  {\bibfnamefont {M.~C.}\ \bibnamefont {Soriano}}, \bibinfo {author}
  {\bibfnamefont {R.}~\bibnamefont {Zambrini}},\ and\ \bibinfo {author}
  {\bibfnamefont {V.}~\bibnamefont {Parigi}},\ }\bibfield  {title} {\bibinfo
  {title} {Experimental memory control in continuous-variable optical quantum
  reservoir computing},\ }\href {https://doi.org/10.1038/s41566-026-01880-9}
  {\bibfield  {journal} {\bibinfo  {journal} {Nature Photonics}\ }\textbf
  {\bibinfo {volume} {20}},\ \bibinfo {pages} {413} (\bibinfo {year}
  {2026})}\BibitemShut {NoStop}%
\bibitem [{\citenamefont {Kornjača}\ \emph {et~al.}(2024)\citenamefont
  {Kornjača}, \citenamefont {Hu}, \citenamefont {Zhao}, \citenamefont {Wurtz},
  \citenamefont {Weinberg}, \citenamefont {Hamdan}, \citenamefont {Zhdanov},
  \citenamefont {Cantu}, \citenamefont {Zhou}, \citenamefont {Bravo},
  \citenamefont {Bagnall}, \citenamefont {Basham}, \citenamefont {Campo},
  \citenamefont {Choukri}, \citenamefont {DeAngelo}, \citenamefont {Frederick},
  \citenamefont {Haines}, \citenamefont {Hammett}, \citenamefont {Hsu},
  \citenamefont {Hu}, \citenamefont {Huber}, \citenamefont {Jepsen},
  \citenamefont {Jia}, \citenamefont {Karolyshyn}, \citenamefont {Kwon},
  \citenamefont {Long}, \citenamefont {Lopatin}, \citenamefont {Lukin},
  \citenamefont {Macrì}, \citenamefont {Marković}, \citenamefont
  {Martínez-Martínez}, \citenamefont {Meng}, \citenamefont {Ostroumov},
  \citenamefont {Paquette}, \citenamefont {Robinson}, \citenamefont
  {Rodriguez}, \citenamefont {Singh}, \citenamefont {Sinha}, \citenamefont
  {Thoreen}, \citenamefont {Wan}, \citenamefont {Waxman-Lenz}, \citenamefont
  {Wong}, \citenamefont {Wu}, \citenamefont {Lopes}, \citenamefont {Boger},
  \citenamefont {Gemelke}, \citenamefont {Kitagawa}, \citenamefont {Keesling},
  \citenamefont {Gao}, \citenamefont {Bylinskii}, \citenamefont {Yelin},
  \citenamefont {Liu},\ and\ \citenamefont {Wang}}]{kornjača2024}%
  \BibitemOpen
  \bibfield  {author} {\bibinfo {author} {\bibfnamefont {M.}~\bibnamefont
  {Kornjača}}, \bibinfo {author} {\bibfnamefont {H.-Y.}\ \bibnamefont {Hu}},
  \bibinfo {author} {\bibfnamefont {C.}~\bibnamefont {Zhao}}, \bibinfo {author}
  {\bibfnamefont {J.}~\bibnamefont {Wurtz}}, \bibinfo {author} {\bibfnamefont
  {P.}~\bibnamefont {Weinberg}}, \bibinfo {author} {\bibfnamefont
  {M.}~\bibnamefont {Hamdan}}, \bibinfo {author} {\bibfnamefont
  {A.}~\bibnamefont {Zhdanov}}, \bibinfo {author} {\bibfnamefont {S.~H.}\
  \bibnamefont {Cantu}}, \bibinfo {author} {\bibfnamefont {H.}~\bibnamefont
  {Zhou}}, \bibinfo {author} {\bibfnamefont {R.~A.}\ \bibnamefont {Bravo}},
  \bibinfo {author} {\bibfnamefont {K.}~\bibnamefont {Bagnall}}, \bibinfo
  {author} {\bibfnamefont {J.~I.}\ \bibnamefont {Basham}}, \bibinfo {author}
  {\bibfnamefont {J.}~\bibnamefont {Campo}}, \bibinfo {author} {\bibfnamefont
  {A.}~\bibnamefont {Choukri}}, \bibinfo {author} {\bibfnamefont
  {R.}~\bibnamefont {DeAngelo}}, \bibinfo {author} {\bibfnamefont
  {P.}~\bibnamefont {Frederick}}, \bibinfo {author} {\bibfnamefont
  {D.}~\bibnamefont {Haines}}, \bibinfo {author} {\bibfnamefont
  {J.}~\bibnamefont {Hammett}}, \bibinfo {author} {\bibfnamefont
  {N.}~\bibnamefont {Hsu}}, \bibinfo {author} {\bibfnamefont {M.-G.}\
  \bibnamefont {Hu}}, \bibinfo {author} {\bibfnamefont {F.}~\bibnamefont
  {Huber}}, \bibinfo {author} {\bibfnamefont {P.~N.}\ \bibnamefont {Jepsen}},
  \bibinfo {author} {\bibfnamefont {N.}~\bibnamefont {Jia}}, \bibinfo {author}
  {\bibfnamefont {T.}~\bibnamefont {Karolyshyn}}, \bibinfo {author}
  {\bibfnamefont {M.}~\bibnamefont {Kwon}}, \bibinfo {author} {\bibfnamefont
  {J.}~\bibnamefont {Long}}, \bibinfo {author} {\bibfnamefont {J.}~\bibnamefont
  {Lopatin}}, \bibinfo {author} {\bibfnamefont {A.}~\bibnamefont {Lukin}},
  \bibinfo {author} {\bibfnamefont {T.}~\bibnamefont {Macrì}}, \bibinfo
  {author} {\bibfnamefont {O.}~\bibnamefont {Marković}}, \bibinfo {author}
  {\bibfnamefont {L.~A.}\ \bibnamefont {Martínez-Martínez}}, \bibinfo
  {author} {\bibfnamefont {X.}~\bibnamefont {Meng}}, \bibinfo {author}
  {\bibfnamefont {E.}~\bibnamefont {Ostroumov}}, \bibinfo {author}
  {\bibfnamefont {D.}~\bibnamefont {Paquette}}, \bibinfo {author}
  {\bibfnamefont {J.}~\bibnamefont {Robinson}}, \bibinfo {author}
  {\bibfnamefont {P.~S.}\ \bibnamefont {Rodriguez}}, \bibinfo {author}
  {\bibfnamefont {A.}~\bibnamefont {Singh}}, \bibinfo {author} {\bibfnamefont
  {N.}~\bibnamefont {Sinha}}, \bibinfo {author} {\bibfnamefont
  {H.}~\bibnamefont {Thoreen}}, \bibinfo {author} {\bibfnamefont
  {N.}~\bibnamefont {Wan}}, \bibinfo {author} {\bibfnamefont {D.}~\bibnamefont
  {Waxman-Lenz}}, \bibinfo {author} {\bibfnamefont {T.}~\bibnamefont {Wong}},
  \bibinfo {author} {\bibfnamefont {K.-H.}\ \bibnamefont {Wu}}, \bibinfo
  {author} {\bibfnamefont {P.~L.~S.}\ \bibnamefont {Lopes}}, \bibinfo {author}
  {\bibfnamefont {Y.}~\bibnamefont {Boger}}, \bibinfo {author} {\bibfnamefont
  {N.}~\bibnamefont {Gemelke}}, \bibinfo {author} {\bibfnamefont
  {T.}~\bibnamefont {Kitagawa}}, \bibinfo {author} {\bibfnamefont
  {A.}~\bibnamefont {Keesling}}, \bibinfo {author} {\bibfnamefont
  {X.}~\bibnamefont {Gao}}, \bibinfo {author} {\bibfnamefont {A.}~\bibnamefont
  {Bylinskii}}, \bibinfo {author} {\bibfnamefont {S.~F.}\ \bibnamefont
  {Yelin}}, \bibinfo {author} {\bibfnamefont {F.}~\bibnamefont {Liu}},\ and\
  \bibinfo {author} {\bibfnamefont {S.-T.}\ \bibnamefont {Wang}},\ }\href
  {https://arxiv.org/abs/2407.02553} {\bibinfo {title} {Large-scale quantum
  reservoir learning with an analog quantum computer}} (\bibinfo {year}
  {2024}),\ \Eprint {https://arxiv.org/abs/2407.02553} {arXiv:2407.02553
  [quant-ph]} \BibitemShut {NoStop}%
\bibitem [{\citenamefont {Abbas}\ and\ \citenamefont
  {Maksymov}(2024)}]{electronics2024}%
  \BibitemOpen
  \bibfield  {author} {\bibinfo {author} {\bibfnamefont {A.~H.}\ \bibnamefont
  {Abbas}}\ and\ \bibinfo {author} {\bibfnamefont {I.~S.}\ \bibnamefont
  {Maksymov}},\ }\bibfield  {title} {\bibinfo {title} {Reservoir computing
  using measurement-controlled quantum dynamics},\ }\bibfield  {journal}
  {\bibinfo  {journal} {Electronics}\ }\textbf {\bibinfo {volume} {13}},\ \href
  {https://doi.org/10.3390/electronics13061164} {10.3390/electronics13061164}
  (\bibinfo {year} {2024})\BibitemShut {NoStop}%
\bibitem [{\citenamefont {Zhu}\ \emph {et~al.}(2025)\citenamefont {Zhu},
  \citenamefont {Ehlers}, \citenamefont {Nurdin},\ and\ \citenamefont
  {Soh}}]{zhu2025}%
  \BibitemOpen
  \bibfield  {author} {\bibinfo {author} {\bibfnamefont {C.}~\bibnamefont
  {Zhu}}, \bibinfo {author} {\bibfnamefont {P.~J.}\ \bibnamefont {Ehlers}},
  \bibinfo {author} {\bibfnamefont {H.~I.}\ \bibnamefont {Nurdin}},\ and\
  \bibinfo {author} {\bibfnamefont {D.}~\bibnamefont {Soh}},\ }\bibfield
  {title} {\bibinfo {title} {Practical few-atom quantum reservoir computing},\
  }\href {https://doi.org/10.1103/wsyq-jyxd} {\bibfield  {journal} {\bibinfo
  {journal} {Phys. Rev. Res.}\ }\textbf {\bibinfo {volume} {7}},\ \bibinfo
  {pages} {023290} (\bibinfo {year} {2025})}\BibitemShut {NoStop}%
\bibitem [{\citenamefont {Llodrà~et al}()}]{llodra_2025}%
  \BibitemOpen
  \bibfield  {author} {\bibinfo {author} {\bibfnamefont {G.}~\bibnamefont
  {Llodrà~et al}},\ }\href@noop {} {}\bibinfo {howpublished} {manuscript in
  preparation}\BibitemShut {NoStop}%
\bibitem [{\citenamefont {Carbonaro}\ \emph {et~al.}(1979)\citenamefont
  {Carbonaro}, \citenamefont {Compagno},\ and\ \citenamefont
  {Persico}}]{CARBONARO197997}%
  \BibitemOpen
  \bibfield  {author} {\bibinfo {author} {\bibfnamefont {P.}~\bibnamefont
  {Carbonaro}}, \bibinfo {author} {\bibfnamefont {G.}~\bibnamefont
  {Compagno}},\ and\ \bibinfo {author} {\bibfnamefont {F.}~\bibnamefont
  {Persico}},\ }\bibfield  {title} {\bibinfo {title} {Canonical dressing of
  atoms by intense radiation fields},\ }\href
  {https://doi.org/https://doi.org/10.1016/0375-9601(79)90445-6} {\bibfield
  {journal} {\bibinfo  {journal} {Physics Letters A}\ }\textbf {\bibinfo
  {volume} {73}},\ \bibinfo {pages} {97} (\bibinfo {year} {1979})}\BibitemShut
  {NoStop}%
\bibitem [{\citenamefont {Trushechkin}\ and\ \citenamefont
  {Volovich}(2016)}]{Trushechkin_2016}%
  \BibitemOpen
  \bibfield  {author} {\bibinfo {author} {\bibfnamefont {A.~S.}\ \bibnamefont
  {Trushechkin}}\ and\ \bibinfo {author} {\bibfnamefont {I.~V.}\ \bibnamefont
  {Volovich}},\ }\bibfield  {title} {\bibinfo {title} {Perturbative treatment
  of inter-site couplings in the local description of open quantum networks},\
  }\href {https://doi.org/10.1209/0295-5075/113/30005} {\bibfield  {journal}
  {\bibinfo  {journal} {Europhysics Letters}\ }\textbf {\bibinfo {volume}
  {113}},\ \bibinfo {pages} {30005} (\bibinfo {year} {2016})}\BibitemShut
  {NoStop}%
\bibitem [{\citenamefont {Cattaneo}\ \emph {et~al.}(2019)\citenamefont
  {Cattaneo}, \citenamefont {Giorgi}, \citenamefont {Maniscalco},\ and\
  \citenamefont {Zambrini}}]{Cattaneo_2019}%
  \BibitemOpen
  \bibfield  {author} {\bibinfo {author} {\bibfnamefont {M.}~\bibnamefont
  {Cattaneo}}, \bibinfo {author} {\bibfnamefont {G.~L.}\ \bibnamefont
  {Giorgi}}, \bibinfo {author} {\bibfnamefont {S.}~\bibnamefont {Maniscalco}},\
  and\ \bibinfo {author} {\bibfnamefont {R.}~\bibnamefont {Zambrini}},\
  }\bibfield  {title} {\bibinfo {title} {Local versus global master equation
  with common and separate baths: superiority of the global approach in partial
  secular approximation},\ }\href {https://doi.org/10.1088/1367-2630/ab54ac}
  {\bibfield  {journal} {\bibinfo  {journal} {New Journal of Physics}\ }\textbf
  {\bibinfo {volume} {21}},\ \bibinfo {pages} {113045} (\bibinfo {year}
  {2019})}\BibitemShut {NoStop}%
\bibitem [{\citenamefont {Vaaranta}\ \emph {et~al.}(2025)\citenamefont
  {Vaaranta}, \citenamefont {Cattaneo},\ and\ \citenamefont
  {Muratore-Ginanneschi}}]{cattaneo_2025}%
  \BibitemOpen
  \bibfield  {author} {\bibinfo {author} {\bibfnamefont {A.}~\bibnamefont
  {Vaaranta}}, \bibinfo {author} {\bibfnamefont {M.}~\bibnamefont {Cattaneo}},\
  and\ \bibinfo {author} {\bibfnamefont {P.}~\bibnamefont
  {Muratore-Ginanneschi}},\ }\bibfield  {title} {\bibinfo {title} {Analytical
  solution of the open dispersive jaynes-cummings model and perturbative
  analytical solution of the open quantum rabi model},\ }\href
  {https://doi.org/10.1103/PhysRevA.111.053717} {\bibfield  {journal} {\bibinfo
   {journal} {Phys. Rev. A}\ }\textbf {\bibinfo {volume} {111}},\ \bibinfo
  {pages} {053717} (\bibinfo {year} {2025})}\BibitemShut {NoStop}%
\bibitem [{\citenamefont {Scala}\ \emph
  {et~al.}(2007{\natexlab{a}})\citenamefont {Scala}, \citenamefont {Militello},
  \citenamefont {Messina}, \citenamefont {Piilo},\ and\ \citenamefont
  {Maniscalco}}]{scala_2007}%
  \BibitemOpen
  \bibfield  {author} {\bibinfo {author} {\bibfnamefont {M.}~\bibnamefont
  {Scala}}, \bibinfo {author} {\bibfnamefont {B.}~\bibnamefont {Militello}},
  \bibinfo {author} {\bibfnamefont {A.}~\bibnamefont {Messina}}, \bibinfo
  {author} {\bibfnamefont {J.}~\bibnamefont {Piilo}},\ and\ \bibinfo {author}
  {\bibfnamefont {S.}~\bibnamefont {Maniscalco}},\ }\bibfield  {title}
  {\bibinfo {title} {Microscopic derivation of the jaynes-cummings model with
  cavity losses},\ }\href {https://doi.org/10.1103/PhysRevA.75.013811}
  {\bibfield  {journal} {\bibinfo  {journal} {Phys. Rev. A}\ }\textbf {\bibinfo
  {volume} {75}},\ \bibinfo {pages} {013811} (\bibinfo {year}
  {2007}{\natexlab{a}})}\BibitemShut {NoStop}%
\bibitem [{\citenamefont {Scala}\ \emph
  {et~al.}(2007{\natexlab{b}})\citenamefont {Scala}, \citenamefont {Militello},
  \citenamefont {Messina}, \citenamefont {Maniscalco}, \citenamefont {Piilo},\
  and\ \citenamefont {Suominen}}]{Scala_2007_2}%
  \BibitemOpen
  \bibfield  {author} {\bibinfo {author} {\bibfnamefont {M.}~\bibnamefont
  {Scala}}, \bibinfo {author} {\bibfnamefont {B.}~\bibnamefont {Militello}},
  \bibinfo {author} {\bibfnamefont {A.}~\bibnamefont {Messina}}, \bibinfo
  {author} {\bibfnamefont {S.}~\bibnamefont {Maniscalco}}, \bibinfo {author}
  {\bibfnamefont {J.}~\bibnamefont {Piilo}},\ and\ \bibinfo {author}
  {\bibfnamefont {K.-A.}\ \bibnamefont {Suominen}},\ }\bibfield  {title}
  {\bibinfo {title} {Cavity losses for the dissipative jaynes–cummings
  hamiltonian beyond rotating wave approximation},\ }\href
  {https://doi.org/10.1088/1751-8113/40/48/015} {\bibfield  {journal} {\bibinfo
   {journal} {Journal of Physics A: Mathematical and Theoretical}\ }\textbf
  {\bibinfo {volume} {40}},\ \bibinfo {pages} {14527} (\bibinfo {year}
  {2007}{\natexlab{b}})}\BibitemShut {NoStop}%
\bibitem [{\citenamefont {Jaeger}(2001)}]{jaeger:techreport2001}%
  \BibitemOpen
  \bibfield  {author} {\bibinfo {author} {\bibfnamefont {H.}~\bibnamefont
  {Jaeger}},\ }\href
  {http://www.faculty.jacobs-university.de/hjaeger/pubs/EchoStatesTechRep.pdf}
  {\emph {\bibinfo {title} {The "echo state" approach to analysing and training
  recurrent neural networks}}},\ \bibinfo {type} {GMD Report}\ \bibinfo
  {number} {148}\ (\bibinfo  {institution} {GMD - German National Research
  Institute for Computer Science},\ \bibinfo {year} {2001})\BibitemShut
  {NoStop}%
\bibitem [{\citenamefont {Yildiz}\ \emph {et~al.}(2012)\citenamefont {Yildiz},
  \citenamefont {Jaeger},\ and\ \citenamefont {Kiebel}}]{jaeger_2012}%
  \BibitemOpen
  \bibfield  {author} {\bibinfo {author} {\bibfnamefont {I.~B.}\ \bibnamefont
  {Yildiz}}, \bibinfo {author} {\bibfnamefont {H.}~\bibnamefont {Jaeger}},\
  and\ \bibinfo {author} {\bibfnamefont {S.~J.}\ \bibnamefont {Kiebel}},\
  }\bibfield  {title} {\bibinfo {title} {Re-visiting the echo state property},\
  }\href@noop {} {\bibfield  {journal} {\bibinfo  {journal} {Neural Networks}\
  }\textbf {\bibinfo {volume} {35}},\ \bibinfo {pages} {1–9} (\bibinfo {year}
  {2012})}\BibitemShut {NoStop}%
\bibitem [{\citenamefont {Boyd}\ and\ \citenamefont {Chua}(1985)}]{boyd_85}%
  \BibitemOpen
  \bibfield  {author} {\bibinfo {author} {\bibfnamefont {S.}~\bibnamefont
  {Boyd}}\ and\ \bibinfo {author} {\bibfnamefont {L.}~\bibnamefont {Chua}},\
  }\bibfield  {title} {\bibinfo {title} {Fading memory and the problem of
  approximating nonlinear operators with volterra series},\ }\href
  {https://doi.org/10.1109/TCS.1985.1085649} {\bibfield  {journal} {\bibinfo
  {journal} {IEEE Transactions on Circuits and Systems}\ }\textbf {\bibinfo
  {volume} {32}},\ \bibinfo {pages} {1150} (\bibinfo {year}
  {1985})}\BibitemShut {NoStop}%
\bibitem [{\citenamefont {Hahto}\ and\ \citenamefont
  {Nokkala}(2025)}]{hahto2025}%
  \BibitemOpen
  \bibfield  {author} {\bibinfo {author} {\bibfnamefont {M.}~\bibnamefont
  {Hahto}}\ and\ \bibinfo {author} {\bibfnamefont {J.}~\bibnamefont
  {Nokkala}},\ }\bibfield  {title} {\bibinfo {title} {Smarter usage of
  measurement statistics can greatly improve continuous variable quantum
  reservoir computing},\ }\href {https://doi.org/10.1088/1367-2630/ae06c4}
  {\bibfield  {journal} {\bibinfo  {journal} {New Journal of Physics}\ }\textbf
  {\bibinfo {volume} {27}},\ \bibinfo {pages} {094510} (\bibinfo {year}
  {2025})}\BibitemShut {NoStop}%
\bibitem [{\citenamefont {Alsing}\ \emph {et~al.}(1992)\citenamefont {Alsing},
  \citenamefont {Guo},\ and\ \citenamefont {Carmichael}}]{alsing_1992}%
  \BibitemOpen
  \bibfield  {author} {\bibinfo {author} {\bibfnamefont {P.}~\bibnamefont
  {Alsing}}, \bibinfo {author} {\bibfnamefont {D.-S.}\ \bibnamefont {Guo}},\
  and\ \bibinfo {author} {\bibfnamefont {H.~J.}\ \bibnamefont {Carmichael}},\
  }\bibfield  {title} {\bibinfo {title} {Dynamic stark effect for the
  jaynes-cummings system},\ }\href {https://doi.org/10.1103/PhysRevA.45.5135}
  {\bibfield  {journal} {\bibinfo  {journal} {Phys. Rev. A}\ }\textbf {\bibinfo
  {volume} {45}},\ \bibinfo {pages} {5135} (\bibinfo {year}
  {1992})}\BibitemShut {NoStop}%
\bibitem [{\citenamefont {Mujal}\ \emph
  {et~al.}(2021{\natexlab{b}})\citenamefont {Mujal}, \citenamefont {Nokkala},
  \citenamefont {Martínez-Peña}, \citenamefont {Giorgi}, \citenamefont
  {Soriano},\ and\ \citenamefont {Zambrini}}]{mujal_2021_b}%
  \BibitemOpen
  \bibfield  {author} {\bibinfo {author} {\bibfnamefont {P.}~\bibnamefont
  {Mujal}}, \bibinfo {author} {\bibfnamefont {J.}~\bibnamefont {Nokkala}},
  \bibinfo {author} {\bibfnamefont {R.}~\bibnamefont {Martínez-Peña}},
  \bibinfo {author} {\bibfnamefont {G.~L.}\ \bibnamefont {Giorgi}}, \bibinfo
  {author} {\bibfnamefont {M.~C.}\ \bibnamefont {Soriano}},\ and\ \bibinfo
  {author} {\bibfnamefont {R.}~\bibnamefont {Zambrini}},\ }\bibfield  {title}
  {\bibinfo {title} {Analytical evidence of nonlinearity in qubits and
  continuous-variable quantum reservoir computing},\ }\href
  {https://doi.org/10.1088/2632-072X/ac340e} {\bibfield  {journal} {\bibinfo
  {journal} {Journal of Physics: Complexity}\ }\textbf {\bibinfo {volume}
  {2}},\ \bibinfo {pages} {045008} (\bibinfo {year}
  {2021}{\natexlab{b}})}\BibitemShut {NoStop}%
\bibitem [{\citenamefont {Bertschinger}\ \emph {et~al.}(2004)\citenamefont
  {Bertschinger}, \citenamefont {Natschl\"{a}ger},\ and\ \citenamefont
  {Legenstein}}]{NIPS2004}%
  \BibitemOpen
  \bibfield  {author} {\bibinfo {author} {\bibfnamefont {N.}~\bibnamefont
  {Bertschinger}}, \bibinfo {author} {\bibfnamefont {T.}~\bibnamefont
  {Natschl\"{a}ger}},\ and\ \bibinfo {author} {\bibfnamefont {R.}~\bibnamefont
  {Legenstein}},\ }\bibfield  {title} {\bibinfo {title} {At the edge of chaos:
  Real-time computations and self-organized criticality in recurrent neural
  networks},\ }in\ \href
  {https://proceedings.neurips.cc/paper_files/paper/2004/file/f8da71e562ff44a2bc7edf3578c593da-Paper.pdf}
  {\emph {\bibinfo {booktitle} {Advances in Neural Information Processing
  Systems}}},\ Vol.~\bibinfo {volume} {17},\ \bibinfo {editor} {edited by\
  \bibinfo {editor} {\bibfnamefont {L.}~\bibnamefont {Saul}}, \bibinfo {editor}
  {\bibfnamefont {Y.}~\bibnamefont {Weiss}},\ and\ \bibinfo {editor}
  {\bibfnamefont {L.}~\bibnamefont {Bottou}}}\ (\bibinfo  {publisher} {MIT
  Press},\ \bibinfo {year} {2004})\BibitemShut {NoStop}%
\bibitem [{\citenamefont {Mackey}\ and\ \citenamefont
  {Glass}(1977)}]{mackey_glass_77}%
  \BibitemOpen
  \bibfield  {author} {\bibinfo {author} {\bibfnamefont {M.~C.}\ \bibnamefont
  {Mackey}}\ and\ \bibinfo {author} {\bibfnamefont {L.}~\bibnamefont {Glass}},\
  }\bibfield  {title} {\bibinfo {title} {Oscillation and chaos in physiological
  control systems},\ }\href {https://doi.org/10.1126/science.267326} {\bibfield
   {journal} {\bibinfo  {journal} {Science}\ }\textbf {\bibinfo {volume}
  {197}},\ \bibinfo {pages} {287} (\bibinfo {year} {1977})}\BibitemShut
  {NoStop}%
\bibitem [{\citenamefont {Llodrà}\ \emph {et~al.}(2025)\citenamefont
  {Llodrà}, \citenamefont {Mujal}, \citenamefont {Zambrini},\ and\
  \citenamefont {Giorgi}}]{llodra2024}%
  \BibitemOpen
  \bibfield  {author} {\bibinfo {author} {\bibfnamefont {G.}~\bibnamefont
  {Llodrà}}, \bibinfo {author} {\bibfnamefont {P.}~\bibnamefont {Mujal}},
  \bibinfo {author} {\bibfnamefont {R.}~\bibnamefont {Zambrini}},\ and\
  \bibinfo {author} {\bibfnamefont {G.~L.}\ \bibnamefont {Giorgi}},\ }\bibfield
   {title} {\bibinfo {title} {Quantum reservoir computing in atomic lattices},\
  }\href {https://doi.org/https://doi.org/10.1016/j.chaos.2025.116289}
  {\bibfield  {journal} {\bibinfo  {journal} {Chaos, Solitons \& Fractals}\
  }\textbf {\bibinfo {volume} {195}},\ \bibinfo {pages} {116289} (\bibinfo
  {year} {2025})}\BibitemShut {NoStop}%
\bibitem [{\citenamefont {Reed}\ \emph {et~al.}(2010)\citenamefont {Reed},
  \citenamefont {Johnson}, \citenamefont {Houck}, \citenamefont {DiCarlo},
  \citenamefont {Chow}, \citenamefont {Schuster}, \citenamefont {Frunzio},\
  and\ \citenamefont {Schoelkopf}}]{reed_purcell1_2010}%
  \BibitemOpen
  \bibfield  {author} {\bibinfo {author} {\bibfnamefont {M.~D.}\ \bibnamefont
  {Reed}}, \bibinfo {author} {\bibfnamefont {B.~R.}\ \bibnamefont {Johnson}},
  \bibinfo {author} {\bibfnamefont {A.~A.}\ \bibnamefont {Houck}}, \bibinfo
  {author} {\bibfnamefont {L.}~\bibnamefont {DiCarlo}}, \bibinfo {author}
  {\bibfnamefont {J.~M.}\ \bibnamefont {Chow}}, \bibinfo {author}
  {\bibfnamefont {D.~I.}\ \bibnamefont {Schuster}}, \bibinfo {author}
  {\bibfnamefont {L.}~\bibnamefont {Frunzio}},\ and\ \bibinfo {author}
  {\bibfnamefont {R.~J.}\ \bibnamefont {Schoelkopf}},\ }\bibfield  {title}
  {\bibinfo {title} {Fast reset and suppressing spontaneous emission of a
  superconducting qubit},\ }\href {https://doi.org/10.1063/1.3435463}
  {\bibfield  {journal} {\bibinfo  {journal} {Applied Physics Letters}\
  }\textbf {\bibinfo {volume} {96}},\ \bibinfo {pages} {203110} (\bibinfo
  {year} {2010})}\BibitemShut {NoStop}%
\bibitem [{\citenamefont {Bronn}\ \emph {et~al.}(2015)\citenamefont {Bronn},
  \citenamefont {Liu}, \citenamefont {Hertzberg}, \citenamefont {Córcoles},
  \citenamefont {Houck}, \citenamefont {Gambetta},\ and\ \citenamefont
  {Chow}}]{bronn_2015}%
  \BibitemOpen
  \bibfield  {author} {\bibinfo {author} {\bibfnamefont {N.~T.}\ \bibnamefont
  {Bronn}}, \bibinfo {author} {\bibfnamefont {Y.}~\bibnamefont {Liu}}, \bibinfo
  {author} {\bibfnamefont {J.~B.}\ \bibnamefont {Hertzberg}}, \bibinfo {author}
  {\bibfnamefont {A.~D.}\ \bibnamefont {Córcoles}}, \bibinfo {author}
  {\bibfnamefont {A.~A.}\ \bibnamefont {Houck}}, \bibinfo {author}
  {\bibfnamefont {J.~M.}\ \bibnamefont {Gambetta}},\ and\ \bibinfo {author}
  {\bibfnamefont {J.~M.}\ \bibnamefont {Chow}},\ }\bibfield  {title} {\bibinfo
  {title} {Broadband filters for abatement of spontaneous emission in circuit
  quantum electrodynamics},\ }\href@noop {} {\bibfield  {journal} {\bibinfo
  {journal} {Applied Physics Letters}\ }\textbf {\bibinfo {volume} {107}},\
  \bibinfo {pages} {172601} (\bibinfo {year} {2015})}\BibitemShut {NoStop}%
\bibitem [{\citenamefont {Laha}\ \emph {et~al.}(2024)\citenamefont {Laha},
  \citenamefont {Yasir},\ and\ \citenamefont {van Loock}}]{laha_2024}%
  \BibitemOpen
  \bibfield  {author} {\bibinfo {author} {\bibfnamefont {P.}~\bibnamefont
  {Laha}}, \bibinfo {author} {\bibfnamefont {P.~A.~A.}\ \bibnamefont {Yasir}},\
  and\ \bibinfo {author} {\bibfnamefont {P.}~\bibnamefont {van Loock}},\
  }\bibfield  {title} {\bibinfo {title} {Genuine non-gaussian entanglement of
  light and quantum coherence for an atom from noisy multiphoton spin-boson
  interactions},\ }\href {https://doi.org/10.1103/PhysRevResearch.6.033302}
  {\bibfield  {journal} {\bibinfo  {journal} {Phys. Rev. Res.}\ }\textbf
  {\bibinfo {volume} {6}},\ \bibinfo {pages} {033302} (\bibinfo {year}
  {2024})}\BibitemShut {NoStop}%
\bibitem [{\citenamefont {Johansson}\ \emph {et~al.}(2013)\citenamefont
  {Johansson}, \citenamefont {Nation},\ and\ \citenamefont {Nori}}]{qutip1}%
  \BibitemOpen
  \bibfield  {author} {\bibinfo {author} {\bibfnamefont {J.}~\bibnamefont
  {Johansson}}, \bibinfo {author} {\bibfnamefont {P.}~\bibnamefont {Nation}},\
  and\ \bibinfo {author} {\bibfnamefont {F.}~\bibnamefont {Nori}},\ }\bibfield
  {title} {\bibinfo {title} {Qutip 2: A python framework for the dynamics of
  open quantum systems},\ }\href
  {https://doi.org/https://doi.org/10.1016/j.cpc.2012.11.019} {\bibfield
  {journal} {\bibinfo  {journal} {Computer Physics Communications}\ }\textbf
  {\bibinfo {volume} {184}},\ \bibinfo {pages} {1234} (\bibinfo {year}
  {2013})}\BibitemShut {NoStop}%
\bibitem [{\citenamefont {Johansson}\ \emph {et~al.}(2012)\citenamefont
  {Johansson}, \citenamefont {Nation},\ and\ \citenamefont {Nori}}]{qutip2}%
  \BibitemOpen
  \bibfield  {author} {\bibinfo {author} {\bibfnamefont {J.}~\bibnamefont
  {Johansson}}, \bibinfo {author} {\bibfnamefont {P.}~\bibnamefont {Nation}},\
  and\ \bibinfo {author} {\bibfnamefont {F.}~\bibnamefont {Nori}},\ }\bibfield
  {title} {\bibinfo {title} {Qutip: An open-source python framework for the
  dynamics of open quantum systems},\ }\href
  {https://doi.org/https://doi.org/10.1016/j.cpc.2012.02.021} {\bibfield
  {journal} {\bibinfo  {journal} {Computer Physics Communications}\ }\textbf
  {\bibinfo {volume} {183}},\ \bibinfo {pages} {1760} (\bibinfo {year}
  {2012})}\BibitemShut {NoStop}%
\bibitem [{\citenamefont {Lambert}\ \emph {et~al.}(2024)\citenamefont
  {Lambert}, \citenamefont {Giguère}, \citenamefont {Menczel}, \citenamefont
  {Li}, \citenamefont {Hopf}, \citenamefont {Suárez}, \citenamefont {Gali},
  \citenamefont {Lishman}, \citenamefont {Gadhvi}, \citenamefont {Agarwal},
  \citenamefont {Galicia}, \citenamefont {Shammah}, \citenamefont {Nation},
  \citenamefont {Johansson}, \citenamefont {Ahmed}, \citenamefont {Cross},
  \citenamefont {Pitchford},\ and\ \citenamefont {Nori}}]{qutip3}%
  \BibitemOpen
  \bibfield  {author} {\bibinfo {author} {\bibfnamefont {N.}~\bibnamefont
  {Lambert}}, \bibinfo {author} {\bibfnamefont {E.}~\bibnamefont {Giguère}},
  \bibinfo {author} {\bibfnamefont {P.}~\bibnamefont {Menczel}}, \bibinfo
  {author} {\bibfnamefont {B.}~\bibnamefont {Li}}, \bibinfo {author}
  {\bibfnamefont {P.}~\bibnamefont {Hopf}}, \bibinfo {author} {\bibfnamefont
  {G.}~\bibnamefont {Suárez}}, \bibinfo {author} {\bibfnamefont
  {M.}~\bibnamefont {Gali}}, \bibinfo {author} {\bibfnamefont {J.}~\bibnamefont
  {Lishman}}, \bibinfo {author} {\bibfnamefont {R.}~\bibnamefont {Gadhvi}},
  \bibinfo {author} {\bibfnamefont {R.}~\bibnamefont {Agarwal}}, \bibinfo
  {author} {\bibfnamefont {A.}~\bibnamefont {Galicia}}, \bibinfo {author}
  {\bibfnamefont {N.}~\bibnamefont {Shammah}}, \bibinfo {author} {\bibfnamefont
  {P.}~\bibnamefont {Nation}}, \bibinfo {author} {\bibfnamefont {J.~R.}\
  \bibnamefont {Johansson}}, \bibinfo {author} {\bibfnamefont {S.}~\bibnamefont
  {Ahmed}}, \bibinfo {author} {\bibfnamefont {S.}~\bibnamefont {Cross}},
  \bibinfo {author} {\bibfnamefont {A.}~\bibnamefont {Pitchford}},\ and\
  \bibinfo {author} {\bibfnamefont {F.}~\bibnamefont {Nori}},\ }\href
  {https://arxiv.org/abs/2412.04705} {\bibinfo {title} {Qutip 5: The quantum
  toolbox in python}} (\bibinfo {year} {2024}),\ \Eprint
  {https://arxiv.org/abs/2412.04705} {arXiv:2412.04705 [quant-ph]} \BibitemShut
  {NoStop}%
\end{thebibliography}%
\end{document}